\documentclass[final,10pt,twocolumn]{IEEEtran}
\usepackage{makeidx}         % allows index generation
\usepackage{graphicx}        % standard LaTeX graphics tool
\usepackage[bottom]{footmisc}% places footnotes at page bottom
\usepackage{ifthen}
\usepackage{subfigure}
\usepackage[usenames,dvipsnames]{color}
\usepackage{algorithm, algorithmicx, algpseudocode}
\usepackage{amsmath, amssymb}
\usepackage{setspace}
\usepackage{url}
\usepackage{balance}
\usepackage{xcolor}
\usepackage{colortbl}
\usepackage{graphics} % for pdf, bitmapped graphics files
\usepackage{graphicx} % for pdf, bitmapped graphics files
\usepackage{epsfig} % for postscript graphics files
\usepackage{amsfonts}
\usepackage{amsthm}
\usepackage{verbatim}
\usepackage{comment}
\usepackage{cuted}
\usepackage{lipsum}
\usepackage[colorlinks=true,linkcolor=magenta,citecolor=blue,urlcolor=cyan,filecolor=red]{hyperref}
\DeclareGraphicsExtensions{.pdf,.png,.jpg,.eps}
\def\ba{\begin{array}}
\def\ea{\end{array}}
\newcommand{\beq}{\begin{equation}}
\newcommand{\eeq}{\end{equation}}
\newcommand{\bq}{\begin{eqnarray}}
\newcommand{\eq}{\end{eqnarray}}
\newcommand{\bqn}{\begin{eqnarray*}}
\newcommand{\eqn}{\end{eqnarray*}}
\newcommand{\bee}{\begin{enumerate}}
\newcommand{\eee}{\end{enumerate}}
\newcommand{\bi}{\begin{itemize}}
\newcommand{\ei}{\end{itemize}}

\newcommand{\barr}{\begin{array}}
\newcommand{\earr}{\begin{array}}
\newcommand{\bseq}{\begin{subequations}}
\newcommand{\eseq}{\end{subequations}}
\newcommand{\bmat}{\begin{bmatrix}}
\newcommand{\emat}{\end{bmatrix}}
\newcommand{\beqs}{\begin{equation*}}
\newcommand{\eeqs}{\end{equation*}}
\newcommand{\bali}{\begin{aligned}}
\newcommand{\eali}{\end{aligned}}

\newcommand{\diag}{\operatorname{diag}}

\newboolean{showcomments}
\setboolean{showcomments}{true}
\newcommand{\ye}[1]{\ifthenelse{\boolean{showcomments}}
{ \textcolor{red}{(Ye:  #1)}}{}}
\newcommand{\slow}[1]{\ifthenelse{\boolean{showcomments}}
{ \textcolor{red}{(Steven:  #1)}}{}}

\newtheorem{proposition}{Proposition}
\newtheorem{assumption}{Assumption}
\newtheorem{theorem}{Theorem}

\newtheorem{lemma}{Lemma}
\newtheorem{corollary}{Corollary}
\newtheorem{remark}{Remark}
\newtheorem{example}{Example}
\begin{document}
\title{Inverse Power Flow Problem}
\author{Ye Yuan,~\IEEEmembership{Senior Member,~IEEE,} Steven H. Low,~\IEEEmembership{Fellow,~IEEE,}\\
Omid Ardakanian,~\IEEEmembership{Member,~IEEE,} Claire J. Tomlin,~\IEEEmembership{Fellow,~IEEE}
\thanks{Ye Yuan is with School of Artificial Intelligence and Automation, Huazhong University of Science and Technology, Wuhan, China. Steven H. Low is with Department of Computer and Mathematical Sciences and Electrical Engineering, California Institute of Technology, Pasadena, United States. Omid Ardakanian is with Department of Computing Science, University of Alberta, Edmonton, Canada. Claire J. Tomlin is with Department of Electrical Engineering and Computer Sciences, University of California, Berkeley, United States. Correspondence to {\tt slow@caltech.edu}.}
\thanks{This version corrects the proof to Proposition~\ref{prop:samehiddennode} and adds details about constructing the admittance matrix of the overall network from its components.} 
%\thanks{This work was supported in part by NSF under CPS:ActionWebs (CNS-0931843) and CPS:FORCES (CNS1239166), by NASA under grants NNX12AR18A and UCSCMCA-14-022 (UARC), by ONR under grants N00014-12-1-0609, N000141310341 (Embedded Humans MURI), and MIT\_5710002646 (SMARTS MURI), and by AFOSR under grants UPenn-FA9550-10-1-0567 (CHASE MURI) and the SURE project.}
}
%\markboth{IEEE Transactions On Automatic Control, Vol. XX, No. Y, 2017}{Yuan et. al.: Using the style file IEEEtran.sty}
\maketitle
\begin{abstract}
This paper formulates an inverse power flow problem which is to infer a nodal admittance matrix (hence the network structure of a power system) from voltage and current phasors measured at a number of buses. We show that the admittance matrix can be uniquely identified from a sequence of measurements corresponding to different steady states when every node in the system is equipped with a measurement device, and a Kron-reduced admittance matrix can be determined even if some nodes in the system are not monitored (hidden nodes). Furthermore, we propose effective algorithms based on graph theory to uncover the actual admittance matrix of radial systems with hidden nodes. We provide theoretical guarantees for the recovered admittance matrix and demonstrate that the actual admittance matrix can be fully recovered even from the Kron-reduced admittance matrix under some mild assumptions. Simulations on standard test systems confirm that these algorithms are capable of providing accurate estimates of the admittance matrix from noisy sensor data.
\end{abstract}
\begin{IEEEkeywords}
Inverse Power Flow Problem, System Identification, Kron Reduction, Phasor Measurement Units.
\end{IEEEkeywords}
%
%\ifCLASSOPTIONpeerreview
%\begin{center} \bfseries EDICS Category: 3-BBND \end{center}
%\fi
% \tableofcontents
% \vspace{0.5in}
%----------------------------------------------------------------------------------------
\section{Introduction}
%----------------------------------------------------------------------------------------t
\IEEEPARstart{T}{he} power industry has witnessed profound changes in recent years. 
These changes are mostly driven by the widespread adoption of distributed energy resources (DER), 
active participation of customers in emerging energy markets, and 
rapid deployment of measurement, communication, and control infrastructure 
resulting in an unprecedented level of visibility and controllability, especially for distribution grids.
%which traditionally had very little instrumentation beyond the substation for monitoring purposes
Despite the increased amount of uncertainty,
these changes offer opportunities for system operators to improve power system stability and efficiency
by leveraging advanced optimization and control techniques. Most of these techniques require the knowledge of the network topology in real time. 
%which has been traditionally lacked in many cases due to the limited visibility into the state of the power system.
%and the operation of new control devices 
%that may continually alter the underlying network topology

The inverse power flow (IPF) problem we define in this paper concerns the estimation 
of the nodal admittance matrix from synchronized measurements of voltage and current phasors (i.e., magnitudes and phase angles)
which can be obtained from phasor measurement units (PMUs) %~\cite{naspi} 
or conventional supervisory control and data acquisition (SCADA) technology.
The IPF problem underlies several crucial smart grid applications,
affecting real-time system operation and long-term planning, 
the most important of which are:

{\it 1) State Estimation} \cite{dehghanpour2018survey} combines the knowledge of the admittance matrix with a set of known state-variables to determine the unknown ones, 
e.g., voltage magnitude and phase angle of unobserved buses, 
thereby building a real-time model of the network for control.
%and economical decisions and to uncover potential operational problems.
%The state estimation can be used to compensate for the transducer error, the time skewness of telemetry data, and the absence of monitoring devices at particular nodes.

{\it 2) Optimization \& Control} \cite{le2017plug} techniques determine 
a sequence of operations that can transition the power system 
from one steady state to another steady state that meets certain stability and efficiency targets.
%These techniques typically require the knowledge of the network topology
% {\color{blue}and} information about the state of the power system.

{\it 3) Event Detection} \cite{Ardakanian2019} concerns identifying
faults, line outages, and other critical events, 
such as transformer tap changes, capacitor and switching operations
from changes in the real-time network model.
%The idea of fault detection is to detect the occurrence of a fault when it happens 
%while fault location concerns pinpointing the fault to within a small section of a distribution feeder.
%The idea of fault progonization is to recognize the dangerous condition 
%but does not draw sufficient current to trip a protective device. 

{\it 4) Cybersecurity} \cite{zhuang2019false} is the problem of identifying
the potential vulnerabilities of a power system 
and designing strategies to protect it from the potential cyber attacks
using telemetry data along with information about its topology.
%\end{itemize}

In this paper, we lay out a theoretical framework for the IPF problem. 
Using the bus injection model (BIM), 
we propose efficient algorithms to identify the admittance matrix. %and analyze their accuracy. 
In particular, we show that when the system has no hidden nodes,
the admittance matrix can be uniquely identified 
from a sequence of complex voltage and current measurements 
corresponding to different steady states.
Should there be some hidden nodes in the network,
we show that a reduced admittance matrix (from Kron reduction \cite{kron}) can be determined;
we develop a method based on graph decomposition, 
maximal clique searching and composition for identifying the admittance matrix 
of the original system for radial networks.
%b) for mesh networks, an algorithm based on low rank and sparse matrix decomposition \cite{Yuan2016inverse}.
%Furthermore, a convex relaxation has been proposed to obtain a computationally efficient algorithm. 
Power flow simulations are performed on the IEEE 14-bus system to illustrate the theoretical results
and evaluate their sensitivity to measurement noise introduced by transducers.
%\footnote{Simulation results can be found in the arxiv version of this paper due to page limit \cite{yuan2016inverse}: \url{https://arxiv.org/abs/1610.06631}.}.

The paper is outlined as follows: after surveying related work in Section~\ref{sec:related},
we formulate the IPF problem and 
propose a solution for the case that the system is fully observable in Section~\ref{sec:nohidden}. 
%We study the fault detection problem in Section~\ref{sec:fd}
When the system has hidden nodes, we propose efficient algorithms to solve the IPF problem 
for radial networks with theoretical guarantee in Section~\ref{sec:hidden}. We evaluate the identification accuracy of the proposed algorithms in Section~\ref{sec:eval} and conclude the paper by presenting directions for future work in Section~\ref{sec:discussion}.
%{\color{blue}We evaluate the identification accuracy of the proposed algorithms in Section~\ref{sec:eval}} and conclude the paper by presenting directions for future work in Section~\ref{sec:discussion}. 
%Due to space limit, solution to the IPF problem for mesh networks and full simulation and experimental validation are presented in \cite{Yuan2016inverse}.
%----------------------------------------------------------------------------------------
\section{Related Work}\label{sec:related}
%----------------------------------------------------------------------------------------

The availability of high-precision, high-sample-rate measurements of transmission and distribution networks in recent years has given impetus to research on topology and admittance identification.  

The IPF problem that identifies both topology and admittance matrix has been studied extensively
in transmission networks~\cite{LiPoorScaglione2013,kekatos,babakmehr,yuan2019data}
as well as distribution grids~\cite{deka,liao} 
using single-phase a.c. and d.c. power flow models.
For example, the topology identification problem is cast as 
a sparse subspace learning problem in~\cite{LiPoorScaglione2013} 
and an efficient algorithm is proposed
to estimate the admittance matrix of the underlying power system
from the measured power injection of different buses.
In~\cite{urban}, the topology of an urban (mesh) distribution network 
is inferred from voltage magnitude, and real/reactive power
measurements carried out by smart meters at the end-nodes. 
A graphical model is built to describe the probabilistic relationships 
between different voltage measurements using lasso.
In~\cite{deka} a graphical learning based approach is developed
to estimate the radial grid topology from nodal voltage measurements.
The learning algorithm is based on conditional independence tests 
for continuous variables over chordal graphs.
An efficient algorithm for topology identification of a power system 
is also proposed in~\cite{babakmehr}
drawing on ideas from compressive sensing and graph theory. 
The authors assume that power and phase angle measurements are available for all nodes.

Different algorithms have been developed in the literature to make this inference, using various
techniques such as weighted least square, maximum-likelihood/maximum-a-posteriori
estimation, minimum spanning tree, sparse recovery, Lasso/Group Lasso, blind identification, 
quickest change detection theory, as well as graphical model learning.  There are three limitations in the current literature that we propose overcome. First, most of the literature focuses on topology identification or change detection, but there is not much work on joint topology and parameter identification, with notable exceptions of
\cite{LiPoorScaglione2013, YuRajagopal2017,Deka2019}. 
Second, most papers require measurements at every node in the network, with the exceptions of
\cite{Zhu2012, Deka2019, Deka2018-TCNS, cavraro2018graph, cavraro2017voltage, li2021distribution}. In particular, \cite{Deka2019} learns the topology with parameters from a stochastic perspective, the true topology can only be found in probability, even when the number of samples is large; \cite{cavraro2018graph, cavraro2017voltage} assume that perturbed data are available (therefore a special inverter is assumed) to identify the network, which could be strong in practice; \cite{li2021distribution} proposed a method based on recursive grouping to estimate the topology and branch impedance for networks that may have hidden nodes, however, without guarantee.

%To our knowledge, no related work addresses the topology identification problem
%when there are hidden states, and this paper is the first one that solves
%such an identification problem with theoretical guarantees.
%
%{\bf Event detection --} 
%Designing algorithms that can quickly detect, identify, and isolate 
%high impedance faults using PMU data is of great importance~\cite{meier}. 
%The state-of-the-art approach is to detect a fault or an outage
%using either an online algorithm,
%which relies on a relatively small number of observations~\cite{dynamicfd,le,sharon,outagedet,topologydet,switchedid}, or
%an offline algorithm
%%which requires a significant amount of data prior to the detection
%~\cite{cavrado,huang2015dynamic}.
%Our approach is more general than these approaches as
%it is capable of identifying \emph{any} type of power system events 
%that induce a change in the admittance matrix of the underlying network 
%in quasi real-time.

%%{\bf Network tomography --} 
%%The topology identification problem with hidden states
%%is similar to electrical impedance tomography,
%%in which the conductivity of a part of the body is inferred from 
%%simultaneous measurements of currents and voltages at the boundary
%%(surface electrode measurements),
%%as both problems concern inferring $Y$ from $\bar{Y}$. 
%%It is known that the tomography problem is feasible only in
%%highly symmetric networks~\cite{inverse1,inverse2}.

%----------------------------------------------------------------------------------------
%----------------------------------------------------------------------------------------
\section{IPF without hidden nodes}\label{sec:nohidden}

%\slow{Edit later. Integrate and summarize this section.}

In this section we study the IPF problem when voltage and current phasor measurements 
are available at every bus in the system.
We formulate the identification problem as a constrained least squares problem
and then convert it to an equivalent unconstrained least squares problem. %which can be analytically solved. 
%The drawback of this method is the requirement of large number of samples and potentially leading to numerical problems. 
%Next, we utilize the prior knowledge that the number of neighbors of every node is much less than the total number of nodes and formulate the identification problem as a constrained $l_0$ problem. We further relax it to a unconstrained $l_1$ optimization and study the exactness property of these formulated optimizations. 
%From the power flow equation or equivalently the Kirchhoff's law, 
%we consider the inverse problem that identifies the model (and corresponding parameters) 
%from data. 
%\slow{Edit later.}
We note that $Y$ has a certain structure that can be exploited when solving the IPF problem---
(a) $Y$ is a symmetric but not Hermitian complex matrix (i.e., $Y\in \mathbb{S}^N$)
and (b) $Y$ encodes the topology of a connected graph (or a connected tree for radial networks).

\subsection{Problem formulation}\label{sec:formulation}
Let $\mathbb{C}$ denote the set of complex numbers, $\mathbb{R}$ the set of real numbers,
and $\mathbb N$ the set of integers. For $A\in \mathbb C^{n\times n}$, Re$(A)$ and Im$(A)$
denote matrices with the real and imaginary parts of $A$, respectively.
%For any set $A \subseteq \mathbb C^n$,  conv$\, A$ denotes the convex hull of $A$.
%For $a\in \mathbb R$, $[a]^+ := \max\{ a, 0 \}$.
%For $a, b \in \mathbb C$, $a\leq b$ means Re$\,\, a \leq \,\,\,$Re$\,\, b$ and 
%Im$\,\, a \leq \,\,\,$Im$\,\, b$.
%We abuse notation to use the same symbol $a$ to denote either a complex number
%Re$\, a + \ii \, $Im$\, a$ or a 2-dimensional real vector $a = $(Re$\, a$, Im$\, a$)
%depending on the context. In general scalar or vector variables are in small letters, e.g. $u, w, x, y, z$.  
%Most power system quantities however are in capital letters, 
%e.g. $S_{jk}, P_{jk}, Q_{jk}, 	I_j, V_j$.
%A variable without a subscript  denotes a vector with appropriate components,
%e.g. $s := (s_j, j=0, \dots, n)$, $S := (S_{jk}, (j,k)\in E)$.   
% For a vector $a = (a_1, \dots, a_k)$, $a_{-i}$ denotes $(a_1, \dots, a_{i-1}, a_{i+1}, a_k)$.
%For vectors $x, y$, $x\leq y$ denotes component-wise inequality.  Matrices are usually in capital letters.   
Let $\mathbb S^n \subseteq \mathbb C^{n\times n}$ be the set of all $n\times n$ complex symmetric (not necessarily Hermitian) matrices. The transpose of a matrix $A$ is denoted $A^T$ and its Hermitian (complex conjugate) transpose is denoted $A^H$.  $A[i,j]$ denotes the element of $A$ located at ith row and jth column. We define $\mathcal{I}$ as the identity matrix with an appropriate dimension and define ${\bf 1}$ as an all-$1$ column vector with an appropriate dimension.
%% A matrix $A$ is Hermitian
%if $A = A^H$.  $A$ is positive semidefinite (or psd), denoted by $A \succeq 0$,
%if  $A$ is Hermitian \emph{and} $x^H A x \geq 0$ for all $x\in \mathbb C^n$;
%in particular if $A\succeq 0$ then by definition $A = A^H$.
%For matrices $A, B$, $A \succeq B$ means $A-B$ is psd. 
%Let $\mathbb S^n$ be the set of all $n\times n$ Hermitian matrices and
%$\mathbb S^n_+$ the set of $n\times n$ psd matrices.\ye{remove some}
%
%\subsection{Preliminary on graph theory}
%A graph $G = (N, E)$ consists of a set $N$ of nodes and a set
%$E \subseteq N\times N$ of edges.   If $G$ is undirected then $(j, k)\in E$ if and
%only if $(k,j)\in E$.   If $G$ is directed then $(j, k)\in E$ only if $(k,j)\not\in E$;
%in this case we will use $(j,k)$ and $j\rightarrow k$ interchangeably to denote
%an edge pointing from $j$ to $k$. 
%We sometimes use $\tilde G = (N, \tilde E)$ to denote a directed graph.
%By ``$j\sim k$'' we mean an edge $(j,k)$ if $G$ is undirected and
%either $j \rightarrow k$ or $k\rightarrow j$ if $G$ is directed.
%Sometimes we write $j\in G$ or $(j,k)\in G$ to mean $j\in N$ or $(j,k)\in E$ 
%respectively.
%A \emph{cycle} $c := (j_1, \dots, j_K)$ is an ordered set of nodes $j_k\in N$
%so that $(j_k, j_{k+1}) \in E$ for $k=1, \dots, K$ with the understanding that
%$j_{K+1} := j_1$.  In that case we refer to a link or a node in the cycle by 
%$(j_k, j_{k+1})\in c$ or $j_k\in c$ respectively.
%
%\subsubsection{Preliminary on power systems}

A power system can be modeled by an  undirected connected graph
$\mathcal{G}=(\mathcal{N}, \mathcal{E})$ 
where $\mathcal{N} := \{1, 2, \ldots, N \}$ represents the set of buses, 
and $\mathcal{E} \subseteq \mathcal{N} \times \mathcal{N}$ 
represents the set of lines, each connecting two distinct buses. 
A bus $j \in \mathcal{N}$ can be a load bus, a generator bus, or a swing bus.
%and its shunt admittance is denoted $y_{j}$. 
Let $V_j$ be the complex voltage at bus $j$ and $s_j$ 
be the net complex power injection (generation minus load) at that bus.
We use $s_j$ to denote both the complex number $p_j + \mathbf{i} q_j$ (
$\mathbf{i}\triangleq\sqrt{-1}$) and the real pair $(p_j, q_j)$ depending on the context. 
For each line $(i,j)\in \mathcal{E}$, we denote its series admittance by $y_{ij}$. 
The bus admittance matrix of this system is denoted $Y$,
which is an $N\times N$ complex-valued matrix whose
off-diagonal elements are $Y_{ij}=-y_{ij}$ 
and diagonal elements are $Y_{ii} = -\sum_{j\neq i} Y_{ij}$, 
assuming that there is no shunt element (this assumption can be relaxed).
Hence, the current injection vector can be expressed as $I = YV$.

%The bus injection model (BIM)\footnote{We use the bus injection model in this paper; 
%however, our approach can be generalized to other models, 
%e.g., the DC power flow model or the Distflow model~\cite{baranwu}.} 
%is defined by the following power flow equations
%describing the Kirchhoff's laws for a given time index, $k\in\{1,\ldots, K\}$: 
%\begin{equation}\label{eq:pfeq}
%s_i(k)=\sum_{j\in \mathcal{N}_i}y^H_{ij}\left(|V_i(k)|^2-V_i(k)V_j^H(k)\right),~\forall i\in \mathcal{N},
%\end{equation}

%where $\mathcal{N}_i$ is the set of nodes directly connected to node $i$.

The formulated IPF problem is: given voltage and current measurements of different steady-states, i.e., $V_i(k)$ and $I_i(k)$ for $k=1,\ldots, K$ and $i\in\mathcal{M} =\{1,\ldots, N\}$, how to recover the true admittance matrix $Y$. Specially, when $\mathcal{M}$ is a subset of $\mathcal{N}$, what part of the true admittance matrix $Y$ can be recovered under what condition.

\subsection{Identification algorithm}
In this section, we consider the case when $\mathcal{M}=\{1,\ldots, N\}$ and propose a solution to the IPF. We will relax this full measurement condition, i.e., when $\mathcal{M}\subset\{1,\ldots, N\}$ in the next section. 

For $k\in\{1,\ldots, K\}$, the Kirchhoff's laws for a given time index yields $I(k) = YV(k)$. Rewriting this formula in vector form for all time indices yields the following equation for a bus $i$:
\begin{equation}\label{eq:kl}
%\underbrace{\begin{bmatrix}
%\frac{s^H_i(1)}{V^H_i(1)}\\
%\frac{s^H_i(2)}{V^H_i(2)}\\
%\vdots\\
%\frac{s^H_i(K)}{V^H_i(K)}
%\end{bmatrix}}_{I_i^{K}}
\underbrace{\begin{bmatrix}
I_i(1)\\
I_i(2)\\
\vdots\\
I_i(K)
\end{bmatrix}}_{I_i^{K}}
=
\underbrace{\begin{bmatrix} 
V_1(1) & V_2(1) & \ldots & V_N(1) \\
V_1(2) & V_2(2) &\ldots & V_N(2)\\
\vdots & \vdots & \ddots & \vdots\\
 V_1(K) & V_2(K) &\ldots & V_N(K)
\end{bmatrix}}_{V^{K}}
\underbrace{\begin{bmatrix}
Y_{i1}\\
Y_{i2}\\
\vdots\\
Y_{iN}
\end{bmatrix}}_{Y_i}.
\end{equation}

The admittance matrix $Y$ can be obtained from solving the optimization problem below:
\begin{align}\label{eq:Yl2}
&\hat{Y}^{K,l_2}\triangleq\arg\min_{Y}   \left\|V^{K} Y-I^{K}\right\|_F \\
\nonumber &\text{s.t.:}~~Y \in \mathbb{S}^N, ~~Y_{ii} = -\sum_{j\neq i} Y_{ij},~~\forall i.
\end{align}
in which $I^K$ is a $K\times N$ matrix, i.e., $I^{K}=\begin{bmatrix} I^{K}_1 & I^{K}_2 & \ldots & I^{K}_N \end{bmatrix}$. 
% Given measurements, $V^K$ and $s^K$, $I^K$ can be computed from eq.~\eqref{eq:kl}. 
%This feasibility problem can be written 
%as a constrained least squares problem using matrix Frobenius norm:  
Define 
\begin{equation*}
\text{vec}(Y)=\begin{bmatrix}Y_{11} & Y_{21} & \ldots & Y_{N1} & Y_{12} & Y_{22}& \ldots & Y_{N N}
\end{bmatrix}^T, 
\end{equation*}
and apply the \textit{vec} operator to the objective function, we obtain:
\begin{align}
&\min_{\text{vec}(Y)\in \mathbb{C}^{N^2\times 1}} \left\|\left(\mathcal{I}\otimes V^{K} \right) \text{vec}(Y)-\text{vec}(I^{K})\right\|_2\\
\nonumber &\text{s.t.:}~~Y \in\mathbb{S}^N,~~Y_{ii} = -\sum_{j\neq i} Y_{ij},~\forall i,
\end{align}
where $\otimes$ is the Kronecker product. This holds because $\text{vec}(ABC)=(C^T\otimes A)\text{vec}(B)$. Let $\text{svec}: \mathbb S^N \rightarrow\mathbb{C}^{(N^2-N)/2\times 1}$ 
be a mapping from a symmetric complex matrix to a complex vector defined as:
\small 
\begin{equation*}
\text{svec}(Y)=\begin{bmatrix}Y_{21} & Y_{31} & \ldots & Y_{N1} & Y_{32} & Y_{42}& \ldots & Y_{N N-1}
\end{bmatrix}^T.
\end{equation*} \normalsize
It can be readily seen that $\text{svec}$ is a bijection for any matrix $Y$ that satisfies a) $Y\in\mathbb{S}^N$ and b) $1^TY=0$. Based on this definition, we have $\text{vec}(Y)= \Gamma \text{svec}(Y)$, 
where $\Gamma\in\mathbb{R}^{N^2\times(N^2-N)/2}$
maps $\text{svec}(Y)$ to the vectorized admittance matrix as illustrated below.
\begin{example}
For the network depicted in Figure~\ref{fig:ex1}, the $\Gamma$ matrix has the following form
\begin{figure}[!]
\centering
\includegraphics[width=0.1\textwidth]{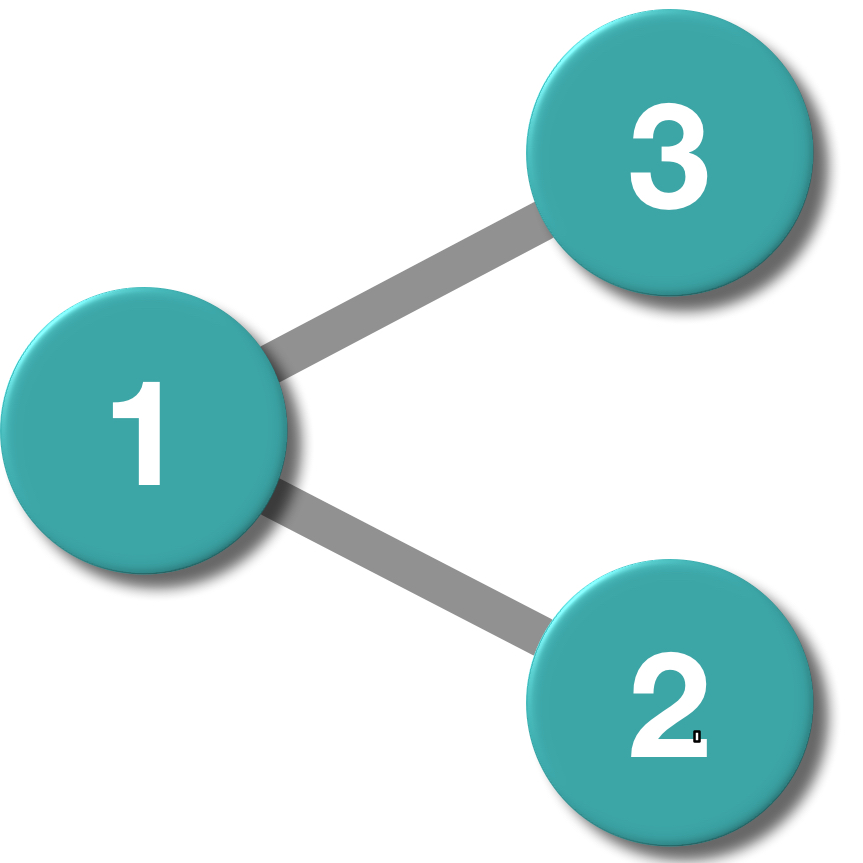}
\caption{An illustrative example of a three-node power system with two lines connecting bus $1$ to buses $2$ and $3$ respectively.}
\label{fig:ex1}
\end{figure}
%\begin{equation}
%\begin{bmatrix}
%Y_{11}\\
%Y_{21}\\
%Y_{31}\\
%Y_{12}\\
%Y_{22}\\
%Y_{32}\\
%Y_{13}\\
%Y_{23}\\
%Y_{33}
%\end{bmatrix}
%=\underbrace{\begin{bmatrix}
%-1 & -1 \\
%1 & 0\\
%0 & 1\\
%1 & 0\\
%-1 & 0\\
%0 & 0 \\
%0 & 1\\
%0 & 0\\
%0 & -1
%\end{bmatrix}
%}_{Q}
%\begin{bmatrix}
%Y_{21}\\
%Y_{31}
%\end{bmatrix}.
%\end{equation}
%
%If one doesn't know the topology of the system, we need to parameterize every single parameters and a different $Q$, i.e.,
{\small\begin{equation*}
\underbrace{\begin{bmatrix}
Y_{11}\\
Y_{21}\\
Y_{31}\\
Y_{12}\\
Y_{22}\\
Y_{32}\\
Y_{13}\\
Y_{23}\\
Y_{33}
\end{bmatrix}}_{\text{vec}(Y)}
=\underbrace{\begin{bmatrix}
-1 & -1 & 0\\
1 & 0 &  0 \\
0 & 1 & 0\\
1 & 0 & 0\\
-1 & 0 & -1 \\
0 & 0  & 1\\
0 & 1 & 0\\
0 & 0 & 1\\
0 &  -1 & -1
\end{bmatrix}
}_{\Gamma}
\underbrace{\begin{bmatrix}
Y_{21}\\
Y_{31}\\
Y_{32}
\end{bmatrix}}_{\text{svec}(Y)}.
\end{equation*}}
\end{example}
Based on the definition of $\Gamma$ in the above equation, 
the constrained optimization problem can be converted to an unconstrained one: 
\begin{equation}\label{eq:l2}
\min_{\text{svec}(Y)\in \mathbb{C}^{(N^2-N)/2\times 1}} \left\|\underbrace{\left(\mathcal{I}\otimes  V^K \right) \Gamma}_{F}\text{svec}(Y)-\text{vec}(I^K)\right\|_2,
\end{equation}
in which $\mathcal{I}$ denotes an identity matrix and $F \triangleq (\mathcal{I} \otimes V^K) \Gamma$. 
%\begin{lemma}
%If $V^{K,H}$ has full column rank, then $I\otimes V^{K,H}$ has full column rank.\end{lemma}
\noindent We define $$\tilde{F}=\begin{bmatrix} \text{Re}(F) & -\text{Im}(F) \\ 
\text{Im}(F) & \text{Re}(F)
\end{bmatrix},~\text{and}~\tilde{b}=\begin{bmatrix}
\text{Re}(\text{vec}(I^K))\\
\text{Im}(\text{vec}(I^K))
\end{bmatrix}.$$
The optimization problem~\eqref{eq:l2} can be written as an unconstrained quadratic program in the real domain:
\begin{equation}
\begin{aligned}
&\min_{\tilde{f}(Y)\in \mathbb{R}^{(N^2-N)\times 1}} \left\|\tilde{F}\tilde{f}(Y) - \tilde{b}\right\|_2,\label{eq:optl1}
\end{aligned}
\end{equation}
in which $\tilde{f}(Y)\triangleq [\text{svec}(Y_R)^T~\text{svec}(Y_I)^T]^T$, $Y_R=\text{Re}(Y)$, and $Y_I=\text{Im}(Y)$.
%
%The following lemma explains how this optimization problem can be solved.
%\begin{lemma}
%If $\left( V^K \right)^H$ has full column rank, then $\tilde{M}$ has full column rank.
%\end{lemma}
%\begin{IEEEproof}
%We can first show that 
%$\mathcal{I}\otimes \left( V^K \right)^H$ has full column rank using the definition of Kronecker product. It is straightforward to show $\tilde{M}$ and $M$ have full column rank. 
%\end{IEEEproof}
This least square problem yields a solution provided that $\tilde{M}$ has full column rank:
\bq\label{eq:fy}
\tilde{f}(Y) =\left(\tilde{F}^T \tilde{F} \right)^{-1}\tilde{F}^T\tilde{b}.
\eq
We compute the solution of the original optimization problem~\eqref{eq:l2} 
from the solution of the optimization problem~\eqref{eq:optl1}
by taking the inverse map of $\tilde{f}$. 
A sufficient condition to guarantee the exactness of 
the solution is that $ V^K$ has full column rank. When $ V^K$ does not have full column rank, we can characterize the part of the admittance matrix that is identifiable (see \cite{Ardakanian2019}).
%\slow{Since $K\geq N$, does the condition ``$\left( V^K \right)^H$ has full column rank'' in Proposition 1 mean $V^K$ is square and invertible? If so, then $Y$ can also be obtained from $V^K Y = I^K$ though the computation is larger-scale - right? Also, to simplify notation, should we say ``$V^K$ has full row rank''?}\ye{these paragraphs have been rewritten}
%
%\slow{So, if we require $V^K$ to have full row rank, but not necessarily invertible, then \eqref{eq:fy} is a pseudoinverse that gives a minimizer to $\|\tilde{M}\tilde{f}(Y) - \tilde{b}\|$.  Right?  If so, is this enough for identifying the Kron-reduced admittance matrix $\bar Y$ in Section III?  Or, do we need enough observations (large enough $K$) to ensure $\tilde{M}\tilde{f}(Y) = \tilde{b}$ is solved exactly?
%}
%\ye{It is the later case, }
\begin{proposition}[Exactness]\label{prop:exact}
If $ V^K$ has full column rank, the optimization problem~\eqref{eq:optl1} has a unique solution given by~\eqref{eq:fy}. 
\end{proposition}
%\begin{IEEEproof}
%The proof can be found in \cite{yuan2016inverse}.
%\end{IEEEproof}
\begin{IEEEproof}
Since $\Gamma\in\mathbb{R}^{N^2\times(N^2-N)/2}$ and has full column rank (this can be checked easily), there exists a matrix $\Gamma^\dag$ such that $\Gamma^\dag \Gamma=\mathcal{I}$. 
For the Kronecker product $\mathcal{I}\otimes V^K \in\mathbb{C}^{KN\times N^2}$, 
$\mathcal{I}\otimes  V^K $ has full column rank when $ V^K$ has full column rank; therefore, $\tilde{F}$ and $F$ have full column rank given the fact that $\text{rank}(\tilde{F})= 2\text{rank}(F)$. 

Finally, we prove by contradiction that if $\tilde{F}$ has full column rank,
the solution to the optimization problem~\eqref{eq:optl1} is unique. 
Suppose there exists $\tilde{f}(Y_1)$ and $\tilde{f}(Y_2)$ ($\tilde{f}(Y_1)\neq\tilde{f}(Y_2)$) 
such that $\tilde{F}\tilde{f}(Y_1)=\tilde{b}$ and $\tilde{F}\tilde{f}(Y_1)=\tilde{b}$, 
then $$\tilde{F}\left(\tilde{f}(Y_1)-\tilde{f}(Y_2)\right)=0,$$
which contradicts the full column rank assumption.
\end{IEEEproof}

\begin{remark}
The approach can be easily extended to the case of nonzero shunt elements
where ${\bf1}^TY\neq 0$ 
by changing the definitions of $\text{svec}(Y)$, $\Gamma$ and $\tilde{f}(\cdot)$. Specifically if $Y$ includes shunt elements then $\text{svec}(Y)$ will include diagonal elements $(Y_{11}, \dots, Y_{NN})$.
\end{remark}

We can add the element-wise positivity constraint to this problem if the conductance and susceptance of each line are positive\footnote{The conductance of a line is always positive, the susceptance can be negative or positive depending on whether the line is inductive or capacitive.}.
\begin{equation}
\begin{aligned}\label{eq:optl1-1}
&\min_{\tilde{f}(Y)\ge0} \left\|\tilde{F}\tilde{f}(Y) - \tilde{b}\right\|_2.
 \end{aligned}
\end{equation}
The above problem is known as nonnegative least squares, which is a convex optimization problem and a global minimizer can be solved using different methods, such as the active set method \cite{wright1999numerical}.
\section{IPF with hidden nodes}\label{sec:hidden}
%----------------------------------------------------------------------------------------

In the previous section we solve the IPF problem when 
voltage and current measurements are available at all buses.
In this section we consider the case when voltage and current measurements are available only at a subset of the buses.
In a distribution system, for example, measurements are typically available at the substation and customer meters but not throughout the grid.

In Section~\ref{subsec:kron}, we show that, in the presence of hidden nodes, the algorithm presented in Section~\ref{sec:nohidden} can identify a Kron-reduced admittance matrix $\bar Y$,
defined in \eqref{eq:defYbar} below, 
for the nodes where measurements are available (Corollary \ref{thm:kron}). In Section~\ref{subsec:id} we show that when the network is a tree then it is indeed possible to uniquely identify the original admittance matrix $Y$ from its Kron reduction under reasonable assumptions. 
%In Section~\ref{subsec:lrs}, we propose a low rank and sparse decomposition method for mesh networks. 

\subsection{Kron-reduced admittance matrix $\bar Y$}
\label{subsec:kron}
%----------------------------------------------------------------------------------------

We call a bus/node a \emph{measured bus/node} if measurements of its voltage and current injection are 
available for identification. We call it a \emph{hidden bus/node} otherwise. 
Let $\mathcal{M}$ and $\mathcal{H}$ represent the set of measured and hidden nodes respectively.
\begin{assumption}\label{ass:0}
We make the following assumptions:
\bi
\item[\ref{ass:0}.1] The underlying graph $\mathcal{G}$ is connected.
\item[\ref{ass:0}.2]Hidden nodes have zero injection $I_i(k)=0$ for all $i\in\mathcal H$.
\ei
\end{assumption}
Note that the second assumption, i.e., the injected current is zero at every hidden node, is reasonable because 
if generators or loads are connected to a node, then the current they inject or draw is typically measured.

Let $H$ be the number of hidden nodes.
Without loss of generality we label the buses so that the first $N-H$ buses are measured and the last $H$ buses are hidden, i.e., 
$\mathcal{M} := \{1, \dots, M\triangleq N-H\}$ and $\mathcal{H} := \{N-H+1, \dots, N\}$.
% $\mathcal{M}_1\cap \mathcal{M}_2=\emptyset$ and $\mathcal{M}_1\cup \mathcal{M}_2=\{1,2,\ldots,N\}$. 
We partition the bus admittance matrix $Y$ into four sub-matrices:
\begin{equation}
\begin{aligned}
Y \ = \ \begin{bmatrix} Y_{11} & Y_{12} \\ Y_{21} & Y_{22} \end{bmatrix}
\ \ &= \ \
\begin{bmatrix} G_{11} & G_{12} \\ G_{21} & G_{22} \end{bmatrix} \ + \ 
\mathbf{i} \begin{bmatrix} B_{11} & B_{12} \\ B_{21} & B_{22} \end{bmatrix}\\
\ &= \ G + \mathbf{i} B. \label{eq:Y=G+iB}
\end{aligned}
\end{equation}
Here
$Y_{11}\in\mathbb S^{N-H}$ describes the connectivity among the measured nodes, 
$Y_{12} = Y_{21}^T\in\mathbb C^{(N-H)\times H}$ the connectivity between the measured 
and the hidden nodes, and $Y_{22}\in\mathbb{S}^{H}$ the connectivity among the hidden nodes.
For $i\in \mathcal{M}$, $(I_i(k), V_i(k), k=1, \dots, K)$ denote the current and voltage measurements at bus $i$ at time $k$.
To simplify notation, we index the entries of $Y_{22}$, not by $i,j = 1,\dots, H$,
but by $i,j=N-H+1, \dots, N$.  We index the entries of $Y_{12}$ by 
$i=1, \dots, N-H$ and $j=N-H+1, \dots, N$ and similarly for $Y_{21} = Y_{12}^T$,
as well as submatrices $G_{ij}, B_{ij}$ in \eqref{eq:Y=G+iB}.
%\slow{Doublecheck for later proofs if this indexing is indeed more convenient, and if so, is consistently used.}\ye{sure, i think it is consistent.}

For $i\in \mathcal{H}$, $I_i(k)=0,~\forall k$, and $V_i(k)$ is the voltage at bus $i$ but is not available for identification. 
To simplify notation, We abuse $V_1(k)$ to denote both the voltage  at bus 1 at time $k$ and the vector of all voltages at measured buses at time $k$, depending on the context; similarly for $V_2(k)$ and $I_1(k)$.
Then
\bq
\begin{bmatrix}
I_1(k)
\\
0
\end{bmatrix}=
\begin{bmatrix} Y_{11} & Y_{12}\\
Y_{21} & Y_{22}
\end{bmatrix}
\begin{bmatrix}
V_1(k)\\
V_2(k)
\end{bmatrix}, \qquad \forall k
\label{eq:I_10V1V2}
\eq
If $Y_{22}$ is invertible then eliminating $V_2$ from \eqref{eq:I_10V1V2} yields a relation  \begin{equation}
I_1(k) \ =\  \bar{Y}V_1(k), \quad\forall k
\label{eq:I1=YV1}
\end{equation}
between currents and voltages at measured nodes 
through the \emph{Kron-reduced admittance matrix} $\bar Y\in\mathbb S^{m}$ defined as:
\bq
\bar{Y} & \triangleq & Y_{11}-Y_{12}Y^{-1}_{22}Y_{12}^T
\label{eq:defYbar}
\eq
for the set of measured nodes. 
In the rest of this subsection we first justify the invertibility of $Y_{22}$ and hence the definition of $\bar Y$. Proposition \ref{prop:exact} then implies that $\bar Y$ can be identified from voltage and current measurements. Moreover $\bar Y$ is the best we can identify for general networks because multiple admittance matrices $Y$ may reduce to the same $\bar Y$. 

%
%\slow{Should we use $I_1^K$, $V_1^K, V_2^K$ here, or is \eqref{eq:I_10V1V2} a description of the Ohm's law over the network but not about measurements? Do we use $(I, V)$ and $(I^K, V^K)$ to denote different things?  Should clarify from beginning.}
%\ye{good point. here we use it as a description of the Ohm's law over the network. Once the equality has been built, one can use similar derivations to do identification using $I_1^K$, $V_1^K, V_2^K$ as in the previous section.}
%

%
%\begin{frame}
%\frametitle{Solvability}
%There are $N-1+hN$ unknown in the equations, in which $h\triangleq\text{Card}\{S\}$ and $(N-h)N$ equations. When $(N-h)N\ge(h+1)N-2$, then we should be able to solve for the unknown variables. It is sufficiently to say that if $h\le \frac{N-1}{2}$, then we should be able to solve for the unknown parameters in $Y$.
%
%\vspace{1cm}
%
%Question: this seems to be quite brutal... Can we use some graph theoretical results? Solvability.
%
%{\color{red} The answer is Yes! The previous deduction was wrong.}
%\end{frame}
%
%\slow{List explicitly Assumption 0?
%\bi
%\item Network is connected.
%\item Hidden nodes have zero injection $I_i(k)=0$ for all $i\in\mathcal M_2$.
%\ei
%}\ye{This is a bit tricky to define connected networks, as we have not introduced graph theory here.}
%

%
\begin{assumption}\label{ass:2}
The admittance matrix $Y\triangleq G+\mathbf{i}B$ defined in \eqref{eq:Y=G+iB} satisfies:
\begin{itemize}
\item[\ref{ass:2}.1] Series impedances of the lines are resistive and inductive: $G[i,j]\le0$ and $B[i,j]\ge0$ for any $i\neq j$;
%
%The off-diagonal elements of $G$ are non-positive and
%the off-diagonal elements of $B$ are non-negative.
\item[\ref{ass:2}.2] Diagonal dominance: $G_{22}[i,i]\ge-\sum_{j\neq i}G_{22}[i,j]$ and $-B_{22}[i,i]\ge\sum_{j\neq i}B_{22}[i,j]$ hold for any $i$.
\end{itemize}
\end{assumption}
%These assumptions mean that the series impedances of the lines are resistive and inductive.
%
% 
% \begin{lemma}[Gershgorin Theorem]
% For a real matrix $A$ and $i \in\{1,\ldots,N\}$, let $R_i = \sum_{j\neq{i}}  \left|A[i,j]\right|$ and $D(A[i,i], R_i)$ be a closed disc centered at $A[i,i]$ with radius $R_i$, then every eigenvalue of a complex matrix $A$  lies within at least one of the Gershgorin discs  $D(A[i,i],R_i)$.  
% \end{lemma}
%
%\begin{lemma}[Rayleigh-Ritz Theorem]
%Let $A\in\mathbb{S}^h$ be a real symmetric matrix with eigenvalues $\{\lambda_1\ge\lambda_2\ge\ldots\ge\lambda_h\}$, then 
%$$\lambda_k = \max \{ \min \{ R_A(x) \mid x \in U \text{ and } x \neq 0 \} \mid \dim(U)=k \},$$
%the Rayleigh-Ritz quotient
%$R_A :\mathbb{C}^h\rightarrow\mathbb{R}$ is defined by
%$R_A(x) = \frac{<Ax, x>}{<x,x>}.$ $\hfill \square$
%\end{lemma}
%

%\slow{Consistent notation: $A_{ij}$ vs $A[i,j]$, or state using the latter when necessary?}
%\ye{$A_{ij}$ denotes block matrix while $A[i,j]$ an element}
\begin{lemma}\label{lemma:P22}
Under Assumption~\ref{ass:0} and \ref{ass:2}, $G_{22}\succeq0$, $-B_{22}\succeq0$ and $G_{22}-B_{22}\succ0$, when $H<N$. 
\end{lemma}
\begin{IEEEproof}
From the Gershgorin Theorem and Assumption~\ref{ass:2}.2, all eigenvalues of the submatrix $G_{22}$ lie in the right-half plane including the origin and all eigenvalues of $B_{22}$ lie in the left-half plane including the origin. Together with the fact that $G_{22}$ and $B_{22}$ are symmetric, we have $G_{22}\succeq0$ and $-B_{22}\succeq0$.

%\slow{Ye: Please check the proof below to make sure the (minor) changes in notation are correct.}\ye{this looks good to me}

This implies that $G_{22}-B_{22}\succeq0$.  
We now show that indeed $G_{22}-B_{22}\succ0$.  Suppose for the sake of contradiction that there exists a nonzero $x\in\mathbb{R}^H$ such that $x^T(G_{22}-B_{22})x = 0$.
Denote by $A_{22}:=G_{22}-B_{22}$ and $A_{21} := G_{21} - B_{21}$ so that 
\bqn
A_{22}[i,i] & = & \sum_{i,j\in\mathcal H: j\neq i} (-A_{22}[i,j]) \ + \ 
    \sum_{i\in\mathcal H, j\in\mathcal M} (-A_{21}[i,j])
\eqn
Then
\begin{align}
%\begin{aligned}
x^T(G_{22}-B_{22})x 
& = \sum_{i,j\in\mathcal H: i\neq j} 
\left( A_{22}[i,j] x_ix_j \ - \ A_{22}[i,j]x_i^2 \right)  \nonumber\\
&\qquad + \sum_{i\in\mathcal H, j\in\mathcal M} (-A_{21}[i,j])x_i^2  \nonumber \\
& =  \sum_{i,j\in\mathcal H: i<j} 
(-A_{22}[i,j]) (x_i - x_j)^2  \nonumber\\
&\qquad + \sum_{i\in\mathcal H, j\in\mathcal M} (-A_{21}[i,j])x_i^2
\label{eq:lemma2.1}
%\end{aligned}
\end{align}
By definition $A_{22}[i,j] = A_{21}[i,j]=0$ if $i\not\sim j$ (i.e., $i$ and $j$ are not adjacent).  
For $i,j\in\mathcal M \cup \mathcal H$, if $i\sim j$  (i.e., $i$ and $j$ are adjacent), then 
$Y_{ij} = G_{ij}+\mathbf{i} B_{ij}\neq 0$, i.e., at least one of $G_{ij}=-g_{ij}\leq 0$ or 
$B_{ij}=-b_{ij}\geq 0$ is nonzero.
%\slow{Define $y_{ij} = g_{ij}+ib_{ij}$?} \ye{can we directly say $G_{ij}\leq 0$ or 
%$B_{ij}\geq 0$ without introducing more variables?}
This implies that $-A_{22}[i,j] = -G_{22}[i,j] + B_{22}[i,j] > 0$ and 
$-A_{21}[i,j] = -G_{21}[i,j] + B_{21}[i,j] > 0$ for all $i\sim j$.  
Therefore, for $x^T(G_{22}-B_{22})x = 0$ in \eqref{eq:lemma2.1}, we must have:
\begin{itemize}
\item[1.] $x_i=x_j$ if $i\sim j$ is a connection between hidden nodes in $\mathcal H$;
\item[2.] $x_i =0$ for any hidden node $i\in\mathcal H$ connected to at least one observed node $j\in\mathcal M$.
\end{itemize}
% We claim that $Y_{22}[i,j]\neq0$ is equivalent to $P_{22}[i,j]-Q_{22}[i,j]\neq0$\footnote{This is because, since $P[i,j]\le0$ and $Q[i,j]\ge0$, the only way $P[i,j]=Q[i,j]$ is when both are zero (i.e., $i$ is not connected to a measured node), a contradiction.}. 
Since the network is connected, for every hidden node $i\in\mathcal H$, there is a path that connects the hidden node to an observed node $k\in\mathcal M$. For all nodes $j$ on this path from $i$ to $k$, the above properties implies that $x_j=0$. Since this is true for all hidden nodes, we have $x=0$, a contradiction.  Hence $G_{22}-B_{22}\succ0$.
\end{IEEEproof}

Recall that the network is connected and has $N$ nodes of which
$H$ are hidden.
\begin{proposition}\label{lemma:invert}
Under Assumptions \ref{ass:0} and \ref{ass:2}, if $H<N$ then $Y_{22}$ is invertible.
\end{proposition}
\begin{IEEEproof}
We prove that $0$ is not an eigenvalue of $Y_{22}$. 
Suppose for the sake of contradiction that there exists a nonzero vector $(v+\mathbf{i}w)$ such that $(G_{22}+\mathbf{i}B_{22})(v+\mathbf{i}w)=0$, i.e., 
\begin{equation*}\label{eq:wv}
\begin{aligned}
G_{22}v-B_{22}w \ = \ 0 \ \ \ \text{ and } \ \ \ 
G_{22}w+B_{22}v \ = \ 0,
\end{aligned}
\end{equation*}
This implies 
\begin{equation}\label{eq:wv1}
\begin{aligned}
(G_{22}+B_{22})v+(G_{22}-B_{22})w&=0,\\
(G_{22}+B_{22})w-(G_{22}-B_{22})v&=0.
\end{aligned}
\end{equation}
Since $G_{22}-B_{22}\succ 0$ by Lemma~\ref{lemma:P22}, we can eliminate
$v$ from \eqref{eq:wv1} to get
{\small\bqn
\left((G_{22}-B_{22})+(G_{22}+B_{22})(G_{22}-B_{22})^{-1}(G_{22}+B_{22})\right)w  =  0.
\eqn}
Multiplying $w^T$ on the left we have
\bqn
w^T(G_{22}-B_{22})w \ + \ 
\hat w^T (G_{22}-B_{22})^{-1} \hat w & = & 0,
\eqn
where $\hat w = (G_{22}+B_{22}) w$.  This contradicts $G_{22}-B_{22}\succ 0$ unless $w=0$.  But $w=0$ implies that $(G_{22}-B_{22})v=0$ from \eqref{eq:wv1}, meaning $v=0$.  This is a contradiction and hence $Y_{22}$ is invertible.
\end{IEEEproof}

Proposition \ref{lemma:invert} guarantees that $Y_{22}$ is invertible and hence the Kron-reduced admittance $\bar Y$ in \eqref{eq:defYbar} is well defined 
under Assumption~\ref{ass:2}.
Moreover, because of \eqref{eq:I1=YV1}, the algorithm in Section~\ref{sec:nohidden} can identify the Kron-reduced admittance matrix $\bar Y$ from voltage and current measurements (Proposition~\ref{prop:exact}).
\begin{corollary}\label{thm:kron}
Suppose Assumptions~\ref{ass:0} and \ref{ass:2} and the condition in  Proposition~\ref{prop:exact} hold. If $H<N$, then the Kron-reduced admittance matrix $\bar{Y}$ 
can be identified from voltage $V_1^K$ and current $I_1^K$ measurements at the measured nodes.
\end{corollary}

An admittance matrix $Y\in\mathbb S^N$ specifies a unique weighted undirected graph $\mathcal G(Y) = (\mathcal{N}(Y), \mathcal{E}(Y))$ with $\mathcal{N}:=\{1, \ldots, N\}$ and $\mathcal{E} \subseteq \mathcal{N}\times \mathcal{N}$ such that there is an edge $(i,j)$ if and only if $Y[i,j]\neq0$.
% We sometimes refer to $Y$ and its underlying graph jointly as the graph
% $\mathcal{G} = (\mathcal{N}, \mathcal{E}, Y)$.
Its Kron reduction $\bar Y$ specifies a unique weighted graph $\bar{\mathcal G}:= \mathcal{G}(\bar{Y}) = (\mathcal M, \bar{\mathcal E})$ that can be obtained from $\mathcal G$ through Algorithm \ref{alg:graphcond}.
%
%
%We say that there is no overlapping between two graphs $\mathcal{G}_1$ and $\mathcal{G}_2$ 
%if $\mathcal{G}_1/\mathcal{G}_2=\mathcal{G}_1$. 
%

\begin{figure}[!]
\centering
\includegraphics[scale=0.5]{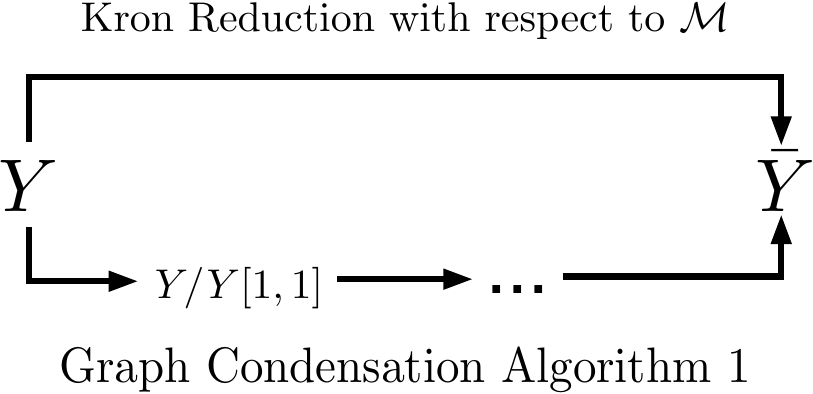}
\caption{Two equivalent schemes to compute the Kron reduced $\bar{Y}$.}
\label{fig:kron2}
\end{figure}

\begin{algorithm}[htbp]
\caption{Graph Condensation Algorithm}
\label{alg:graphcond}
\begin{algorithmic}[1]
	\State \textbf{Input:} a graph $\mathcal{G}=(\mathcal{N}, \mathcal{E})$ with $N$ nodes 
	and a set of measured nodes $\mathcal{M}=\{1,2,\ldots,M\}$ and admittance matrix $Y$
	\For {$v=M+1 : N$}
	\State Remove hidden node $v$ from $\mathcal{N} = \mathcal{N} - \{v\}$ and all edges from $\mathcal{E}$ that are incident on $v$;
	\State For all node pairs $w$ and $l$ that are neighbors of $v$, add an edge between $w$ and $l$ to $\mathcal{E}$;  
	\State Update the admittance matrix $Y = Y/Y[i,i]$ using eq.~\eqref{eq:kron1node}. 
	\EndFor 
    \State \textbf{return} $\bar{\mathcal{G}}=\mathcal{G}$ and $\bar{Y}=Y$. 
\end{algorithmic}
\end{algorithm}
%
%We apply Kron reduction to remove the hidden nodes with degree $2$. 
Each iterative step in the algorithm removes a hidden node $i\in \mathcal H$ and connects all its neighbors to each other. This step can be represented algebraically as an update on the admittance matrix to compute its Schur complement of $Y[i,i]$.
Specifically we can partition possibly after permutation an admittance matrix $Y$ of an appropriate dimension into the form:
$$Y=\begin{bmatrix} 
Y(i,i) & Y(i,i]\\
Y(i,i]^T & Y[i,i]
\end{bmatrix},$$
where $Y[i,i]\in\mathbb{C}$ is the $i$th diagonal elements of $Y$,  and  $Y(i,i),Y(i,i]$ are shown in Eq.~\eqref{eq:Yii}.
\begin{figure*}	
	{\small\begin{align} \label{eq:Yii}
		Y(i,i) &= \begin{bmatrix}
			Y[1,1] & \ldots & Y[1,i-1] & Y[1,i+1] & \ldots & Y[1,N]\\
			\vdots & \ddots & \vdots & \vdots & \ddots & \vdots \\
			Y[i-1,1] & \ldots & Y[i-1,i-1] & Y[i-1,i+1] & \ldots & Y[i-1,N]\\
			Y[i+1,1] & \ldots & Y[i+1,i-1] & Y[i+1,i+1] & \ldots & Y[i+1,N]\\
			\vdots & \ddots & \vdots & \vdots & \ddots & \vdots \\
			Y[N,1] & \ldots & Y[N,i-1] & Y[N,i+1] & \ldots & Y[N,N]
		\end{bmatrix},~ Y(i,i] = \begin{bmatrix}
		Y[1,i] \\	\vdots \\	Y[i-1,i] \\	Y[i+1,i] \\	\cdots \\	Y[N,i]
	\end{bmatrix}.
	\end{align}}
\end{figure*}
Then each iterative step of Algorithm \ref{alg:graphcond} updates the admittance matrix by ($Y[i,i]$ is always invertible due to Proposition~\ref{lemma:invert}):
\begin{equation}\label{eq:kron1node}
Y/Y[i,i] = Y(i,i) - Y(i,i]Y^{-1}[i,i]Y(i,i]^T. 
\end{equation}
This step is repeated until all the hidden nodes are removed from the original graph, 
producing the Kron-reduced graph 
 $\bar{\mathcal{G}}=\{\mathcal{M}, \bar{\mathcal{E}}\}$ \cite{kron} and its admittance matrix $\bar{Y}$ (shown in Fig.~\ref{fig:kron2}).
% \ye{we newly added the figure and cite results above.}

%\slow{Is Figure ~\ref{fig:kron2} used in later section?  If not, not sure if we need to include the figure?
%\begin{figure}[!]
%\centering
%\includegraphics[scale=0.55]{figure/kron.pdf}
%\caption{Two equivalent schemes to compute the Kron reduced $\bar{Y}$.}
%\label{fig:kron2}
%\end{figure}
%} \ye{I will then remove it for now}

%\begin{remark}
%Lemma~\ref{lemma:invert} guarantees the invertibility of $Y_{22}$ in a connected graph when the number of hidden nodes is less than $N$. Hence, $\bar{Y}$ is identifiable from input-output data. 
%\end{remark}

Given an admittance matrix $Y$, each partition $(Y_{11}, Y_{12}, Y_{22})$ in \eqref{eq:Y=G+iB} defines uniquely a Kron-reduced matrix $\bar Y := \bar Y(Y_{11}, Y_{12}, Y_{22})$ given by \eqref{eq:defYbar}.
This mapping is clearly not injective in general, i.e., given an $M\times M$ symmetric matrix $\bar Y\in\mathbb S^M$ (possibly with $\bar Y \textbf{1} = 0$) 
there are generally multiple $N\times N$ symmetric matrices $Y$ that can be partitioned into (non-unique) $(Y_{11}, Y_{12}, Y_{22})$ whose Kron reductions are the given $\bar Y$, as long as $N\geq M$.

\begin{example}\label{ex:equi}
Consider a (Kron-reduced) admittance matrix for a two-node network ($\theta\neq0$):
\begin{equation*}
\bar{Y} \ \ = \ \ \begin{bmatrix}
\theta & -\theta \\
-\theta & \theta 
\end{bmatrix}.
\end{equation*}
The following $3\times 3$ admittance matrice $Y$ with the given partition has $\bar Y$ as its Kron reduction:
\bqn
Y \ := \ \left[ \begin{array}{c|c} Y_{11} & Y_{12} \\\hline
    Y_{12}^T & Y_{22} \end{array} \right]
\ = \ \left[
\begin{array}{c c | c}
\theta+\theta' & -\theta & -\theta'  \\
-\theta & \theta & 0 \\ \hline
-\theta' & 0 & \theta'
\end{array} \right],
\eqn
for arbitrary nonzero $\theta'$.
The underlying network is shown in Fig.~\ref{fig:ex1}
with Node~3 as the hidden node, so that $\bar Y$ corresponds to the Kron-reduced admittance matrix for Nodes~1 and 2 in Fig.~\ref{fig:ex1}.
In this case the hidden node has degree 1. 
Another $3\times 3$ admittance matrices $\tilde{Y}$ that also has $\bar Y$ as its Kron reduction is:
\bqn
\tilde Y \ := \ \left[ \begin{array}{c|c} \tilde{Y}_{11} & \tilde{Y}_{12} \\\hline
    \tilde{Y}_{12}^T & \tilde{Y}_{22} \end{array} \right]
\ = \ \left[
\begin{array}{c c | c}
\theta_1 & 0 & -\theta_1 \\
0 & \theta_2 & -\theta_2 \\ \hline
-\theta_1 & -\theta_2 & \theta_1+\theta_2
\end{array} \right],
\eqn
as long as $(\theta_1, \theta_2)$ satisfies
\bqn
\frac{\theta_1\theta_2}{\theta_1+\theta_2} & = & \theta.
\eqn
The network underlying $\tilde Y$ is isomorphic to that in Fig.~\ref{fig:ex1} with Node~1 being the hidden node, so that $\bar Y$ is the Kron-reduced admittance matrix for Nodes~2 and 3 in Fig.~\ref{fig:ex1}.
In this case the hidden node has degree 2.
\end{example}

Example \ref{ex:equi} shows that in general only the Kron-reduced admittance matrix $\bar Y$ is identifiable from measurements at the measured nodes. For arbitrary networks it is impossible to identify the original admittance matrix $Y$ whose Kron reduction yields $\bar Y$. We next show the surprising result that, when the underlying network is a tree and every hidden nodes has a degree $\ge3$, then the original admittance matrix $Y$ can indeed be discovered even in the presence of hidden nodes.

\subsection{Radial networks: exact identification}\label{subsec:id}

%\slow{Check later throughout to make sure it's clear what is known and unknown at each step of the process, e.g., ID of measured/hidden nodes, submatrices of $Y$, etc.}\ye{sure!}

Consider a radial network and suppose we have identified a Kron-reduced admittance matrix $\bar Y$ from partial voltage and current measurements.
In this section we develop a novel algorithm to compute the original admittance matrix $Y$ from $\bar Y$ under the following additional assumptions.
\begin{assumption}\label{ass:degree2} 
The admittance matrix $Y$ satisfies:
\begin{itemize}
\item[\ref{ass:degree2}.1] The underlying graph $\mathcal G(Y)$ is a tree. 
\item[\ref{ass:degree2}.2] Every hidden node has a degree $\ge 3$. \item[\ref{ass:degree2}.3] There is no shunt element in $Y$, i.e., $Y\textbf{1}=0$. 
\end{itemize}
\end{assumption}
\begin{remark}
Assumption \ref{ass:degree2}.2 is a necessary condition for identification as shown in Example~\ref{ex:equi} where the hidden node has a degree 1 or 2.
\end{remark}

\begin{remark}
Assumption \ref{ass:degree2}.3 can be further relaxed as demonstrated in the full version \cite{yuan2016inverse}.
\end{remark}

We start with some definitions.
Consider an undirected graph $\mathcal{G} = (\mathcal{N}, \mathcal{E})$ where $\mathcal{N} := \{1, \dots, N\}$ is the set of nodes and $\mathcal E \subseteq \mathcal N\times \mathcal N$ is the set of edges.
% , weighted by the (off-diagonal entries of the) admittance matrix $Y$ satisfying Assumption~\ref{ass:degree2}.3.
% Two nodes $j$ and $k$ are \emph{adjacent} if $(j,k) \in \mathcal{E}$. 
% Similar to the adjacency matrix of $\mathcal{G}$, $Y\in\mathbb S^N$ contains a nonzero complex number for each edge in $\mathcal{E}$. 
A \emph{complete} graph is one in which all nodes are adjacent. 
A subgraph of $\mathcal{G}$ is a graph $\mathcal{G}' = (\mathcal{N}', \mathcal{E}')$ with $\mathcal{N}'\subseteq \mathcal{N}$ and $\mathcal{E}' \subseteq \mathcal{E}$. 
A \emph{clique} of $\mathcal{G}$ is a complete subgraph of $\mathcal{G}$. 
A \emph{maximal clique} of $\mathcal{G}$ is a clique that is not a subgraph of another clique of $\mathcal{G}$.  
We say $\mathcal{G}$ is a \emph{tree} if 
there is exactly one path between every two nodes. A \emph{forest} is a disjoint union of trees.
%
% Given a symmetric admittance matrix $Y\in\mathbb S^N$, we can define its corresponding graph $\mathcal{G} = (\mathcal{N}, \mathcal{E}, Y)$ with $\mathcal{N}:=\{1, \ldots, N\}$ and $\mathcal{E} \subseteq \mathcal{N}\times \mathcal{N}$ such that for any pair of nodes $(k,j)$, there exists an edge if and only if $Y[k,j]\neq0$.

%\ye{I added new definitions about hidden boundary nodes here}

For our purposes, an admittance matrix $Y$ is a complex symmetric matrix (we usually assume $Y$ satisfies Assumption \ref{ass:degree2}.3).
%\ye{to delete: Given any $N\times N$ admittance matrix $Y$, let $\mathcal G(Y):=(\mathcal N(Y), \mathcal E(Y))$ denote the graph where $\mathcal N(Y) := \{1, \dots, N\}$ and there is an edge $(i,j)\in\mathcal E(Y)$ if and only if $Y[i,j]\neq 0$.}  
We sometimes refer to $\mathcal G(Y)$ as the \emph{underlying graph of $Y$} and write $\mathcal G:=(\mathcal N, \mathcal E)$ when $Y$ is clear from the context.
Consider two $N\times N$ admittance matrices $Y_1$ and $Y_2$. We define two functions of $(Y_1, Y_2)$ and their underlying graphs.
First $Y_3 := Y_1+Y_2$ is also an admittance matrix and its underlying graph $\mathcal G(Y_3) = (\mathcal N(Y_3), \mathcal E(Y_3))$ is the graph with the same set of nodes and edges in both graphs, i.e., $\mathcal E(Y_3) := \mathcal E(Y_1) \cup \mathcal E(Y_2)$.
 When the matrices are clear from the context, we also denote the function $Y_3 = Y_1+Y_2$ by $\mathcal G_3 = \mathcal G_1 \oplus \mathcal G_2$. 
 Note that if $Y_1$ and $Y_2$ satisfy Assumption \ref{ass:degree2}.3, so does $Y_3$.
Second define the $N\times N$ matrix $Y_4 := Y_1\backslash Y_2$ as a function of $(Y_1, Y_2)$ by:
\bqn
Y_4[i,j] & := & \left\{ \begin{array}{ll}
    Y_1[i,j] & \text{ if }  i\sim j \text{ and } (i,j)\not\in\mathcal E_2 \\
    -\sum_j Y_4[i,j]  & \text{ if }  i = j \\
    0 & \text{ otherwise}
    \end{array} \right.
\eqn
The underlying graph $\mathcal G(Y_4)$ is a subgraph of $\mathcal G(Y_1)$ where edges in $\mathcal G(Y_2)$ have been removed.  When the matrices are clear from the context, we also denote the function $Y_4= Y_1\backslash Y_2$ by $\mathcal G_4 = \mathcal G_1/\mathcal G_2$. Note that $Y_4$ satisfies Assumption \ref{ass:degree2}.3 by definition.

Fix an (unknown) admittance matrix $Y$ and assume its underlying graph $\mathcal G := \mathcal G(Y)$ is a tree. Suppose its Kron-reduced admittance matrix $\bar Y$ and its underlying graph $\bar{\mathcal G} := \mathcal G(\bar Y)$ are given. For example $\bar Y$ is obtained according to Corollary \ref{thm:kron} from partial voltage and current measurements at measured nodes in $\mathcal M$. 

Next, we will propose a recursive algorithm to recover $Y$ from $\bar Y$. Specifically, We can decompose $\bar{\mathcal G}$ to two graphs $\mathcal G_1$ and $\mathcal G_2$ ($\bar Y_1$ and $\bar Y_2$ correspondingly) with distinct properties in Section~\ref{sec:decomposition}. Secondly, we further introduce a partition of $Y$ in Section \ref{sec:partition} and a corresponding parameterization of $Y$ in Section \ref{sec:parameterization}. Thirdly, we can compute these parameters from known quantity in $\bar Y$ in Section \ref{sec:computation}. Finally, the overall recursive algorithm to recover $Y$ is proposed in \ref{sec:algorithm}.  

\subsubsection{Decomposition of $\bar{\mathcal G}$}\label{sec:decomposition}
Let $\bar{\mathcal E_1}$ denote the subset of all edges of $\bar{\mathcal G}$ that are between measured nodes in the original graph $\mathcal G$, and $\bar{\mathcal E_2}$ denote the subset of all edges of $\bar{\mathcal G}$ that have been added by Step 4 of the graph condensation Algorithm \ref{alg:graphcond} when hidden nodes are removed from $\mathcal G$.
By definition of $\bar{\mathcal E_1}, \bar{\mathcal E_2}$, we have  $\bar{\mathcal G} = (\mathcal M, \bar{\mathcal E_1} \cup \bar{\mathcal E_2})$. 
% Note that we only know $\mathcal E_1 \cup \mathcal E_2$ but not individual $\mathcal E_1$ and $\mathcal E_2$. We now present a method to identify them. 
\begin{lemma}\label{lemma:1}
Under Assumption \ref{ass:0} and \ref{ass:degree2}, $\bar{\mathcal E_1} \cap \bar{\mathcal E_2} = \emptyset$.
\end{lemma}
\begin{IEEEproof}
If there exists an edge $(i,j)\in \bar{\mathcal E_1} \cap \bar{\mathcal E_2}$, then $(i,j)$ must be an edge in the original graph $\mathcal G$ and nodes $i$ and $j$ must also be connected through a path consisting of only hidden nodes. This creates a loop and contradicts that $\mathcal G$ is a tree.  Hence $\bar{\mathcal E_1} \cap \bar{\mathcal E_2} = \emptyset$.  
\end{IEEEproof}

Since $\bar{\mathcal G} = (\mathcal M, \bar{\mathcal E_1} \cup \bar{\mathcal E_2})$, Lemma \ref{lemma:1} motivates decomposing $\bar{\mathcal G}$ into two subgraphs, $\mathcal{G}_1:= (\mathcal{M}, \bar{\mathcal{E}_1})$ and $\mathcal{G}_2:=(\mathcal{M}, \bar{\mathcal{E}_2})$, both defined on $\mathcal M$ of measured nodes but with disjoint edge sets. While the graph $\bar{\mathcal G}:=\mathcal G(\bar Y)$ is defined by the Kron-reduced admittance matrix $\bar Y$, at this point the graphs $\mathcal G_1$, $\mathcal G_2$ are only defined in terms of the graph $\bar{\mathcal G}$ (in fact in terms of $\mathcal G$) and are not associated with any admittance matrices.
Define the matrices:
\begin{subequations}\label{eq:yg1yg2}
\bq
\bar Y_1 & := & Y_{11} \ - \ \text{diag}\{{\bf 1}^TY_{11}\}
\label{eq:yg1} \\
\bar Y_2 & := & \text{diag}\{{\bf 1}^TY_{11}\}
\ - \ Y_{12}Y_{22}^{-1}Y_{12}^T.
\label{eq:yg2}
\eq
\end{subequations}
The key observation, stated in the next result, is that $\mathcal G_1$, $\mathcal G_2$ have simple structures, that the matrices defined in \eqref{eq:yg1yg2} are indeed admittance matrices, 
and that $\mathcal G_1$, $\mathcal G_2$ are the underlying graphs of these admittance matrices. Even though we do not know the submatrices $Y_{11},~Y_{12},~Y_{22}$ of $Y$, the simple structures of $\mathcal G_1$, $\mathcal G_2$ allow us to compute $\bar Y_1,~\bar Y_2$ as we will see.

%\begin{assumption}\label{ass:Y22diagonal}
%There is no direct connection between two hidden nodes. 
%\end{assumption}
%Assumption~\ref{ass:Y22diagonal} will be relaxed later. 
\begin{theorem}[Separability]
\label{thm:separability}
Suppose the admittance matrix $Y$ satisfies Assumptions \ref{ass:0}, \ref{ass:2} and \ref{ass:degree2}. 
Then 
\begin{itemize}
\item[1.] $\mathcal{G}_1$ is a forest.
% and $\mathcal{G}_1 =\mathcal{G}_c/ \mathcal{G}_2$.
\item[2.] $\mathcal{G}_2=\oplus_i \mathcal{C}_i$ for some  $\mathcal C_i$ are edge-disjoint maximal cliques each with more than 2 nodes.\footnote{Strictly speaking, each $\mathcal C_i$ is a subgraph of $\mathcal G_2$ with $\mathcal M$ as its node set. It consists of a single maximal clique and the remaining isolated nodes in $\mathcal M$. We will abuse notation and use $\mathcal C_i$ to both refer to this subgraph of $\mathcal G_2$ and to the maximal clique in $\mathcal C_i$.}
\item[3.] $\mathcal G_1 = \mathcal G(\bar Y_1)$ and $\mathcal G_2 = \mathcal G(\bar Y_2)$ so that 
$\bar{\mathcal G} = \mathcal G_1 \oplus \mathcal G_2$.
\end{itemize}
\end{theorem}
\begin{IEEEproof}
For the first assertion, $\mathcal{G}_1$ is a forest since it is a subgraph of the tree $\mathcal{G}$.
For the second assertion $\mathcal{G}_2$ is a collection of maximal cliques $\mathcal{C}_i$ due to Step 4 of the graph condensation Algorithm \ref{alg:graphcond}. To show that the maximal clique (in each) $\mathcal{C}_i$ is of size at least 3, suppose $\mathcal{C}_i$ consists of $m_i$ (measured) nodes and, in the original graph $\mathcal G$, these $m_i$ measured nodes ``surround'' $h_i$ hidden nodes, i.e., the neighbors of each of these hidden nodes are either hidden nodes or nodes in $\mathcal{C}_i$. 
Let $d_j$ denote the degrees of hidden nodes $j=1, \dots, h_i$.  These $m_i+h_i$ nodes form a (connected) subtree of $\mathcal G$ with exactly $m_i+h_i-1$ edges. Since $m_i$ of these edges are between measured and hidden nodes and $h_i-1$ edges are between hidden nodes, we must have 
$\sum_{j=1}^{h_i} d_j  = m_i + 2(h_i-1)$
and hence $m_i = 2 + \sum_{i=1}^{h_i}(d_i-2)$.
Since $h_i\geq 1$ and $d_i\geq 3$ (Assumption \ref{ass:degree2}.2), we have $m_i\geq 3$.
To show that $\mathcal{C}_i$ and $\mathcal{C}_j$ are edge-disjoint, suppose for the sake of contradiction that there is an edge $(k,l)$ in both $\mathcal{C}_i$ and $\mathcal{C}_j$.  By the definition of $\mathcal G_2$, $(k,l)$ is not an edge in the original graph $\mathcal G$.  Since nodes $k, l$ are both in $\mathcal C_i$, there is a path from $k$ to $l$ in $\mathcal G$ that consists of only hidden nodes connected to measured nodes in the maximal clique $\mathcal{C}_i$.  Since nodes $k, l$ are both in $\mathcal C_j$, there is disjoint path from $k$ to $l$ in $\mathcal G$ that consists of a set of hidden nodes connected to  nodes in $\mathcal{C}_j$. These two paths form a loop in $\mathcal G$, a contradiction.  Hence $\mathcal{C}_i$ and $\mathcal{C}_j$ do not share any edge in $\mathcal G_2$.

For the third assertion, note that the matrix $\bar Y_1$ defined in \eqref{eq:yg1} is symmetric and hence an admittance matrix. 
The diagonal entry $Y_{11}[i,i]$ of $Y_{11}$ is the negative sum of the off-diagonal entries of the $i$th row/column of the original admittance matrix $Y$ (plus any shunt element at bus $i$), so that the $i$th entry of ${\bf 1}^TY_{11}$ is equal to the $i$th row sum of $Y_{12}$ (plus any shunt element at bus $i$).  Hence $\bar Y_1$ satisfies Assumption \ref{ass:degree2}.3 if $Y$ does.  Moreover, by the definition of $\mathcal G_1$, the edges in $\mathcal E_1$ correspond exactly to the off-diagonal entries of $Y_{11}$ that are nonzero.  
This implies that the graph $\mathcal G(\bar Y_1)$ that underlies the admittance matrix in \eqref{eq:yg1} is indeed $\mathcal G_1$.

The matrix $\bar Y_2$ defined in \eqref{eq:yg2} is also symmetric and hence is an admittance matrix. 
If $Y$ satisfies Assumption \ref{ass:degree2}.3, then
%\footnote{\slow{Doublecheck that topology identification remains OK when Assumption \ref{ass:degree2}.3 is relaxed.}\ye{for here, we consider zero shunt element case, later we also discussed nonzero shunt element case, where we can not recover the diagonal elements}}
\begin{align*}
{\bf 1}^TY_{11} = - {\bf 1}^TY_{12}^T,\qquad {\bf 1}^TY_{12} = -{\bf 1}^TY_{22}.
\end{align*}
This implies
\begin{equation*} 
\text{diag}\{{\bf 1}^TY_{11}\}=\text{diag}\{{\bf 1}^TY_{12}Y_{22}^{-1}Y_{12}^T\},
\end{equation*}
i.e., $\bar Y_2$ defined in \eqref{eq:yg2} satisfies Assumption \ref{ass:degree2}.3 when $Y$ does. 
%\ye{please see the revised version above: still working on it for an easier proof. Maybe  I miss something, I am just wondering whether we have already proven it from Lemma 1, $\bar{Y}=\bar{Y}_1+\bar{Y}_2$ and above results?}

Next we show that $\mathcal G_2 = \mathcal G(\bar Y_2)$.
% To show that, we need to show that $(i,j)\in \mathcal E_2$ (or equivalently, $\bar Y_2[i,j] \neq 0$) if and only if the (off-diagonal) $(i,j)$-entry $\left(Y_{12}Y_{22}^{-1}Y_{12}^T\right)[i,j]\neq 0$, where $\mathcal E_2$ is defined to be the set of edges produced by Step 4 of Algorithm ~\ref{alg:graphcond} or $\mathcal E(\bar Y)\setminus \mathcal E_1$. 
From \eqref{eq:yg1yg2}
\bqn
\bar Y & = & Y_{11} \ - \ Y_{12}Y_{22}^{-1}Y_{12}^T \ \ = \ \ 
\bar Y_1 \ + \ \bar Y_2,
\eqn
and hence $\bar{\mathcal G} \ := \ \mathcal G(\bar Y) \ = \ \mathcal G_1 \oplus \mathcal G_2$ with  $\mathcal{G}(\bar Y_1) =\mathcal G_1$.  Therefore we have $\mathcal{G}(\bar Y_2) =\mathcal G_2$. This concludes the proof. 
\end{IEEEproof}

\begin{remark}
From the third assertion, we have shown that, once $\mathcal G_1$ and $\mathcal G_2$ are obtained from $\mathcal G$, $\bar{Y}_1$ and $\bar{Y}_2$ defined in \eqref{eq:yg1yg2} can be obtained. 
\end{remark}
% such that $\mathcal{G}_c=\mathcal{G}_1\cup \mathcal{G}_2$.

%Graph-theoretically, $G_1$ ($Y_{11}$) corresponds to a tree, and $G_2$ ($Y_{12}Y_{22}^{-1}Y_{12}^T$) gives a union of fully-connected graphs.

%\begin{comment}
%\begin{lemma}
%For any two nodes in a radial network, they either have at most one shared neighbor 
%or have no common neighbor if they are adjacent. 
%\end{lemma}
%\begin{IEEEproof}
%This can be easily proved as it contradicts the tree topology.
%\end{IEEEproof}
%\end{comment}

%
%We can relax the Assumption~\ref{ass:Y22diagonal} 
%that there is no connection between hidden nodes. 
%When there exists a connection between hidden nodes, 
%we can still recover the network using Algorithm~\ref{alg:graph}. 
%
%%%\begin{corollary}
%%%Given a radial network that satisfies Assumption \ref{ass:degree2} and if there is no direct connection between hidden nodes, the reduced graph $\mathcal{G}_2$ can be decomposed to a number of cliques $\mathcal{C}_1,\ldots, \mathcal{C}_k$ (with the number of nodes $\ge 3$) 
%%%and a tree $\mathcal{T}$, such that $\mathcal{C}_1\cap \mathcal{C}_2=\{0\}$ 
%%%and the following statements hold:
%%%\begin{itemize}
%%%\item[1.] $\mathcal{G}_1=\mathcal{T}$;
%%%\item[2.] $\mathcal{G}_2=\cup_i \mathcal{C}_i$; 
%%%\item[3.] the number of cliques and the number of hidden states are equal, i.e., $k=h$.
%%%\end{itemize}
%%%\end{corollary}

%
%Finding the maximum clique of a graph is termed the clique problem and is NP-complete in general. 
There are many algorithms for solving the clique problem, such as the Bron-Kerbosch algorithm, which we adopt in Algorithm~\ref{alg:graph}.
%\begin{algorithm}[!]
%\caption{Bron-Kerbosch Algorithm}
%\label{alg:bk}
%\begin{algorithmic}[2]
% \State BronKerbosch $(R,P,X)$:
%       \If {$P$ and $X$ are both empty:}
%           \State report $R$ as a maximal clique
%           \EndIf
%       \For {each vertex $v$ in $P$:}
%           \State BronKerbosch $(R \cup {v}, P \cap N(v), X \cap N(v))$
%          $$P := P / {v}, X := X \cup {v}$$
%        \EndFor
%           \end{algorithmic}
%\end{algorithm}
           
\begin{algorithm}[htbp]
\caption{Graph Decoupling Algorithm}
\label{alg:graph}
\begin{algorithmic}[1]
    \State \textbf{Input:} a condensed graph $\bar{\mathcal{G}}$
	\State Set $\mathcal{G}'=\bar{\mathcal{G}}$, $i=1$.
	\While{$\mathcal{G}'$ has a clique with more than two nodes}
	\State Use Bron-Kerbosch Algorithm to find a clique ($\ge3$ nodes, together with other isolating nodes) $\mathcal{C}_i$ 
	in $\mathcal{G}'$, 
	\State Let $\mathcal{G}'=\mathcal{G}'/\mathcal{C}_i$, $i=i+1$,
	\EndWhile
    \State \textbf{return} $\mathcal{G}_2= \oplus_i \mathcal{C}_i $, $\mathcal{G}_1=\bar{\mathcal{G}}/ \mathcal{G}_2$  and the corresponding $\bar Y_1$ and  $\bar Y_2$.
\end{algorithmic}
\end{algorithm}

%\begin{remark}
%\end{remark}
%----------------------------------------------------------------------------------------
%\subsubsection{Recovering the admittance matrix $Y$}\label{ssec:y22}

\subsubsection{Partition of $Y$}\label{sec:partition}
Next we propose an algorithm to obtain $Y_{11}$, $Y_{22}$ and $Y_{12}$, and therefore the original admittance matrix $Y$.

The decomposition of $\bar{\mathcal G}$ into $\mathcal{G}_1$ and $\mathcal{G}_2$ guaranteed by Theorem \ref{thm:separability} allows us to partition the set $\mathcal M$ 
% $\mathcal M_1 =: \mathcal M_{1b} \cup \mathcal M_{1i}$ 
into a subset of \emph{internal} measured nodes that are not connected to any hidden nodes and a disjoint subset of \emph{boundary} measured nodes that connect to some hidden nodes. 
We can similarly partition $\mathcal H$ into a subset of \emph{internal} hidden nodes that are not connected to any measured nodes and the disjoint subset of \emph{boundary} hidden nodes that connect to some measured nodes.  The decomposition of $\bar{\mathcal G}$ into $\mathcal{G}_1$ and $\mathcal{G}_2$ identifies only the types of measured nodes, but not those of hidden nodes.
We can hence arrange the original admittance matrix $Y$ into the following structure (only the upper triangular submatrix is shown as $Y$ is symmetric):

\begin{subequations}
\bq\small
Y = \begin{bmatrix} 
    \begin{array}{c|c} Y_{11} & Y_{12} \\ \hline  & Y_{22} 
    \end{array} \end{bmatrix}
 =: 
\begin{bmatrix}
\begin{array}{c c | c c}
Y_{11,11} & Y_{11,12} &  0 & 0 \\
          & Y_{11,22} &  Y_{12,21} & 0 \\ \hline
  & &  Y_{22,11} & Y_{22,12} \\
  & &         & Y_{22,22}
\end{array}
\end{bmatrix}.
\label{eq:Ypartition.2a}
\eq\normalsize
Here, for $Y_{11}$, the submatrix $Y_{11,11}$ corresponds to connectivity among the internal measured nodes, $Y_{11,22}$ corresponds to connectivity among the boundary measured nodes, and 
$Y_{11,12}$ corresponds to connectivity between the internal and boundary measured nodes.  
Similarly, for $Y_{22}$, the submatrix $Y_{22,11}$ corresponds to connectivity among the boundary hidden nodes, $Y_{22,22}$ to that among the internal hidden nodes, and $Y_{22,12}$ to that between the internal and boundary hidden nodes.  
The submatrix $Y_{12,21}$ corresponds to connectivity between the set of boundary measured nodes and the set of boundary hidden nodes.
Denote the inverse $Y_{22}^{-1}$ by:
\bqn
X_{22} & := & Y_{22}^{-1} \ \ =: \ \ 
\begin{bmatrix} X_{22,11} & X_{22,12}  \\ X_{22,12}^T & X_{22,22} \end{bmatrix}.
\eqn
We have
\bq
\begin{aligned}
\bar Y  &=  Y_{11} \ - \ Y_{12}Y_{22}^{-1}Y_{12}^T  \\
& =  
\begin{bmatrix} Y_{11,11} & Y_{11,12} \\ Y_{11,12}^T & Y_{11,22} \end{bmatrix}
 - 
\begin{bmatrix} 0 & 0 \\ 0 &  Y_{12,21}X_{22,11}Y_{12,21}^T \end{bmatrix},
\label{eq:Y11Y12}
\end{aligned}
\eq
\label{eq:Ypartition.2}
\end{subequations}
where $X_{22,11}=(Y_{22,11}-Y_{22,12}Y_{22,22}^{-1}Y_{22,12}^T)^{-1}$  from \eqref{eq:Ypartition.2a} and the Woodbury formula. Specifically, given the definition of Schur complement $$\det (Y_{22,11}-Y_{22,12}Y_{22,22}^{-1}Y_{22,12}^T) \det Y_{22,22} = \det Y_{22},$$ and from the invertibility of $Y_{22}$ (shown in Proposition~\ref{lemma:invert}), the right-hand side of the above equation is nonzero. Therefore $\det (Y_{22,11}-Y_{22,12}Y_{22,22}^{-1}Y_{22,12}^T)$ cannot be zero, and as a result, the invertibility of $X_{22,11}$ can be guaranteed.

\begin{figure*}[!]
\centering
\includegraphics[width=0.8\textwidth] {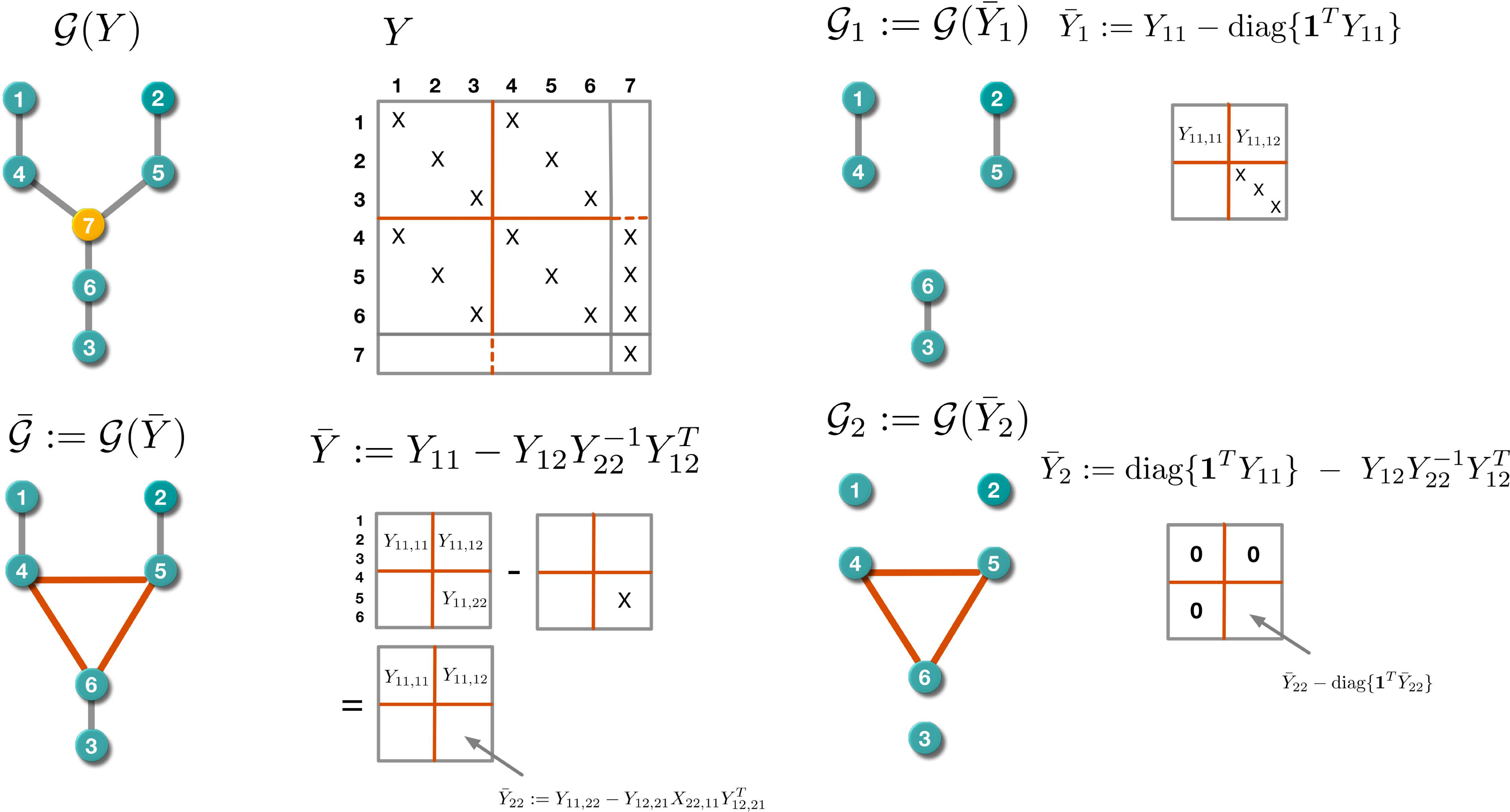} 
\caption{Illustration of admittance matrices and their underlying graphs: $(Y, \mathcal G(Y))$, $(\bar Y, \bar{\mathcal G})$, $(\bar Y_1, \mathcal G_1)$, and $(\bar Y_2, \mathcal G_2)$.}
\label{fig:201909-Example}
\end{figure*}

Since we can compute $\bar Y$ from partial voltage and current measurements, we can identify submatrices $Y_{11,11}$ and $Y_{11,12}$ for internal measured nodes from $\bar Y$ according to \eqref{eq:Y11Y12}.  
The edges in $\mathcal E_1$ correspond to the off-diagonal entries of $[Y_{11,11} \ \ Y_{11,12}]$ as well as $Y_{11,12}^T$, and they form a forest (Theorem \ref{thm:separability}).  The edges in $\mathcal E_2$ correspond to the off-diagonal entries of $Y_{11,22} - Y_{12,21}X_{22,11}Y_{12,21}^T$, and they form a collection of cliques. Recall that both $\mathcal G_1$ and $\mathcal G_2$ have $\mathcal M$ as their node set; see the example in Fig.~\ref{fig:201909-Example}.

%\slow{In Figure \ref{fig:201909-Example}, the lower-right submatrix in $\bar Y_1$ is not zero; it has nonzero diagonal entries.}\ye{that is correct}

In the rest of this subsection we focus on identifying the remaining submatrices $Y_{11,22}$, $Y_{12,21}$ as well as $Y_{22}$ (or specifically, $Y_{22,11},~Y_{22,12},~Y_{22,22}$) of $Y$.  For this purpose we assume without loss of generality that all measured nodes are boundary measured nodes, i.e., the rows and columns corresponding to submatrices $Y_{11,11}$ and $Y_{11,12}$ as well as their contributions to the diagonal entries of $Y_{11,22}$ have been removed from $Y$.  Then
\bq
Y  = \begin{bmatrix} 
    \begin{array}{c|c} Y_{11} & Y_{12} \\ \hline  & Y_{22} 
    \end{array} \end{bmatrix}
 =: 
\begin{bmatrix}
\begin{array}{c | c c}
Y_{11,22} &  Y_{12,21} & 0 \\ \hline
  &  Y_{22,11} & Y_{22,12} \\
  &         & Y_{22,22}
\end{array}
\end{bmatrix}.
\label{eq:Ypartition.2b}
\eq
Our goal is to identify $Y$ in \eqref{eq:Ypartition.2b} given its Kron-reduction: 
\bqn
\bar Y & = &  
Y_{11,22} - Y_{12,21} X_{22,11} Y_{12,21}^T.
\eqn

Theorem \ref{thm:separability}.2 implies that the underlying Kron-reduce graph $\mathcal G(\bar Y)$ is a disjoint collection of maximal cliques $\mathcal{C}_i$ among boundary measured nodes.
By \emph{hidden nodes in a maximal clique $\mathcal C_i$} of the Kron-reduced graph $\bar{\mathcal G}$, we mean the nonempty set of hidden nodes in the original graph $\mathcal G$ that are connected either to the measured nodes in $\mathcal C_i$ or other hidden nodes in $\mathcal C_i$. A measured node can be in multiple cliques $\mathcal C_i$ though $\mathcal C_i$ are edge-disjoint
(Theorem \ref{thm:separability}.2).
%\slow{Ye: In Lemma \ref{lemma:10}, do you mean ``A measured node is connected to exactly one hidden node in any clique $\mathcal C_i$ of which it is a member.''}\ye{exactly, this has been corrected.}

\begin{lemma}\label{lemma:10}
Suppose the admittance matrix $Y$ satisfies Assumptions \ref{ass:0}, \ref{ass:2} and \ref{ass:degree2}.
A measured node can connect to only one hidden node in any cliques $\mathcal{C}_i$ of which it is a member.
\end{lemma}
\begin{IEEEproof}
If a measured node connects to more than one hidden node in a maximal cliques $\mathcal{C}_i$, 
there exists a loop since there is a path between any two hidden nodes in $\mathcal{C}_i$, hence a contradiction. 
% and there is no path between nodes in different cliques. 
\end{IEEEproof}

We further assume, without loss of generality, that $\mathcal G(\bar Y)$ consists of a single clique; 
otherwise, we can repeatedly apply Algorithm~\ref{alg:Y22} below to each clique separately to determine the corresponding submatrices and then combine them to obtain $Y_{22}$ and $Y_{12}$. The general case has been discussed and elaborated in the Appendix.
%\slow{But see a question below on a measured node connected to two hidden nodes and form two cliques.} \ye{Similar to your later question, this is under the assumption that there is one clique without loss of generality.  It can be easily generalized to more than one clique after repeating the procedure and given the definition of $\oplus$.}

\begin{remark}\label{remark:3}
With this further assumption,
Lemma \ref{lemma:10} guarantees that $Y_{12}$ has exactly one nonzero element in each row. 
\end{remark}

\subsubsection{Parameterization of $Y$}\label{sec:parameterization}
Recall that there are $M$ (boundary) measured nodes, indexed by $1, \dots, M$, so that $Y_{11,22}$ in \eqref{eq:Ypartition.2b} is $M\times M$.  Suppose there are $H_b$ boundary hidden nodes, indexed by $M+1, \dots, M+H_b$, and
$H_i := H-H_b$ internal hidden nodes, indexed by $M+H_b+1, \dots, M+H$.  Then $Y_{22,11}$ in \eqref{eq:Ypartition.2b} is $H_b\times H_b$ and $Y_{22,22}$ is $H_i\times H_i$.
Suppose each measured node $i\in\{1, \dots, M\}$ is connected to the hidden node $h(i)\in\{M+1, \dots, M+H_b\}$ by a line with series admittance $y_{ih(i)}$.  From Remark \ref{remark:3} we know there is a unique $h(i)$ for each $i$, but voltage and current measurements only identify the identity of each measured node $i$, but not the hidden node $h(i)$ it is connected to (nor the values of $H, H_b, H_i$).  The next result suggests a method to identify all measured nodes that are connected to the same boundary hidden node.

\begin{proposition}\label{prop:samehiddennode}
Suppose the admittance matrix $Y$ satisfies Assumptions \ref{ass:0}, \ref{ass:2} and \ref{ass:degree2}.
Two measured nodes $i$ and $j$ are connected to the same hidden node if and only if the off-diagonal entries of rows $i$ and $j$ of $\bar Y_2$ are proportional, i.e., there exists $\gamma(i,j)\neq 0$ such that
\bqn
\frac{\bar Y_2[i,k]}{\bar Y_2[j,k]} & = & \gamma(i,j), \qquad k\neq i, j, \ k = 1, \dots, M.
\eqn
\end{proposition}
\begin{IEEEproof}
As discussed above, each row of $Y_{12,21}$ has exactly a single nonzero entry.  For 
each measured node $i$, let $h(i)$ denote the hidden node to which $i$ is adjacent
and $y_{ih(i)}$ denotes the series admittance of line $(i,h(i))$.  Then 
\begin{align*}
Y_{12,21} & \ =: \ 
\begin{bmatrix} - y_{1 h(1)} e_{h(1)}^{\sf T} \\ \vdots \\ - y_{M h(M)} e_{h(M)}^{\sf T} \end{bmatrix}
\end{align*}
where the unit vector $e_i\in\{0, 1\}^{H_b}$ is the column vector with a single 1 in the $i$th entry and 0 elsewhere.
Denote the $(i,j)$th entry of $X_{22,11}$ by $\beta_{ij}$.  
Then the $i$th row of matrix $Y_{12,21} X_{22,11} Y_{12,21}^{\sf T}$ in %\eqref{eq:defbarY2}
\eqref{eq:yg2} is
\begin{align*}
y_{ih(i)} \left[ \beta_{h(i)h(1)} y_{1h(1)} 		% \ \beta_{h(i)h(2)} y_{2h(2)} 
	\ \cdots \  \beta_{h(i)h(M)} y_{Mh(M)} \right].
\end{align*}

The {\bf necessity} of the proposition can be proven as follows: if $i$ and $j$ are connected to the same hidden node, i.e., $h(j) = h(i)$, then $b_{h(j)}^{\sf T} = b_{h(i)}^{\sf T}$.  Therefore, for $k\neq i, j$, we have from \eqref{eq:yg2}
\begin{align*}
\bar Y_2[i, k] & \ = \ \gamma(i,j)\, \bar Y_2[j,k]
\end{align*}
for $\gamma(i,j) := y_{ih(i)}/y_{jh(j)}$.  

We now present the {\bf sufficiency} proof. Fix $i\neq j$, and suppose there exists $\gamma'(i,j)\neq 0$ such that 
$\bar Y_2[i, k] = \gamma'(i,j) \, \bar Y_2[j,k]$ for all $k\neq i, j$.  Then 
\begin{align}
\gamma \beta_{h(i)h(k)} & \ = \ \beta_{h(j)h(k)}, 	\quad k\neq i,j
% y_{ih(i)} \beta_{h(i)h(k)} y_{kh(k)} & \ = \ \gamma' \, y_{jh(j)} \beta_{h(j)h(k)} y_{kh(k)},
\label{eq:beta_ij}
\end{align}
where $\gamma := y_{ih(i)}/\left( \gamma'(i,j) y_{jh(j)} \right)$.
We have to prove that $h(i) = h(j)$.

Suppose $h(i)\neq h(j)$.
Assume without loss of generality that $h(i) = 1$ and $h(j) = 2$ (corresponding 
to nodes $M+1$ and $M+2$ respectively).  By Assumptions \ref{ass:degree2}.3,
$Y$ is symmetric and hence $\bar Y$, $X_{22} := Y_{22}^{-1}$, and $X_{22,11}$
are symmetric. Then \eqref{eq:beta_ij} means that the matrix $X_{22,11}$ is of 
the form
\begin{subequations}
\begin{align}
X_{22,11} & \ = \ \left[ \begin{array} {c c | c c c}
\beta_{11} & \beta_{12} & \beta_{13} & \cdots & \beta_{1H_b} \\
\beta_{12} & \beta_{22} & \gamma \beta_{13} & \cdots & \gamma\beta_{1H_b} \\ \hline
\beta_{13} & \gamma \beta_{13} & \beta_{33} & \cdots & \beta_{3H_b} \\
\vdots & \vdots & \vdots & \ddots & \vdots \\
\beta_{1H_b} & \gamma \beta_{1H_b} & \beta_{3H_b} & \cdots & \beta_{H_bH_b} 
\end{array} \right]  
\label{eq:X_2211.2a}
\\
& \ =: \ \begin{bmatrix} A_{11} & A_{12} \\ A_{12}^{\sf T} & A_{22} \end{bmatrix}.
\label{eq:X_2211.2b}
\end{align}
In particular rank$(A_{12})\leq 1$.
\label{eq:X_2211.2}
\end{subequations}
Suppose more than one measured nodes are connected to each of the hidden nodes 
$h(i)$ and $h(j)$.  Specifically let $h(i) = h(i')$ and $h(j) = h(j')$ for distinct $i'$, $j'$.  
Then letting $k = i'$ and then $k=j'$ in \eqref{eq:beta_ij} implies that 
\begin{align*}
\beta_{h(i)h(i)} & \ = \ \beta_{h(i)h(i')} \ = \ \frac{1}{\gamma} \, \beta_{h(j)h(i')} 
	\ = \ \frac{1}{\gamma}\, \beta_{h(j)h(i)}	\\
\beta_{h(j)h(j)} & \ = \ \beta_{h(j)h(j')} \ = \ \gamma \, \beta_{h(i)h(j')}
	\ = \ \gamma\, \beta_{h(i)h(j)}.
\end{align*}
This implies that $\beta_{21} = \beta_{12} = \gamma\beta_{11}$ and $\beta_{22} = \gamma \beta_{12}$
in \eqref{eq:X_2211.2a}, i.e., the first two rows of $X_{22,11}$ are linearly dependent.
This contradicts the invertibility of $X_{22,11}$ and hence $h(i)=h(j)$ if more than one measured
nodes are connected to each of $h(i)$ and $h(j)$.

Suppose then at least one of $h(i)$, $h(j)$ is connected to a single measured node 
(i.e., $i$ or $j$) 
so that $\gamma \beta_{h(i)h(k)} = \beta_{h(j)h(k)}$ in \eqref{eq:beta_ij} may not hold for 
$k=i$ or $k=j$ or both.
Let 
 \begin{align}
X_{22,11}^{-1} & \ =: \ 
\begin{bmatrix} B_{11} & B_{12} \\ B_{12}^{\sf T} & B_{22} \end{bmatrix}
\label{eq:X_2211^-1}
\end{align}
where $B_{11}\in\mathbb C^{2\times 2}$ and $B_{22}\in\mathbb C^{(H_b-2)\times (H_b-2)}$.

%We now explain the mistake in the sufficiency proof of \cite[Proposition 3]{YuanLow2022}.
%A key property we will need is that rank$(B_{12})\leq 1$.  Recall that \eqref{eq:beta_ij} implies 
%rank$(A_{12})\leq 1$ in \eqref{eq:X_2211.2b}.  The proof in \cite{YuanLow2022}
%claims that $A_{11}$ must be nonsingular for $X_{22,11}$ to be nonsingular, 
%but $X_{22,11}$ can be nonsingular even if both $A_{12}$ and $A_{11}$ are of rank 1.
%If $A_{11}$ is indeed nonsingular, then we can express $X_{22,11}^{-1}$ in terms of the Schur 
%complement of $A_{11}$.  From \eqref{eq:A-1.a}, the off-diagonal block $B_{12}$ of 
%$X_{22,11}^{-1}$ is
%\begin{align*}
%B_{12} & \ = \ - A_{11}^{-1} \, A_{12} \, \left( X_{22,11}/A_{11} \right)^{-1}
%\end{align*}
%Therefore rank$(B_{12})\leq \text{rank}(A_{12})\leq 1$.

We now prove rank$(B_{12})\leq 1$ through direct calculation.
%, without assuming the nonsingularity of $A_{11}$.  
We then derive a contradiction first for the case when 
rank$\left( B_{12} \right) = 0$ and then when rank$\left( B_{12} \right) = 1$.
This implies that $h(i)=h(j)$, i.e., $i$ and $j$ are connected to the same (boundary)
hidden node.

\vspace{0.1in}\noindent
\textbf{Proof of rank$(B_{12})\leq 1$.}
From \eqref{eq:X_2211.2} and \eqref{eq:X_2211^-1}, the entries of $B_{12}$ are given by: 
for $k\geq 3$,
\begin{subequations}
\begin{align}
[B_{12}]_{1k} &   % \ = \ \frac{(-1)^{k+1}}{\text{det}(X_{22,11})}\, M_{k1}
	\ = \ \frac{(-1)^{k+1}}{\text{det}(X_{22,11})}\, 
	\left| \begin{array}{c | c} \beta_{12} & b^{\sf T} \\ \beta_{22} & \gamma b^{\sf T} \\ \hline
		\gamma c_{-k} & B_{-k}  \end{array} \right| 
\\
[B_{12}]_{2k} &   % \ = \ \frac{(-1)^{k}}{\text{det}(X_{22,11})}\, M_{k2}
	\ = \ \frac{(-1)^{k}}{\text{det}(X_{22,11})}\, 
	\left| \begin{array}{c | c} \beta_{11} & b^{\sf T} \\ \beta_{12} & \gamma b^{\sf T} \\ \hline
		c_{-k} & B_{-k}  \end{array} \right| 
\end{align}
where $b^{\sf T}$ is the first row of $A_{12}$,
\label{eq:B_12.1}
\end{subequations}
$c_{-k}$ is the $H_b-3$ dimensional (column) vector which is the first column
of $A_{12}^{\sf T}$ with its $k$th entry removed, and 
$B_{-k}$ is the $(H_b-3)\times (H_b-2)$ matrix which is $A_{22}$ with its $k$th
row removed.
For ease of reference we state the following calculation as a lemma without proof.
\begin{lemma}
\label{lemma:det.1}
Given any scalars $\alpha, \beta, \gamma\in\mathbb C$, any column vectors
$b, c$, and any matrix $B$ of matching sizes, we have the following determinant
\begin{align*}
\left| \begin{array}{c | c} \alpha & b^{\sf T} \\ \beta & \gamma b^{\sf T} \\ \hline
	c & B   \end{array} \right| & \ = \ (\gamma\alpha-\beta) 
	\left| \begin{array}{c} b^{\sf T} \\ B   \end{array} \right|. 
\end{align*}
\end{lemma}

Applying Lemma \ref{lemma:det.1} to \eqref{eq:B_12.1} we have
\begin{align*}
[B_{12}]_{1k} &  \ = \ (-1)^{k+1} \, a_1 \, d_{k},	&  k & \geq 3
\\
[B_{12}]_{2k} & \ = \  (-1)^k \, a_2\, d_{k},		&  k & \geq 3
\end{align*}
where 
\begin{align*}
a_1 & := \frac{\gamma\beta_{12} - \beta_{22}}{\text{det}(X_{22,11})}, &
a_2 & := \frac{\gamma\beta_{11} - \beta_{12}}{\text{det}(X_{22,11})}, &
d_{k} & := \left| \begin{array}{c} b^{\sf T} \\ B_{-k}   \end{array} \right| 
\end{align*}
i.e., $B_{12}$ is of the form
\begin{align}
B_{12} & \ = \ \begin{bmatrix}
	a_1 d_3 & -a_1 d_4 & \cdots & (-1)^{H_b-1} a_1 d_{H_b-2}  \\
	-a_2 d_3 & a_2 d_4 & \cdots & (-1)^{H_b} a_2 d_{H_b-2} \end{bmatrix}
\label{eq:B_12}
\end{align}
This shows that rank$(B_{12})\leq 1$.

\vspace{0.1in}\noindent
\textbf{Case 1: rank$(B_{12}) = 0$.}
The key observation is that, $X_{22,11}^{-1} = Y_{22}/Y_{22,22}$ 
is the admittance matrix of the Kron-reduced network consisting of only boundary hidden nodes where two boundary
hidden nodes $h(k)$ and $h(k')$ are adjacent if and only if, in the original network, $h(k)$
and $h(k')$ are either adjacent or are connected by a path with only internal hidden nodes
(even though $Y_{22}$ may not have zero row sums because each diagonal entry of $Y_{22}$
may include line admittances that can be interpreted as a shunt admittance in the Kron-reduced network).
Note that $h(k)$ and $h(k')$ cannot both be adjacent and connected by a path with only 
internal hidden nodes because, otherwise, there is a loop in the original network.
Then $B_{12}$ describes the connectivity of nodes $h(i)$, $h(j)$ with other boundary
hidden nodes in the Kron-reduced network.

When rank$\left( B_{12} \right) = 0$ (either $a_1 = a_2 = 0$ or $d_k=0$ for all $k\geq 3$
in \eqref{eq:B_12}), 
then $h(i)$ and $h(j)$ are connected at most to each other in the Kron-reduced network 
represented by $Y_{22}/Y_{22,22}$.  If either $h(i)$ or $h(j)$ is connected to a single 
measured node, this violates Assumptions \ref{ass:degree2}.2 that every hidden node has degree at least 3
in the original (not Kron-reduced) network.

\vspace{0.1in}\noindent
\textbf{Case 2: rank$(B_{12}) = 1$.}
Then either $a_1>$ or $a_2>0$, or both.
 Suppose one of $h(i)=1$ and $h(j)=2$, say, $h(i)$ is connected to more than one measured
 node (and $h(j)$ is connected to a single measured node).  The argument above shows 
 that $\gamma \beta_{h(i)h(i)} = \beta_{h(j)h(i)}
 = \beta_{h(i)h(j)}$ and hence $a_2 = 0$ and $[B_{12}]_{2k} = 0$ for all $k\geq 3$.
 This means that $h(j)=2$ is connected only to a single measured node and at most the
boundary hidden node $h(i)$ in the Kron-reduced network represented by $Y_{22}/Y_{22,22}$,
 violating Assumptions \ref{ass:degree2}.2 that every hidden node has degree at least 3 in the original
network.

 Suppose then both $h(i)$ and $h(j)$ are connected to a single measured node each.
 Then $a_1>0$ and $a_2>0$ because otherwise, either $[B_{12}]_{1k} = 0$ or
 $[B_{12}]_{2k} = 0$ for all $k\geq 3$, violating Assumptions \ref{ass:degree2}.2.  But \eqref{eq:B_12} implies
 that $[B_{12}]_{1k} \neq 0$ if and only if $[B_{12}]_{2k} \neq 0$, i.e., $h(i)$ and $h(j)$
 are connected to exactly the same set of boundary hidden nodes in the Kron-reduced
 network represented by $Y_{22}/Y_{22,22}$.  If $h(i)$ and $h(j)$ are connected to each
 other and no common boundary hidden node or if they are connected to only 1 common
 hidden node (and not to each other), then Assumptions \ref{ass:degree2}.2 is violated.  If they are connected to 2 or 
 more common boundary hidden nodes, then there must be a loop in the original 
 (not Kron-reduced) network, violating Assumptions \ref{ass:degree2}.1.  
 
 This completes the proof of sufficiency.

\end{IEEEproof}

Note that if there are only $M=3$ measured nodes then Assumptions \ref{ass:degree2}.1 and \ref{ass:degree2}.2 imply that all of them must be connected to the same boundary hidden node. 

Given the Kron-reduced admittance matrix $\bar Y_2$, Proposition \ref{prop:samehiddennode} allows us to group together the (boundary) measured nodes that are connected to the same (boundary) hidden node. This also identifies the number of boundary hidden nodes, even though we do not know (yet) the number or identity of internal hidden nodes nor the connectivity among the nodes.
We can re-arrange the submatrix matrix $Y_{12,21}$  into a form easier for identification.  
Specifically let measured nodes $1, \dots, k_1$ be connected to hidden node $M+1$, measured nodes $k_1+1, \dots, k_2$ to hidden node $M+2$, $\dots$, measured nodes $k_{H_b-1}+1, \dots, k_{H_b}:=M$ to hidden node $M+H_b$.  
Note that Proposition \ref{prop:samehiddennode} yields the values for $H_b$ and $(k_1, k_2, \dots, k_{H_b}=M)$ even though it provides no information about the value of $H$, the total number of hidden nodes.
To simplify notation, denote the series admittance $y_{i h(i)}$ of line $(i,h(i))$ by $y_i$. Then $Y_{12} = \begin{bmatrix} Y_{12,21} & 0 \end{bmatrix}$ where $Y_{12,21}$ is $M\times H_b$ and can be arranged into the following simple form:

{\small\begin{align*}
Y_{12,21} &= 
\begin{bmatrix}
-y_1  &  0  &  \cdots  &  0  \\
\vdots & \vdots & \vdots  & \vdots \\ 
-y_{k_1} &  0  &  \cdots  &  0  \\
0  &  -y_{k_1+1}  &  \cdots  &  0  \\
\vdots & \vdots & \vdots  & \vdots \\ 
0  &  -y_{k_2}  &  \cdots  &  0  \\
\vdots & \vdots & \vdots  & \vdots \\ 
0  &  0  &  \cdots   &   -y_{k_{H_b}}
\end{bmatrix}\\
 &=:  
\begin{bmatrix}
-\hat y_1 & 0 & \cdots & 0 \\
0  &  -\hat y_2  & \cdots & 0 \\
\vdots & \vdots & \ddots & \vdots \\
0 & 0 & \cdots & -\hat y_{H_b}
\end{bmatrix},
\end{align*}}

\noindent 
where for $i=1, \dots, H_b$, $\hat y_i$ is a $(k_i-k_{i-1})$-dimensional column vector corresponding to $k_i - k_{i-1}$ measured nodes that are connected to the hidden node $M+i$.
%where $\hat y_j$ are the column vectors:
%\bq
%\hat y_j & := &  \begin{bmatrix}  y_{k_{j-1}+1} u_{i_{k_{j-1}+1}}  \\ \vdots \\ y_{k_j} u_{i_{k_j}}  \end{bmatrix},
%\qquad j = 1, \dots, s \ \ \ \ \ (\text{with } k_0:=0).
%\label{eq:1HG:haty}
%\eq
Since $Y$ has zero row sum by Assumption \ref{ass:degree2}.3,
the diagonal matrix $\text{diag}\{{\bf 1}^TY_{11}\} = \text{diag}\{{\bf 1}^TY_{11,22}\} = \text{diag}(y_i = y_{ih(i)}, i=1, \dots, M)$. 
We have
\begin{align*}
	&Y_{12}Y_{22}^{-1} Y_{12}^T \\
	= &\text{diag}(\hat y_j)\, X_{22,11} \, \text{diag}(\hat y_j^T) \\
	=&\begin{bmatrix}
		\beta_{11}\, \hat y_1 \hat y_1^T & \beta_{12}\, \hat y_1 \hat y_2^T & \cdots & \beta_{1H_b}\, \hat y_1 \hat y_{H_b}^T \\
		\beta_{21}\, \hat y_2 \hat y_1^T & \beta_{22}\,  \hat y_2 \hat y_2^T & \cdots & \beta_{2H_b}\,  \hat y_2 \hat y_{H_b}^T \\
		\vdots & \vdots & \ddots & \vdots \\
		\beta_{H_b 1}\,  \hat y_{H_b} \hat y_1^T & \beta_{H_b 2}\,  \hat y_{H_b} \hat y_2^T & \cdots & \beta_{H_b H_b}\, \hat y_{H_b} \hat y_{H_b}^T
	\end{bmatrix}.
\end{align*}	
%\begin{align*}
%Y_{12}Y_{22}^{-1} Y_{12}^T &= \text{diag}(\hat y_j)\, X_{22,11} \, \text{diag}(\hat y_j^T)\\
%&=\begin{bmatrix}
%	\beta_{11}\, \hat y_1 \hat y_1^T & \beta_{12}\, \hat y_1 \hat y_2^T & \cdots & \beta_{1H_b}\, \hat y_1 \hat y_{H_b}^T \\
%	\beta_{21}\, \hat y_2 \hat y_1^T & \beta_{22}\,  \hat y_2 \hat y_2^T & \cdots & \beta_{2H_b}\,  \hat y_2 \hat y_{H_b}^T \\
%	\vdots & \vdots & \ddots & \vdots \\
%	\beta_{H_b 1}\,  \hat y_{H_b} \hat y_1^T & \beta_{H_b 2}\,  \hat y_{H_b} \hat y_2^T & \cdots & \beta_{H_b H_b}\, \hat y_{H_b} \hat y_{H_b}^T
%\end{bmatrix}.
%\end{align*}

\noindent Then the admittance matrix corresponding to the graph $\mathcal G_2$ in Theorem \ref{thm:separability} is:
\begin{align}
\bar Y_2  =&  	
\begin{bmatrix}
	\text{diag}(\hat y_1)  & 0 & \cdots & 0 \\
	& \text{diag}(\hat y_2)  & \cdots & 0 \\ 
	%	 &  & M_3 & \cdots & 0 \\
	&    & \ddots & \vdots  \\
	& &  & \text{diag}(\hat y_M)
\end{bmatrix} \nonumber  \\
&-  \begin{bmatrix}
	\beta_{11}\, \hat y_1 \hat y_1^T & \beta_{12}\, \hat y_1 \hat y_2^T & \cdots & \beta_{1H_b}\, \hat y_1 \hat y_{H_b}^T \\
	& \beta_{22}\, \hat y_2 \hat y_2^T & \cdots & \beta_{2H_b}\, \hat y_2 \hat y_{H_b}^T \\
	&  & \ddots & \vdots \\
	&   &  & \beta_{H_b H_b}\, \hat y_{H_b} \hat y_{H_b}^T
\end{bmatrix}.
\label{eq:1HG:barY.1}
\end{align}
Recall that we have already identified the Kron-reduced admittance matrix $\bar Y_2$, i.e., we know every entry of $\bar Y_{2}$ on the left-hand side of \eqref{eq:1HG:barY.1}. 
We now explain how to identify  $(\beta_{ij}, i,j=1,\dots, H_b)$ and $(y_i = y_{ih(i)}, i=1, \dots, M)$ on the right-hand side of \eqref{eq:1HG:barY.1}.  In particular, $(y_i = y_{ih(i)}, i=1, \dots, M)$ yields $Y_{12}$ of the original admittance matrix $Y$.
%
%\slow{My email to you in the last few days have been bounced: please take your time, no hurry. When all the technical details are sorted out, we can revisit the structure and decide how best to present the results, what to emphasize/deemphasize, etc.}
%
%\slow{How about the following flow of the algorithm, and presentation, from this point on:
%\begin{enumerate}
%\item Start with the knowledge of $\bar Y(N:=M) := \bar Y_2$ in (21), $M$ ($\bar Y_2$ is $M\times M$), and the total number $H_b$ of hidden boundary nodes.  
%\item Let $k$ be number of hidden nodes each of which is connected to at least 2 measured nodes. 
%For each such hidden node $j$, identify all the line admittances $y_i = y_{ih(i)}$ such that $h(i)=j$ as well as $\beta_{ij}$.
%\item Update expanded Kron-reduced admittance from $\bar Y(N)$ to $\bar Y(N+k)$.  Repeat Steps 2 and 3 until $Y = \bar Y(M+H)$ has been identified.
%\item Theorem: (i) In Step 2, if $Y = \bar Y(M+H)$ has not been identified then $k\geq 1$. Hence the procedure will continue and stop when $N+k = M+H$. (ii) When it stops, $Y = \bar Y(M+H)$ is correctly identified.
%\end{enumerate}
%}
%\ye{This is a great structure, I have changed the structure accordingly below}
%

\subsubsection{Computation of parameters in $Y$}\label{sec:computation}
%\noindent\textbf{Identification of $(y_i = y_{ih(i)}, i=1, \dots, M)$.}
Let $\bar Y_{2, k_1}$ be the diagonal submatrix consisting of the first $k_1$ rows and columns of $\bar Y_2$ corresponding to the first $k_1$ measured nodes connected to the first hidden node $M+1$: 
\begin{align}
\bar Y_{2, k_1}  := & 	
	\text{diag}(\hat y_1) \ \ - \ \  
\beta_{11}\, \hat y_1 \hat y_1^T 
\nonumber \\
 =&  \begin{bmatrix}
	y_1  & \cdots & 0 \\
	  &   \ddots & \vdots  \\
	 &  & y_{k_1} 
	 \end{bmatrix} - 
\beta_{11}
\begin{bmatrix} y_1 \\ \vdots \\ y_{k_1} \end{bmatrix}
\begin{bmatrix} y_1 & \cdots & y_{k_1} \end{bmatrix}.
\label{eq:1HG:barY.2}
\end{align}
We first explain how to identify $(\beta_{11}$, $\hat y_1)$ on the right-hand side of \eqref{eq:1HG:barY.2} from the knowledge of $\bar Y_{2,k_1}$ on the left-hand side of \eqref{eq:1HG:barY.2}. 
The identification of other $\beta_{ii}, \hat y_i$ corresponding to $k_i - k_{i-1}$ measured nodes connected to the hidden node $M+i$ from the diagonal blocks 
$\bar Y_{2, k_i} := \text{diag}(y_{k_{i-1}-1}, \dots, y_{k_i}) - \beta_{ii}\, \hat y_i \hat y_i^T$ can be done similarly. 
%We next explain how to identify non-diagonal blocks of $\bar Y_2$ by considering the following scenarios.

\emph{Case 1: $k_1\geq 2$.} In this case, hidden node $M+1$ is connected to two or more measured nodes indexed by $i=1, \dots, k_1$. Consider the first two measured nodes and the corresponding $2\times 2$ principal submatrix of $Y_{2,k_1}$: for $i,j = 1, 2$
\bq
\bar Y_{2,k_1}[i,j] & = & \left\{ \begin{array}{lcl}
	  y_i  - \beta_{11} \, y_{i}^2 & \text{ if } & i=j\\
	- \beta_{11}\, y_{i} y_{j} & \text{ if } & i \neq j
	 \end{array}  \right.
\label{eq:1HG:barY.3b}
\eq
This leads to the following equations in $(\beta_{11}, y_1, y_2)$:
\begin{align*}
y_1-\beta_{11}y_1^2 \ &= \ \bar{Y}_{2,k_1}[1,1] \ =: \ a_1\\
- \beta_{11}y_1y_2 \ &= \ \bar{Y}_{2,k_1}[1,2] \ =: \ a_2\\
y_2-\beta_{11}y_2^2 \ &=\ \bar{Y}_{2,k_1}[2,2] \ =: \ a_3
\end{align*}
yielding:
\begin{equation}  \label{eq:1HG:barY.3b.y1y2b11}
\begin{aligned}
	y_1 &=  \frac{a_1a_3 - a_2^2}{a_2+a_3}, \quad y_2 = \frac{a_1a_3 - a_2^2}{a_1+a_2}, \\
	\beta_{11} &=  -\frac{a_2(a_1+a_2)(a_2+a_3)}{\left(a_1a_3 - a_2^2\right)^2}.
\end{aligned}	
\end{equation}
To identify other $(y_j, j>2)$, note that
\bqn
- \beta_{11}\, y_1\, y_j & = & Y_{2, k_1}[1,j], \qquad j = 3, \dots, k_1
\eqn
yielding
\bqn
y_j & = & -\frac{Y_{2, k_1}[1,j]}{\beta_{11}\, y_1},
\eqn
where $\beta_{11}$ and $y_1$ are given by \eqref{eq:1HG:barY.3b.y1y2b11}. Once $\hat y_1, \dots, \hat y_{k_j}$ are found, we can calculate from off-diagonal entries of $\bar Y_2$ all $\beta_{ij}$ from \eqref{eq:1HG:barY.1}.

\emph{Case 2:}
Once we have {recovered} the coefficients for hidden boundary nodes with at least two connections to measured nodes in Case 1, next, we can treat these recovered hidden nodes as measured nodes and repeat the above procedure until no hidden node is left. A key step is to construct a new Kron reduced matrix once parts of the admittance matrix have been found. Let the original $Y$ have the following partition as in \eqref{eq:Ypartition.2b}:
\begin{equation*}
Y = \begin{bmatrix}
\begin{array}{c | c c}
Y_{11,22} &  Y_{12,21} & 0 \\ \hline
  &  Y_{22,11} & Y_{22,12} \\
  &         & Y_{22,22}
\end{array}
\end{bmatrix}.
\end{equation*}
%where $Y_{11,11}$ { represents the admittances} between a subset of measured nodes that are connected to hidden nodes with at least $2$ connections to measured nodes, $Y_{11,22}$ represents {admittances} between the remaining measured nodes.
%\slow{In (19), $Y_{22,11}$ corresponds to the set of boundary hidden nodes and $Y_{22,22}$ the set of internal hidden nodes.  So in (27), should both $Y_{11,11}$ and $Y_{11,22}$ be connected to $Y_{22,11}$?}\ye{this is a different partition than the one in (27). }
The Kron reduced admittance matrix can be decomposed to $\bar Y_1$ and $\bar Y_2$. Specifically, $\bar Y_2$ has the following form:
\begin{align*}
	\bar{Y}_2 =& \text{diag} \{{\bf 1}^TY_{11,22} \}- Y_{12,21} X_{22,11} Y_{12,21}^T\\ 
	=& \text{diag} \{{\bf 1}^TY_{11,22} \}- \begin{bmatrix}
		Y_{12,21} & 0 
			\end{bmatrix} \begin{bmatrix}
		Y_{22,11} & Y_{22,12}\\
		Y^T_{22,12} & Y_{22,22} 
	\end{bmatrix}^{-1}\begin{bmatrix}
		Y_{12,21}^T \\
		0 
	\end{bmatrix}.	
\end{align*}
Based on the results in Case 1, one can recover $\text{diag} \{{\bf 1}^TY_{11,22} \}$, $Y_{12,21}$ and $X_{22,11}$. Since $\bar Y_1$ is known from Algorithm~\ref{alg:graph}, $\text{diag} \{{\bf 1}^TY_{11,22} \}$ allows us to compute $Y_{11}$ from the equality \eqref{eq:yg1} and the partition in \eqref{eq:Ypartition.2b}:
\begin{equation}\label{eq:Y11}
Y_{11}  = \bar Y_1 + \text{diag} \{{\bf 1}^TY_{11,22} \}.
\end{equation}
%Hence the entire rows and columns of $Y$ corresponding to the boundary measured nodes are known after \eqref{eq:Y11}. We can then focus on the submatrices $Y_{22,11},~Y_{22,12},~Y_{22,22}$ corresponding to only the boundary and internal hidden nodes, i.e., we can reduce the unknown admittance matrix $Y$ to the new smaller admittance matrix in the red box below:
%\slow{ It's not obvious what ``hidden boundary nodes are discovered and the corresponding coefficients are identified'' mean. Need to connect the identification of $(\beta_{ij}, y_i)$ in Step 1 above to this. Perhaps explain of the original $Y$ matrix in \eqref{eq:Ypartition.2b}, what entries are known after Step 1 above, how that determine the new Kron-reduced admittance $\bar Y$, etc.}
Hence the entire rows and columns of $Y$ corresponding to the boundary measured nodes are known after \eqref{eq:Y11}. We can then focus on the submatrices $Y_{22,11},~Y_{22,12},~Y_{22,22}$ corresponding to only the boundary and internal hidden nodes, i.e., we can reduce the unknown admittance matrix $Y$ to the new smaller admittance matrix below, which amounts to restricting attention to the subgraph without the boundary measured nodes.
\begin{equation*}
Y = \begin{bmatrix}
\begin{array}{ c | c}
   Y_{22,11} & Y_{22,12} \\\hline
       & Y_{22,22}
\end{array}
\end{bmatrix}.
\end{equation*}
The Kron reduced admittance matrix of this new (unknown) admittance matrix $Y$ can then be obtained from the knowledge of $X_{22,11}$:
\begin{equation*}
\bar Y := Y_{22,11} - Y_{22,12} Y_{22,22}^{-1} Y_{22,12}^T = X_{22,11}^{-1}.
\end{equation*}
Moreover, we have identified the set of boundary hidden nodes. Applying Theorem~\ref{thm:separability}, Algorithm~\ref{alg:graph} and Proposition~\ref{prop:samehiddennode} to this new $\bar Y$ allows us to identify a set of internal hidden nodes to which this set of boundary hidden nodes are connected. Moreover, we can treat the set of boundary hidden nodes as boundary measured nodes and the newly identified internal hidden nodes as boundary hidden nodes. Therefore, even though we do not know the number or the identity of the \emph{remaining} internal hidden nodes, we can partition the new (unknown) admittance matrix $Y$ into the form at the beginning of Case 2 and therefore repeat the computation on this new (smaller) admittance matrix recursively, strictly reducing the number of internal hidden nodes in each iteration until the set of internal hidden nodes becomes null.

\emph{Case 3:} For any hidden node that connects to one or zero measured node, these hidden nodes will eventually have more than one connection to measured nodes once the other hidden nodes {have been recovered} and therefore can be recovered. It is easy to show that there will never exist a scenario that all the hidden nodes have at most $1$ connection to measured nodes for a tree graph. To see this, note that for any clique, $H\ge M$ as every hidden node connects to a different measured node. On one hand, the sum of all hidden nodes' degrees is greater than $3H$ {under Assumption~\ref{ass:degree2}}. On the other hand,  it is at most $2(H-1)+M$, which is twice the sum of all edges between hidden nodes and the number of connections between hidden nodes and measured nodes. However, $2(H-1)+M<3H$, a contradiction.

\emph{Case 4:}
 If all hidden nodes are hidden boundary nodes, i.e., $Y_{22}= Y_{22,11}$, then $Y_{22} = X_{22,11}^{-1}$ and hence the entire admittance matrix $Y$ can be identified. If there are hidden nodes that are not hidden boundary nodes, we can treat hidden boundary nodes as measured nodes now and repeat the above procedure based on Case 2. 

\subsubsection{Overall recursive algorithm}\label{sec:algorithm}
The overall identification procedure is summarized in Algorithm \ref{alg:Y22}\footnote{For notational simplicity, we assume without loss of generality that all measured nodes are boundary measured nodes. Yet, this assumption can easily be relaxed.}. 
\begin{algorithm}[!]
\caption{Recover $Y$ from $\bar{Y}$}
\label{alg:Y22}
\begin{algorithmic}[1]
%	\For {$i = 1: h$}
	\State \textbf{Input:} $\bar Y_1$ and $\bar Y_2$
%	\State Let $\hat{Y}_{22} = \bar Y_2$
	\For{each pair of nodes $(j,k)$}
	\State Compute $\gamma[j,k]$ from $\bar{Y}_2$. 
	\EndFor
%\For {$j,k = 1: h$}
%	\If{$\beta[j,k]=1$}
%	$$\hat{Y}_{11}[j,j] = \bar{Y}_{11}[j,j] -  \bar{Y}_{11}[j,k]/\alpha[j,k]$$
%	$$\hat{Y}_{11}[k,k] = \bar{Y}_{11}[k,k] -  \bar{Y}_{11}[k,j]\times\alpha[j,k]$$
%	\EndIf
	\State Solve for $\text{diag}\{{\bf 1}^TY_{11,22} \}$, $Y_{12,21}$ and $X_{22,11}$ from \eqref{eq:1HG:barY.2}, set $\hat{Y} = \begin{bmatrix} \bar Y_1+ \text{diag}\{{\bf 1}^TY_{11,22} \} &  Y_{12,21} \\ Y_{12,21}^T &X_{22,11}^{-1}\end{bmatrix}:=\begin{bmatrix} \hat Y_{11}  &  \hat Y_{12} \\ \hat Y_{12}^T & \hat Y_{22}\end{bmatrix}$
	and set $\bar{Y}_{2} = X_{22,11}^{-1}$
	\If{the graph corresponding to $\bar{Y}_2$, i.e., $\mathcal{G}(\bar{Y}_2)$ is not radial}
	\For{each pair of nodes $(j,k)$}
	\State Compute $\gamma[j,k]$ from $\bar{Y}_2$. 
	\EndFor
       	\State Solve for $\text{diag}\{{\bf 1}^TY_{11,22} \}$, $Y_{12,21}$ and $X_{22,11}$ from \eqref{eq:1HG:barY.2} and set
$\hat{Y} = \begin{bmatrix} \hat{Y}_{11} &  \hat{Y}_{12} & 0 \\ \hat{Y}_{12}^T & \text{diag}\{{\bf 1}^TY_{11,22} \} & Y_{12,21} \\ 0 & Y_{12,21}^T & X_{22,11}^{-1}\end{bmatrix}$. 
        \State Set
        $$\hat{Y}_{11}=\begin{bmatrix} \hat{Y}_{11} &  \hat{Y}_{12} \\ \hat{Y}_{12}^T & \text{diag}\{{\bf 1}^TY_{11,22} \}
        \end{bmatrix}, ~ \hat{Y}_{12} = \begin{bmatrix} 0\\Y_{12,21}\end{bmatrix}.
	$$
	\State Set  $\bar{Y} = X_{22,11}^{-1}$ and apply Algorithm~\ref{alg:graph} to obtain $\bar Y_1$ and $\bar Y_2$.
	 \EndIf
	%and compute the weight of the links connecting observed nodes to hidden nodes
	%by parameterizing a matrix $X$. 
%	\State Solve for $\text{diag}\{{\bf 1}^TY_{11}\}=\hat{Y}_{12}X_{22}\hat{Y}^T_{12}+\bar Y_2$ and $\hat{Y}_{11}$ according to \eqref{eq:Y11}
%%%%%	\State Repeat the above process and treat $\hat{Y}_{22}$ as $\bar{Y}$
%	\State Obtain $\hat{Y}_{22,11}, \hat{Y}_{22,12}, \hat{Y}_{22,22}$ such that 
%	$\hat{Y}_{22} = \hat{Y}_{22,11}- \hat{Y}_{22,12} \hat{Y}_{22,22}^{-1}\hat{Y}_{22,12}^T.$
%        \State Set $\hat{Y}_{22}=\hat{Y}_{22,22}$
%        \If{the graph corresponding to $\hat{Y}_{22,22}$ contains a loop}
%        \State Repeat the process from Line $14$ to $16$
%        \Else 
        \State \textbf{return} $Y=\hat{Y}$ 
%        \EndIf
\end{algorithmic}
\end{algorithm}

\section{Simulations}\label{sec:eval}
%----------------------------------------------------------------------------------------

%\slow{Move simulations to later section and include simulations with hidden nodes?}
%\ye{sure, the issue is that we don't have real simulations with hidden nodes except some examples in the appendix using symbols.}

\subsection{Simulated data}
In this section we implement the proposed algorithms in MATLAB
and evaluate their identification accuracy
by performing simulations in MATPOWER~\cite{matpower}. 
The optimization problems are solved using the CVX toolbox~\cite{cvx}.
We run power flow analysis on the IEEE $14$-bus test system, 
representing a portion of a power system in the Midwestern U.S., 
which has $14$ buses, $11$ aggregated loads, and $5$ generators, 
$3$ of which are synchronous compensators used for reactive power support~\cite{ieee14}. 
To validate our identification algorithms in three-phase distribution systems, 
we have also performed simulations on IEEE test distribution networks, 
which are not presented here (see our previous work~\cite{ardakanian2016event} 
for the identification results in distribution systems).

%\begin{figure}[!]
%\center
%\includegraphics[width=.47\textwidth]{figure/14bus.jpg}
%\caption{The IEEE $14$-bus test system~\cite{ieee14}.}
%\label{fig:14}
%\end{figure}

We assume that PMUs are installed at selected buses and 
that they can precisely measure the voltage and current magnitudes and phase angles 
that we obtain from power flow calculations, unless stated otherwise.
For each scenario, we run $100$ steady state simulations, each pertaining to a time slot, 
to determine the voltage and current magnitude and phase angle of every bus,
while varying the real and reactive power demand of the loads across the time slots.
Specifically, for a given time slot, 
the real and reactive power consumption of a constant PQ load 
are computed by multiplying a scaling factor 
drawn from a uniform distribution over the interval $[0.8, 1.2]$ by the real and reactive power consumption data provided in~\cite{ieee14}.
We obtain the admittance matrix of this system 
using a built-in function of the power flow simulator.
It turns out that the absolute values of nonzero complex elements 
of the admittance matrix are between $1.86$ and $40.06$,
reminding the readers that a complex number's absolute value 
is its distance from zero in the complex plane.

\begin{figure}[!]
\center
\includegraphics[width=.4\textwidth]{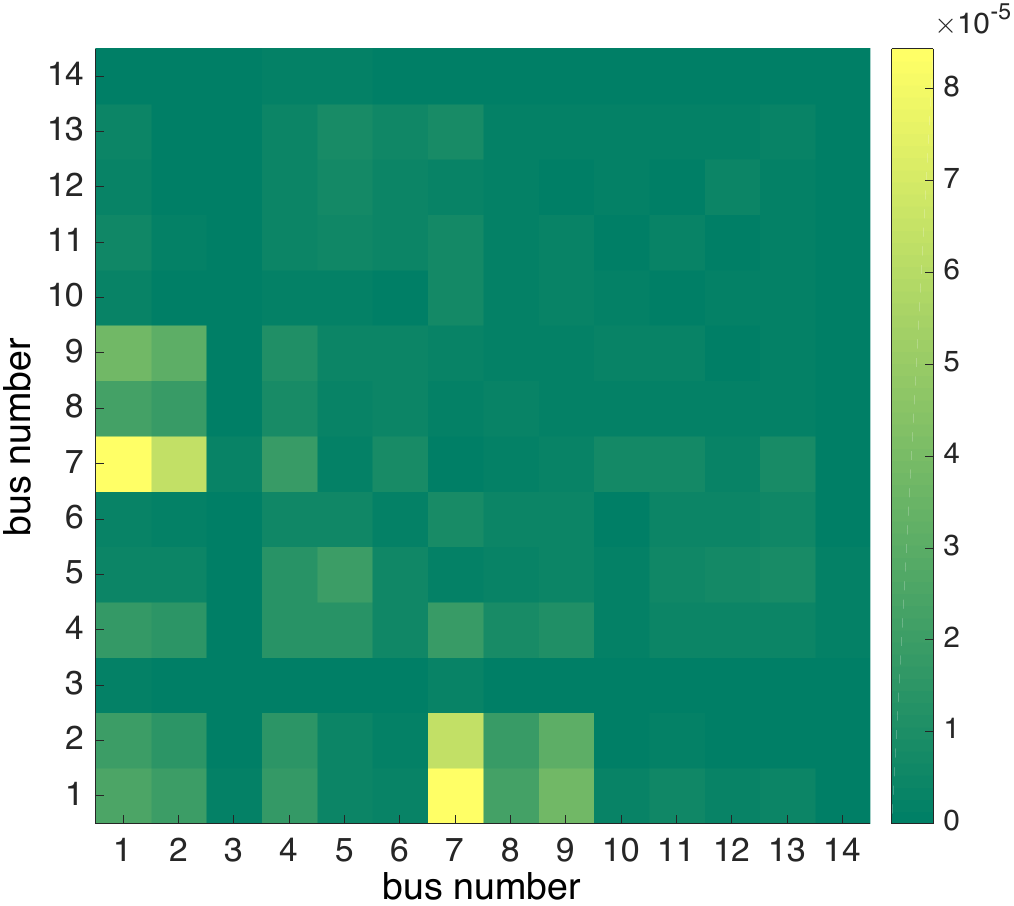}
\caption{The identification error when there is no hidden state. The color of a cell located at row $i$ and column $j$ represents the value of $|Y[i,j]-\hat{Y}[i,j]|$. It can be seen that the identification error using measurements of $15$ time slots ($K=15$) is quite small compared to the absolute value of the elements of $Y$.
}
\label{fig:basewonoise}
\end{figure}

\begin{figure}[!]
\center
\includegraphics[width=.4\textwidth]{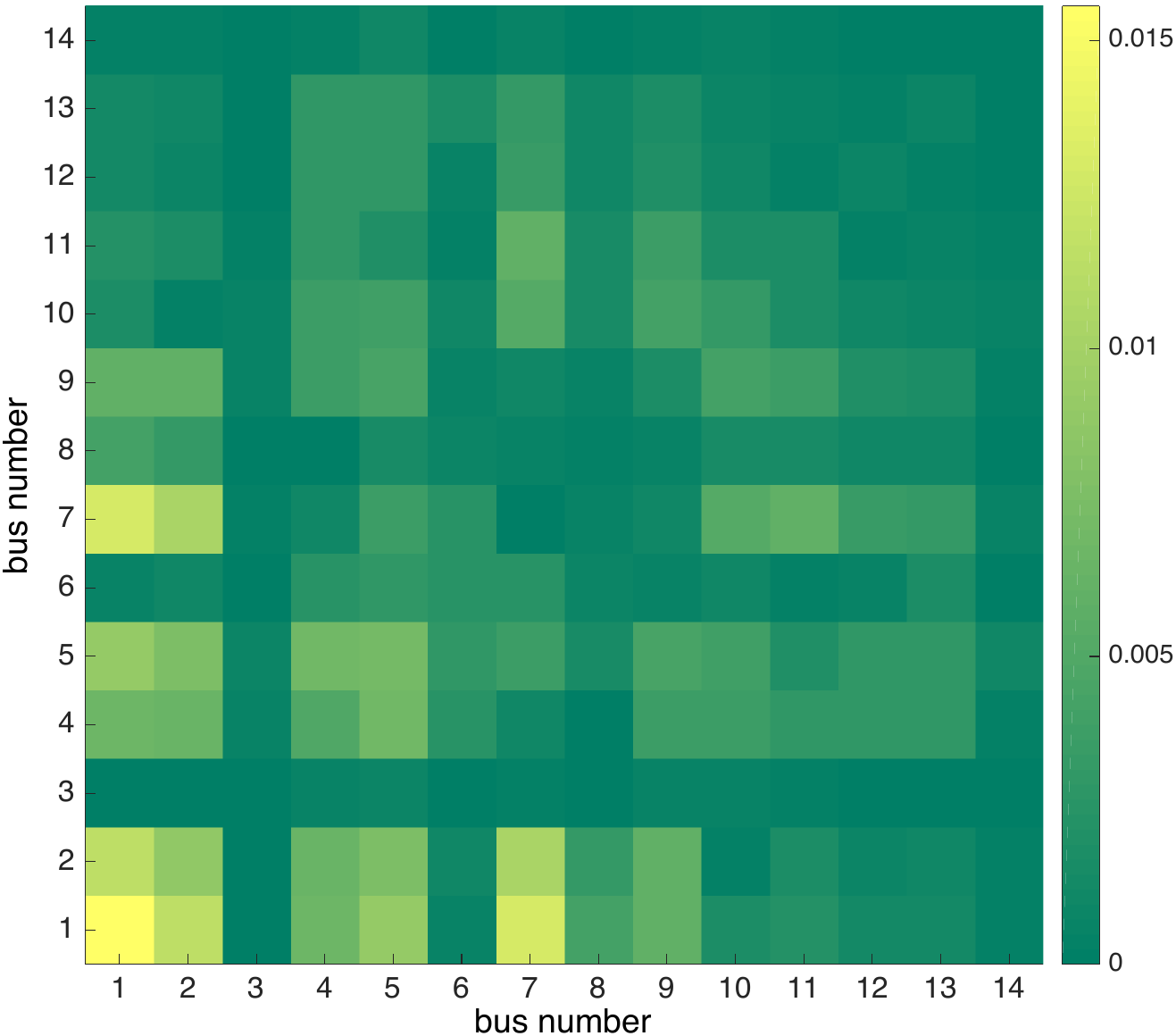}
\caption{The identification error when white Gaussian noise is added to both complex voltage and current measurements. Similar to the previous case, the errors are sufficiently small.}
\label{fig:basewnoise}
\end{figure}

We first consider the scenario that every bus is equipped with a PMU.
Assuming that the self admittance of bus $7$, i.e., the transformer bus, is known,
Fig.~\ref{fig:basewonoise} shows the identification error, 
defined as $|Y-\hat{Y}|$,
and the vertical color bar indicates the mapping of data values into colors.
Hence, the color of a cell located at row $i$ and column $j$
represents the value of $|Y[i,j]-\hat{Y}[i,j]|$.
It can be seen that the identification error 
using measurements of $15$ time slots ($K=15$) is quite small 
compared to the absolute value of the elements of $Y$,
and this error does not vary much if we use more observations.
We next analyze the sensitivity of the proposed algorithm 
to the measurement error which is typically introduced by the transducers.
To this end, white Gaussian noise with a signal-to-noise ratio of $125$
is added to both complex voltage and current measurements.
The signal-to-noise ratio is chosen such that the measurement accuracy
lies within the reported range for existing PMU technology.
Fig.~\ref{fig:basewnoise} shows the absolute identification error for this case.
Similar to the previous case, the errors are sufficiently small.
In general, we observe that the identification error increases as 
we decrease the signal-to-noise ratio 
and it becomes really large when the signal-to-noise ratio drops below $100$;
at this point we say that the $Y$ matrix cannot be identified from data.

\subsection{Experimental data}
\begin{figure}[!]
	\centering
	\includegraphics[width=.3\textwidth]{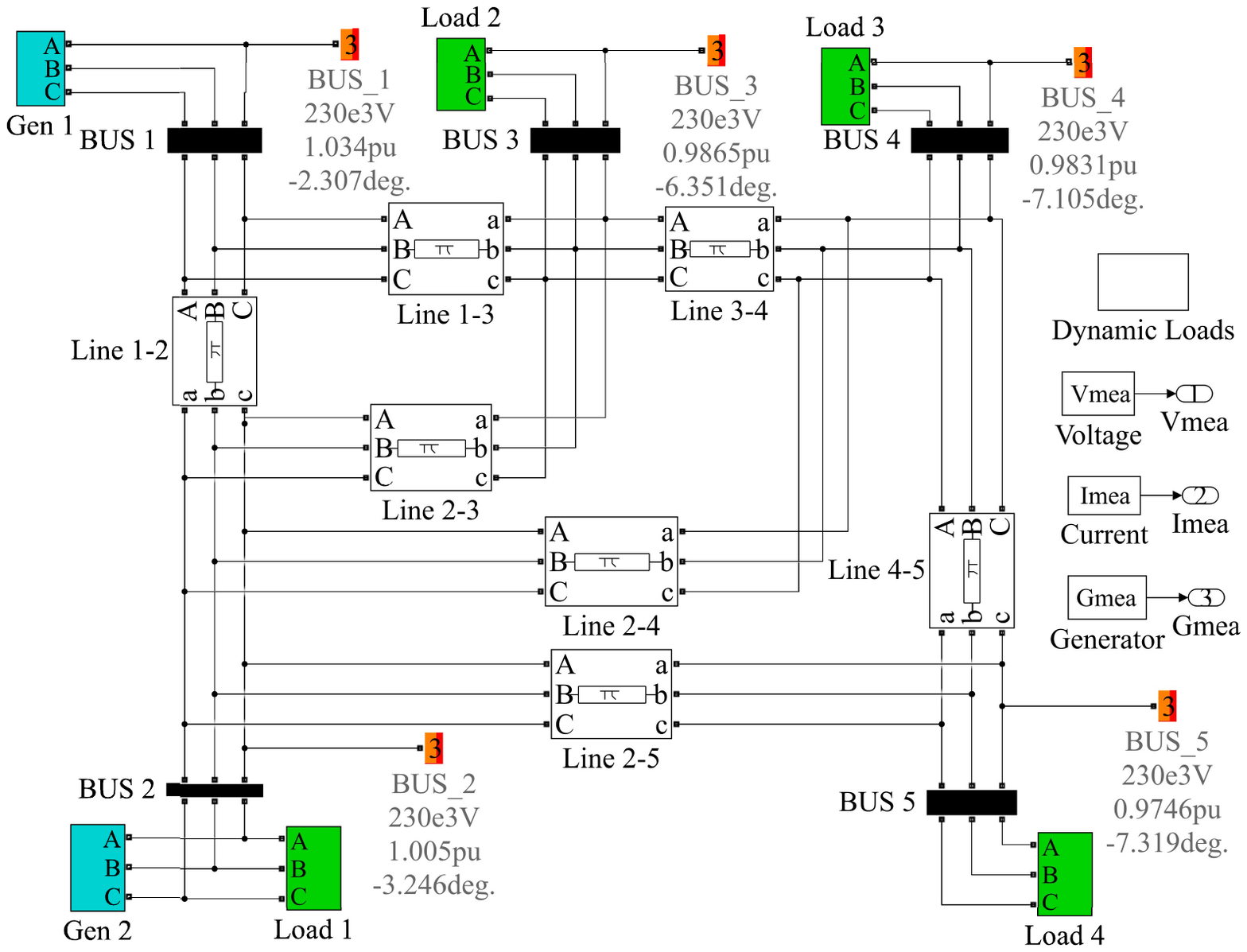}
		\includegraphics[width=.3\textwidth]{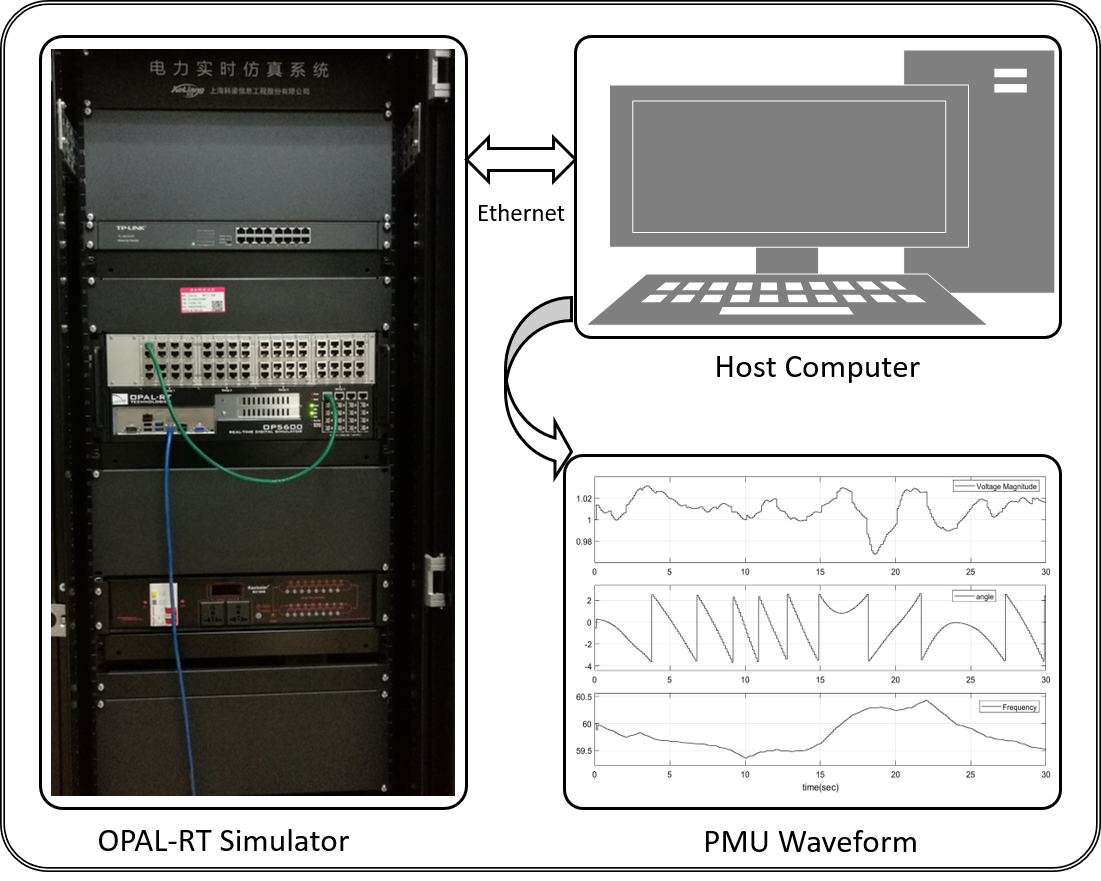}
	\caption{Left: The network topology of the 5-bus benchmark systems. Right: The OPAL-RT real-time simulation platform that are used for data generation.}
	\label{fig:5busmodel}
\end{figure}

\begin{figure}[htbp]
	\centering
	\includegraphics[width=.4\textwidth]{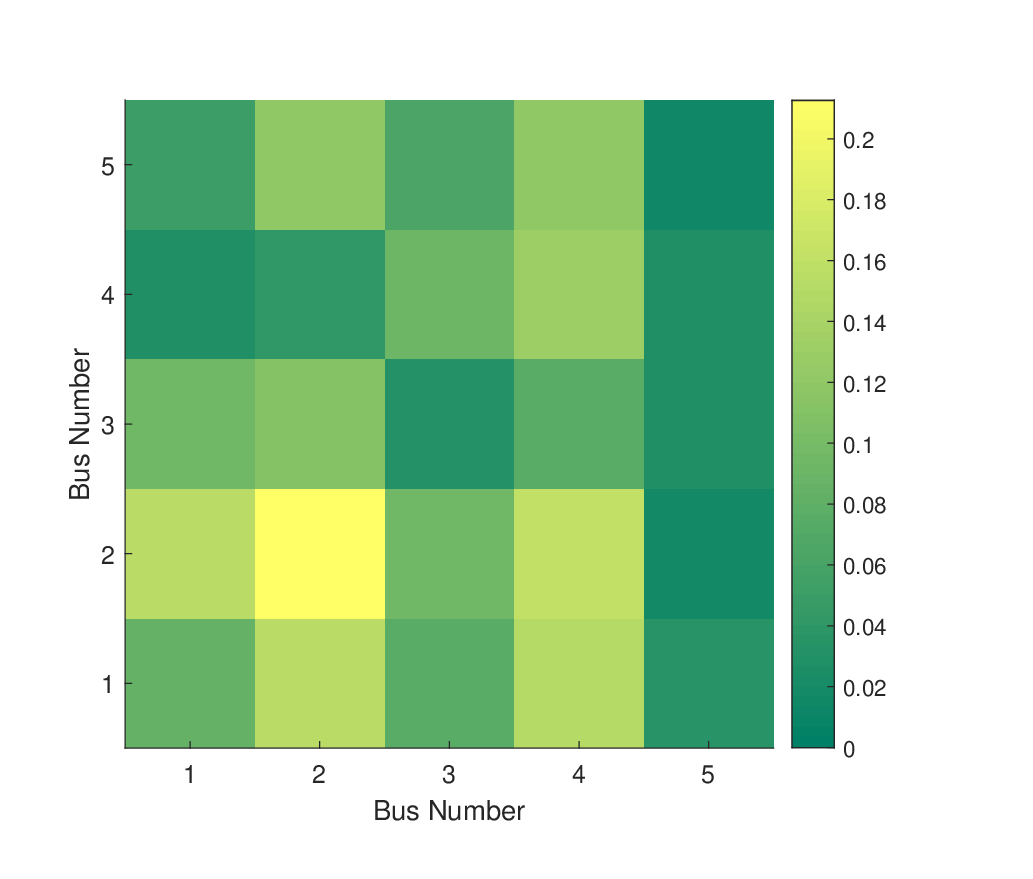}
	\caption{The identified error of admittance matrix along buses for the 5-bus benchmark example we tried. }
	\label{fig:5busErr}
\end{figure}

Next, we test the proposed IPF algorithm on the IEEE 5-bus benchmark shown in the left panel of Fig.~\ref{fig:5busmodel}. The network contains $4$ dynamic loads, $2$ generators and $7$ transmission lines. The system admittance matrix is obtained from ~\cite{IEEE5bus} with the absolute values of nonzero complex elements 
of the admittance matrix are between $3.80$ and $31.32$. Simulation is performed on \textsc{OPAL-RT} real time simulator~\cite{opal} in our lab shown in the right panel of Fig.~\ref{fig:5busmodel}. The generator is represented by a sixth order model with turbine governors and excitation systems. Assume that the sampling frequency of PMU is $20$ Hz and the dynamic loads change over time. We collect PMU data (nodal voltages and injected currents) of all buses for $40$ seconds. 
Note that the frequency fluctuates with changes of the load demands, which is consistent with field measurements; 
PMU has $0.49\%$ average total vector error in simulation. 
Fig.~\ref{fig:5busErr} shows the identification error, defined as $|Y-\hat{Y}|$ (element-wise absolute value), and the vertical color bar indicates the mapping of data values into colors. One can see that the IPF is capable of identifying the topology correctly and estimating system parameters with small error from noisy data with the maximum parameter estimation error  $0.2126$. 

%The least squares (LS) method is utilized for inverse power flow. 
%Two metrics, topology success rate and normalized root-mean-square error (\text{nRMSE}) of estimated parameters, are used for evaluation,
%\[ \text{nRMSE} =\frac{\|Y_{id}-Y\|_F}{\|Y\|_F}, \]
%where $Y_{id}$ is the identified admittance matrix.
%The topology success rate is the ratio of correct identified number of lines to the total.

%%
%\begin{table}[htbp]
%	\begin{center}
%		\centering
%		\caption{The identified results.}
%		\begin{tabular}{c|c|c}
%			\noalign{\global\arrayrulewidth1pt}\hline\noalign{\global\arrayrulewidth0.8pt}
%			\rowcolor{mygray}
%			Method & topology success rate &  nRMSE \\
%			\hline
%			LS & $100\%$  &	0.0089 \\	
%			\hline
%		\end{tabular}%
%		\label{tab:5busresults}%
%	\end{center}
%\end{table}%

% \begin{figure}[!]
%	\centering
%	\includegraphics[width=.3\textwidth]{figure/device1.png}
%	\caption{The OPAL-RT real-time simulation platform that are used for data generation.}
%	\label{fig:platform}
%\end{figure}

\subsection{Hidden node case}
\begin{figure}[!]
\centering
\includegraphics[width=.4\textwidth]{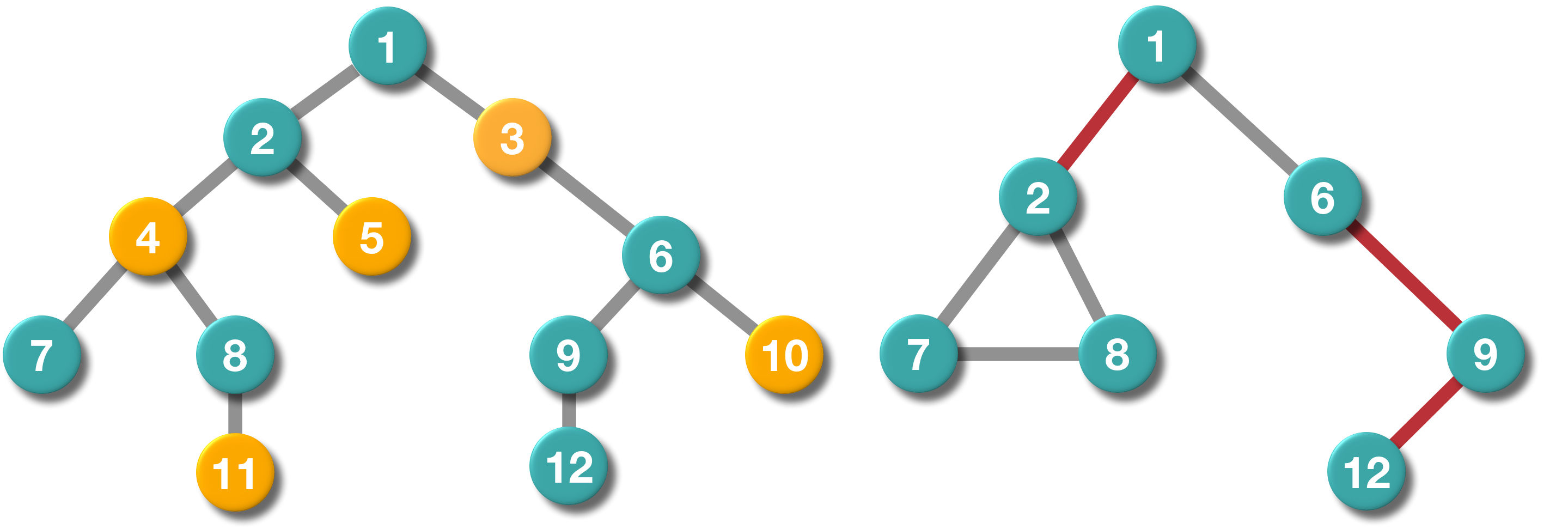}
\caption{An example of a $12$-bus network. 
\textbf{Left:} The original network topology. 
Green circles denote the observed nodes while yellow ones denote the hidden nodes. 
\textbf{Right:} The topology of the condensed graph obtained from Kron reduction. 
The red lines denote the edges in $\mathcal{G}_1$ 
and the grey ones denote the ones in $\mathcal{G}_2$.}
\label{fig:ex4}
\end{figure}

Given a graph shown in Fig.~\ref{fig:ex4} (left), 
if sensors are deployed at nodes $\mathcal{M}=\{1, 2, 6, 7, 8, 9, 12\}$, we can use Kron reduction to obtain the graph shown in Fig.~\ref{fig:ex4} (right). 
In this example, the hidden nodes are $\mathcal{H}=\{3, 4, 5, 10, 11\}$, 
from which nodes $\{3, 5, 10, 11\}$ have degree less than $3$, 
and therefore, cannot be identified from data. 
We now illustrate how the proposed algorithm can
identify the actual admittance matrix including Node $4$. 

\begin{figure}[!]
\centering
\includegraphics[width=0.5\textwidth]{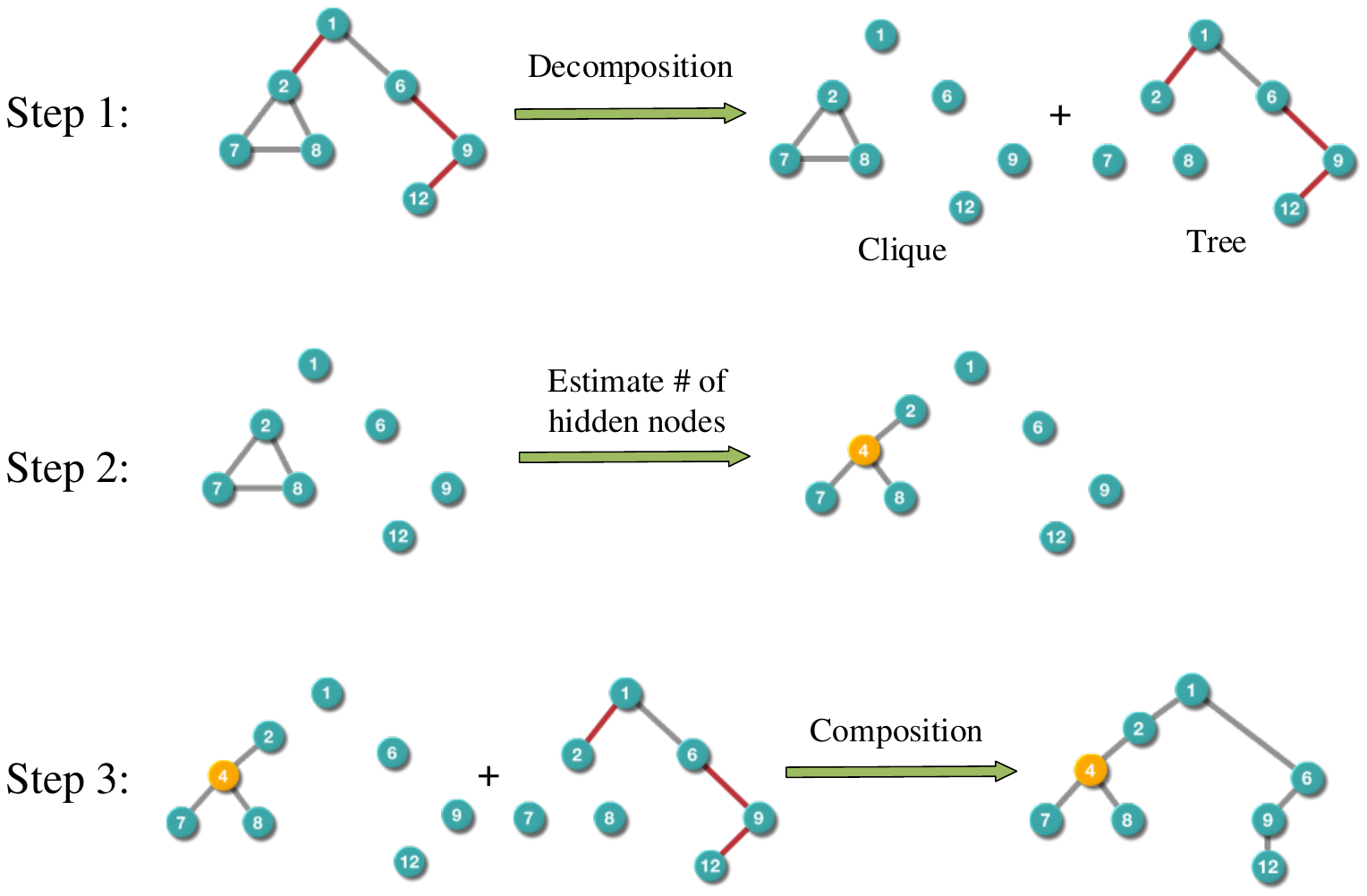}
\caption{A step-by-step illustration of the proposed algorithm. Step $1$ is described in Algorithm~\ref{alg:graph} that one separates the graph to a number of cliques and a forest. Step $2$ is described in Algorithm~\ref{alg:Y22} that one recovers  the admittance submatrix from every clique. Step $3$ combines all the recovered graphs (together with admittance submatrices) to recover the whole admittance matrix.}
\label{fig:ex3}
\end{figure}

The first step is to decompose the graph corresponding to the estimated $\bar{Y}$ matrix to two graphs using Algorithm~\ref{alg:graph}, one is a collection of cliques and the other one is a tree and some isolated nodes. The second step is to apply Algorithm~\ref{alg:Y22} to identify the original admittance matrices for all cliques. The final step is to take the union of the cliques and the graph obtained in the second step. These steps are shown in Fig.~\ref{fig:ex3}. 

Indeed, even the original graph does not satisfy the assumption on the degree of hidden nodes. We can recover a graph with smallest number of nodes that is a) consistent with $\bar{Y}$ and b) satisfies Assumption~\ref{ass:degree2}.

Consider another graph with the following topology in Fig.~\ref{fig:new_topology}, the admittance matrix has the following form\footnote{For notational simplicity, we used $Y_{ij}$ to denote the $Y[i,j]$ element in this section.}: 
\begin{figure}[!]
\centering
\includegraphics[width=.3\textwidth]{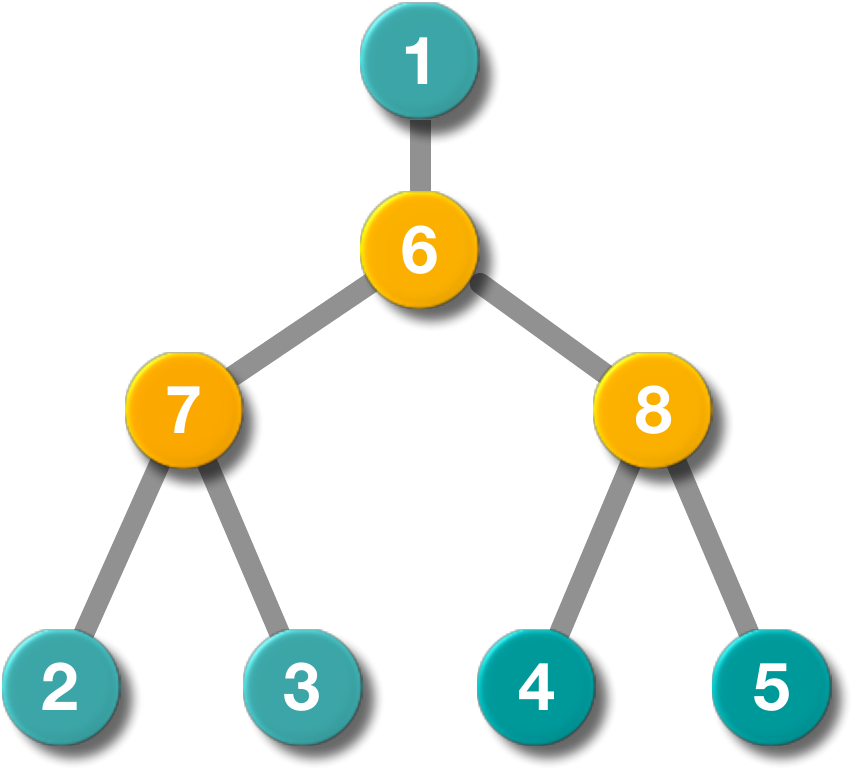}
\caption{An example of a $8$-bus network, where $5$ nodes are measured, and there exists three hidden nodes.}
\label{fig:new_topology}
\end{figure} 
$$Y=\begin{bmatrix}
Y_{11} & 0 & 0 & 0 & 0 & -Y_{16}  & 0 & 0 \\
& Y_{22} & 0 & 0 & 0 & 0 & -Y_{27}& 0\\
& & Y_{33} & 0 & 0 & 0 & -Y_{37} & 0\\
& & & Y_{44} & 0 & 0 & 0 & -Y_{48}\\
& & & &Y_{55}  & 0 & 0 & -Y_{58}\\
& & & &  & Y_{66} & -Y_{67} & -Y_{68}\\
& & & &  &  &  Y_{77} & 0\\
& & & &  &  &  & Y_{88}
\end{bmatrix}.$$
Through direct computation, we have 
{\tiny\begin{align*}
&\bar{Y}=\\
&\begin{bmatrix} 
Y_{11}-X_{11}Y_{16}^2 & -X_{12} Y_{16}Y_{27} & -X_{12}Y_{16}Y_{37} & -X_{13}Y_{16}Y_{48} & -X_{13}Y_{16}Y_{58}\\
& Y_{22}-X_{22}Y_{27}^2 & -X_{22}Y_{27}Y_{37} & -X_{23}Y_{27}Y_{48} & -X_{23}Y_{27}Y_{58}\\
& & Y_{33}-X_{22}Y_{37}^2  & -X_{22}Y_{37}Y_{48} & - X_{23}Y_{37}Y_{58}\\
& & & Y_{44}-X_{33}Y_{48}^2 & - X_{33}Y_{48}Y_{58}\\
& & & & Y_{55}-X_{33}Y_{58}^2
\end{bmatrix}.
\end{align*}}

For the `zero shunt element' case, we follow Algorithm~\ref{alg:Y22}. 
For starters, we compute $\gamma$
$$
\gamma=\begin{bmatrix} 
* & 0 & 0 & 0 & 0\\
& *& 1 & 0 & 0\\
& & * & 0 & 0\\
& & & *  & 1\\
& & & & *  
\end{bmatrix}.
$$
It means that nodes $2$ and $3$ are connected to the same hidden node, and 
nodes $4$ and $5$ are connected to the same hidden node. 
Thus, it is straightforward to compute $Y_{22}, Y_{33}, Y_{44}, Y_{55}, X_{22}, X_{23}, X_{33}$ from $\bar{Y}$ and furthermore $Y_{27}=-Y_{22}, Y_{37}=-Y_{33}, Y_{48}=-Y_{44}, Y_{58}=-Y_{55}$. 
Next, we obtain a ``new'' $\bar{Y}$ by including two new hidden boundary nodes as measured nodes in eq.~\eqref{eq:sinu_Ybar}. 
\begin{figure*}
\begin{equation}\label{eq:sinu_Ybar}
\bar{Y}_{\text{new}}
=\begin{bmatrix}
Y_{\text{new},11}  & 0  &    \begin{bmatrix}X_{12}Y_{16} & X_{13}Y_{16}\end{bmatrix}  \begin{bmatrix} X_{22},  X_{23}\\ X_{23}, X_{33}
\end{bmatrix}^{-1}\\
       & \text{diag}\{Y_{22},Y_{33},Y_{44}, Y_{55}\} &        \begin{bmatrix} 
        -Y_{27} & 0 \\
        -Y_{37} & 0 \\
                0 & -Y_{48} \\
              0& -Y_{58}\\
        \end{bmatrix}\\
   & &     
        \begin{bmatrix} X_{22},  X_{23}\\ X_{23}, X_{33}
\end{bmatrix}^{-1}
\end{bmatrix},
\end{equation}
\end{figure*}
where $\begin{bmatrix}X_{12}Y_{16} & X_{13}Y_{16}\end{bmatrix}$ can be computed from the corresponding row of $\bar{Y}$ and
\begin{align*}
\bar{Y}_{\text{new},11} &= Y_{11}-X_{11}Y_{16}^2 \\
&+\begin{bmatrix}X_{12}Y_{16} & X_{13}Y_{16}\end{bmatrix}  \begin{bmatrix} X_{22},  X_{23}\\ X_{23}, X_{33}
\end{bmatrix}^{-1}\begin{bmatrix}X_{12}Y_{16} \\ X_{13}Y_{16}\end{bmatrix}.
\end{align*}
It is noted that nodes $2,3,4,5$ have no connections to non hidden boundary nodes, then  they can be removed.    
$${\small
\bar{Y}_{\text{new}}
%=\begin{bmatrix}
%Y_{\text{new},11}   &    \begin{bmatrix}X_{12}Y_{16} & X_{13}Y_{16}\end{bmatrix}  \begin{bmatrix} X_{22},  X_{23}\\ X_{23}, X_{33}
%\end{bmatrix}^{-1}\\
%       &     
%        \begin{bmatrix} X_{22},  X_{23}\\ X_{23}, X_{33}
%\end{bmatrix}^{-1}
%\end{bmatrix}
=\begin{bmatrix}
Y_{11}-Y_{66}^{-1}Y_{16}^2   &    Y_{66}^{-1}Y_{16}\begin{bmatrix}Y_{67} &  Y_{68}\end{bmatrix}.
\\
       &     
      \begin{bmatrix}Y_{77} & 0 \\ 0 & Y_{88}\end{bmatrix}- Y_{66}^{-1}      \begin{bmatrix}Y_{67} \\  Y_{68}\end{bmatrix}
\begin{bmatrix}Y_{67} &  Y_{68}\end{bmatrix}
\end{bmatrix}}.
$$
The last equality due to matrix inverse lemma. We  can repeat the procedure, by noting that the size of $\bar{Y}_{\text{new}}$ is three, and furthermore solve for $Y_{11}, Y_{77}, Y_{88}, Y_{66}, Y_{67}, Y_{68}, Y_{16}$, therefore recover the original admittance matrix $Y$.

\section{Conclusions}\label{sec:discussion}
%----------------------------------------------------------------------------------------

This paper presents a framework for the inverse power flow problem 
which identifies the admittance matrix of a power system from 
synchronized voltage and current measurements pertaining to a subset of its buses. 
The algorithms proposed in this work can identify the graph topology together with its associated admittance matrix with guarantee for radial networks; and it can further jointly address state estimation and topology identification problems with theoretical guarantees, if certain conditions are met.
These findings are supported by high-fidelity power flow simulations performed on standard test systems. 

In future work, we plan to extend our framework to three phase power flow models, which takes the mutual coupling between phases into account, develop efficient algorithms for identifying the admittance matrix of radial distribution systems with few measurement nodes, and analyze the sensitivity of the identification results to non-stationary measurement errors. 

\section{Acknowledgement} 
We thank Dr. Wei Zhou (Huazhong University of Science and Technology) for useful discussion. Ye Yuan was supported by the National Natural Science Foundation of China under Grant 92167201.

\appendix
Section~\ref{sec:hidden} describes in detail how to identify each maximal clique in isolation and claims that admittance matrices obtained from these maximal cliques can be put together to construct the admittance matrix $Y$ of the entire network. Here, we elaborate on how this is done.

Theorem~\ref{thm:separability} implies that the graph $\mathcal G(\Bar Y)$ underlying the Kron-reduced admittance matrix $\bar Y$ contains a collection of edge-disjoint maximal cliques $\{ \mathcal C_j(\Bar Y) \}$.  The boundary observed nodes in each $\mathcal C_j(\Bar Y)$ are connected to each other through a tree $\mathcal T_j(Y)$ of hidden nodes in the original network $\mathcal G(Y)$.  
% The tree $\mathcal T_j(Y)$ terminates in boundary observed nodes in $\mathcal C_j(\Bar Y)$.
As mentioned previously, every boundary observed node in $\mathcal C_j(\Bar Y)$ is connected to exactly one hidden node in $\mathcal T_j(Y)$ for, otherwise, there exists a loop in $\mathcal G(Y)$, contradicting that $\mathcal G(Y)$ is a tree.

Theorem~\ref{thm:separability} does not explicitly describe how these maximal cliques are connected in the graph ${\mathcal G}_2$.  Two maximal cliques can be connected in exactly one of three ways, corresponding to how the boundary observed nodes in the original graph $\mathcal G(Y)$ are connected through internal observed or hidden nodes.  For each of these three cases, we now explain how to derive the (Kron-reduced) admittance matrix of every maximal clique and how to put them together to construct the overall admittance matrix $Y$ after each maximal clique has been identified in isolation.

\begin{figure}[htbp]
	\centering
 \subfigure [Disconnected maximal cliques] {
	\includegraphics[width=0.2\textwidth] {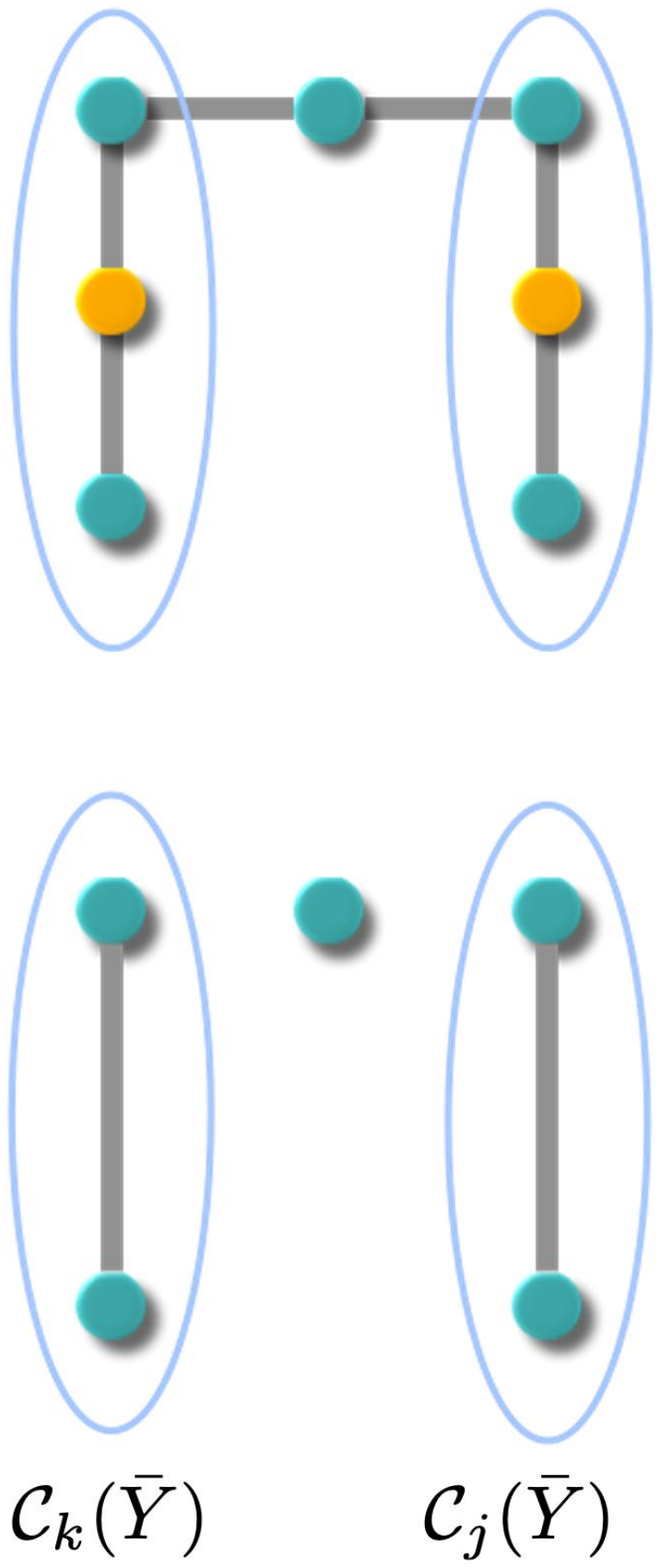}
 }
 \qquad
\subfigure [Connected by a line in $Y_{11,22}$] {
	\includegraphics[width=0.2\textwidth] {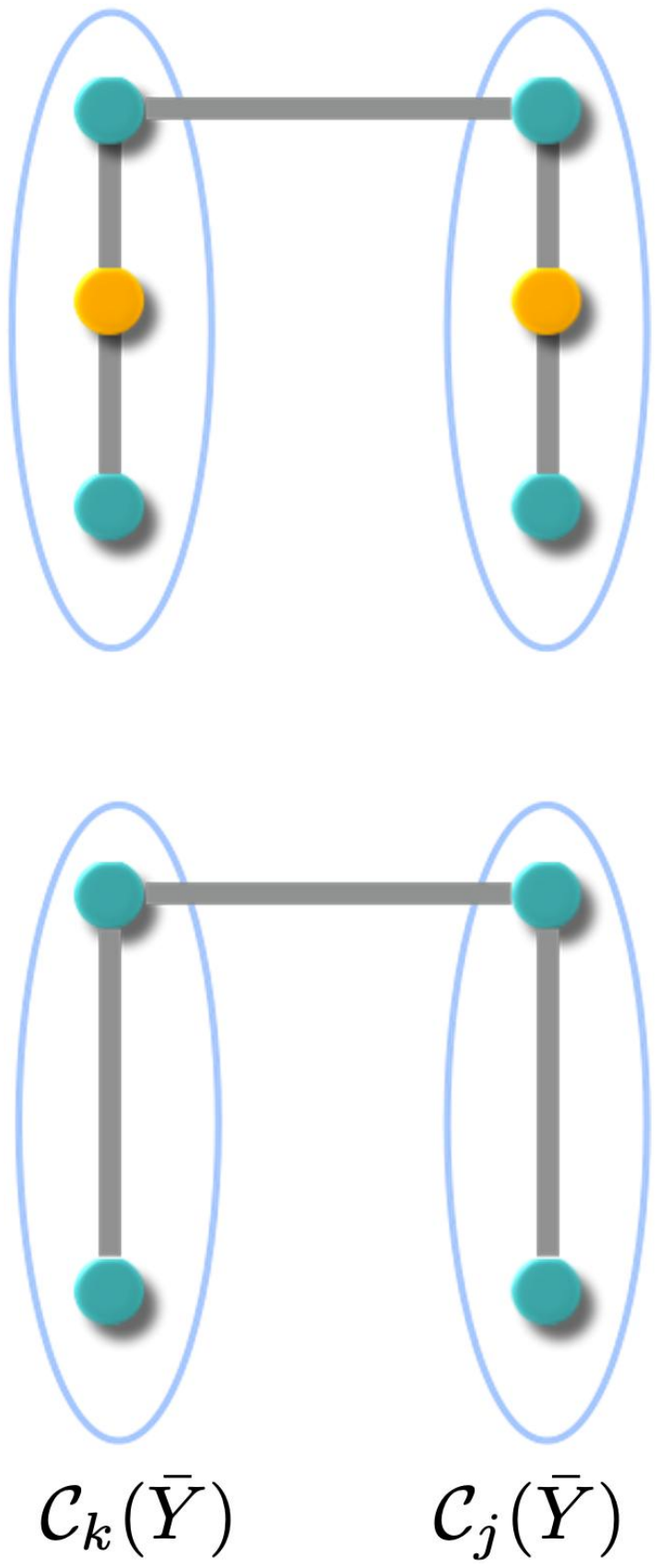}
 }
 \quad\quad
 \subfigure [Connected by a boundary observed node] {
	\includegraphics[width=0.3\textwidth] {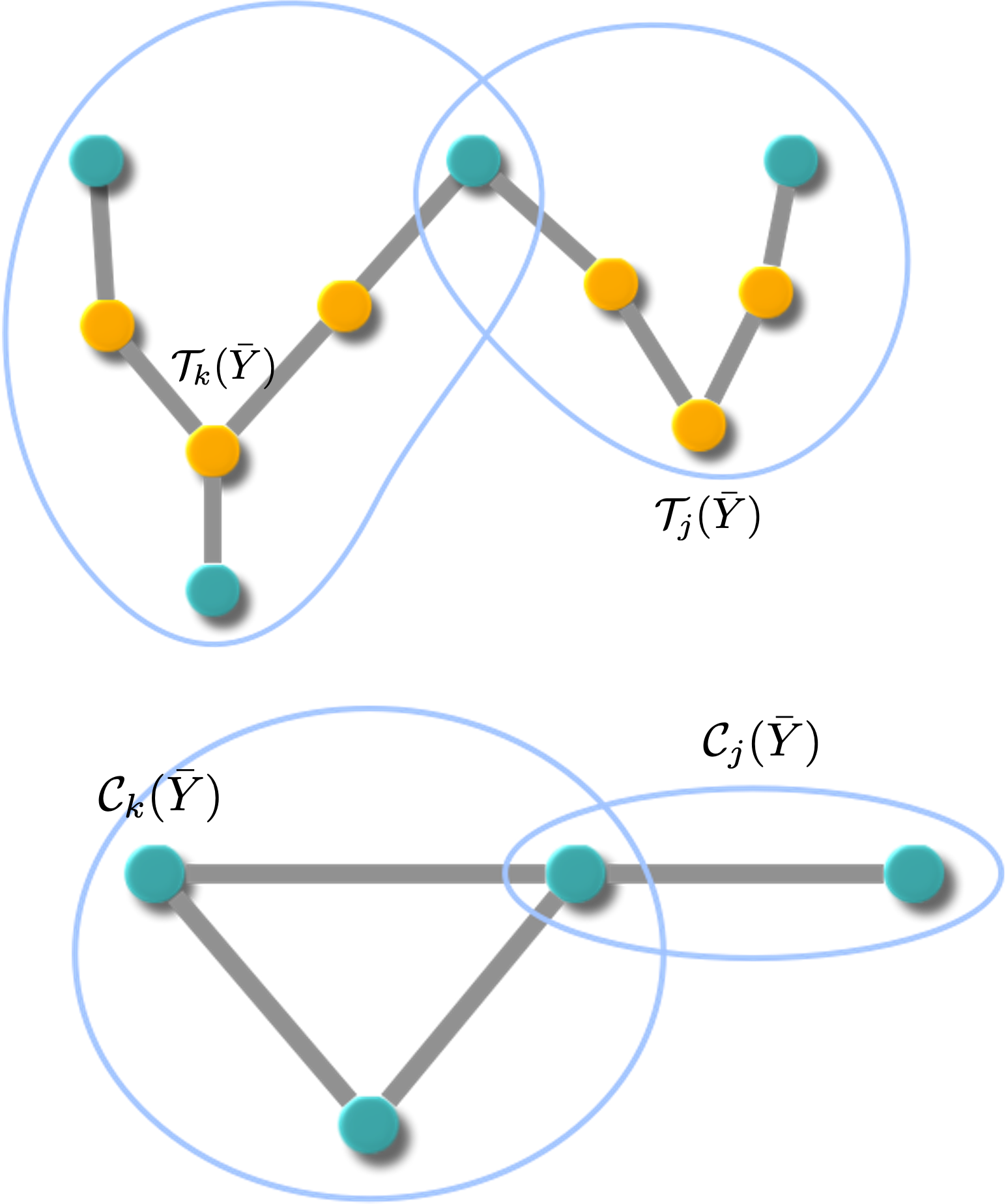}
 }
	\caption{Examples of maximal cliques in $\Bar{\mathcal G}_2$.
	Shaded nodes are measured nodes and unshaded nodes are hidden nodes.
	(These examples do not satisfy  Assumptions \ref{ass:degree2}.2).}
	\label{fig:MaximalCliques}
	\end{figure}

Consider two maximal cliques $\mathcal C_j(\bar Y)$ and $\mathcal C_k(\bar Y)$ in the graph ${\mathcal G}_2$ in Theorem~\ref{thm:separability}. The three ways they can be connected are illustrated in Figure \ref{fig:MaximalCliques}. A single graph $\mathcal G(Y)$ may contain maximal cliques spanning cases 1, 2, and 3.
%%%%%%%%%%%%%%%%%%%%%%%%%%%%%%%%%%%%%%%%%%%%%%%%%%%%%
%%%%%%%%%%%%%%%%%%%%%%%%%%%%%%%%%%%%%%%%%%%%%%%%%%%%%

\vspace{0.1in}\noindent
\textbf{Case 1: Figure \ref{fig:MaximalCliques}(a).}
$\mathcal C_j(\bar Y)$ and $\mathcal C_k(\bar Y)$ are disconnected when they are separated 
	in the original graph $\mathcal G(Y)$ by \emph{internal} observed nodes ($Y$ in equation 
	(18) assumes there are no internal observed nodes).
	The Kron-reduced matrix $\Bar Y$ takes the form ($\times$ denotes nonzero entries):
\begin{align}
\Bar Y & \ = \ \left[ \begin{array} {ccccc}
\times & \times & & \times &  \\ \cline{2-5}
\times & \multicolumn{1}{|c}{ \cellcolor[rgb]{1,0.5,0.3}{\times} } & \cellcolor[rgb]{1,0.5,0.3} \times & &  \multicolumn{1}{c|} {} \\ 
  & \multicolumn{1}{|c}{ \cellcolor[rgb]{1,0.5,0.3} \times } & \cellcolor[rgb]{1,0.5,0.3} \times & &  \multicolumn{1}{c|} {} \\ 
\times & \multicolumn{1}{|c} {} & & \cellcolor[rgb]{0, 0.6, 1} \times & \multicolumn{1}{c|}{ \cellcolor[rgb]{0, 0.6, 1} \times } \\ 
& \multicolumn{1}{|c}{} & & \cellcolor[rgb]{0, 0.6, 1} \times & \multicolumn{1}{c|} {\cellcolor[rgb]{0, 0.6, 1} \times} \\ \cline{2-5}
	\end{array} \right]
% \ =: \ 
% \begin{bmatrix} \bar Y_{11,11} & \multicolumn{4}{c} {\bar Y_{11,12}} \\
% \multirow{2}{*} { \bar{Y}_{11,12}^{ T} } & \multicolumn{2}{c}{\bar Y_{11}} &  \multicolumn{2}{c}{\bar Y_{12} } \\ 
% &  \multicolumn{2}{c}{\bar Y_{12}^{ T}} &  \multicolumn{2}{c}{Y_{22} \end{bmatrix}}
\label{eq:barY.case1}
\end{align}
where the submatrix of $\bar Y$, shaded in red and denoted by
$W_{11}$, corresponds to the maximal clique $\mathcal C_j(\bar Y)$ and the
submatrix of $\bar Y$, shaded in blue and denoted by
$W_{22}$, corresponds to $\mathcal C_k(\bar Y)$.  The first rows and columns
correspond to the internal measured nodes and their connection to the two blocks of
boundary measured nodes.
The admittance matrix $\bar Y^j$ of the maximal clique $\mathcal C_j(\bar Y)$ (i.e., if 
$\mathcal C_j(\bar Y)$ were the entire graph) is:
\begin{subequations}
\begin{align}
	\bar Y^j & \ := \ W_{11} \ - \ \diag\left( \textbf 1^{ T} W_{11} \right),
\end{align}
i.e., the diagonal entries of $W_{11}$ is modified so that $\bar Y^j$ has zero
row/column sums.  Here $\textbf 1$ denotes a vector of all 1s of appropriate size.
Similarly the admittance matrix $\bar Y^k$ of the $\mathcal C_k(\bar Y)$ 
is:
\begin{align}
	\bar Y^k & \ := \ W_{22} \ - \ \diag\left( \textbf 1^{ T} W_{22} \right).
\end{align}
\label{eq:barYjbarYk}
\end{subequations}
	
For each maximal clique $\mathcal C_i(\bar Y)$ in isolation, Section~\ref{sec:hidden} explains how to identify the admittance matrix $Y^i$ from its 
 Kron reduction $\bar Y^i$.	
The admittance matrix $Y^i$ represents the topology and line admittances
of the tree $\mathcal T_i(\bar Y)$ of hidden nodes and the observed nodes in $\mathcal C_i(\bar Y)$.   
%
%
%\begin{comment}
%For the example in 
%	Figure \ref{fig:MaximalCliques}(a), suppose the admittance matrix $Y^j$ 
%	identified from its Kron-reduced matrix $\bar Y^j$ 
%	(of the form given in equation (18) of \cite{YuanLow2022}) is:
%\begin{subequations}
%\begin{align}
%	Y^j & 
%%	\ := \ \left[ \begin{array}{c | c}
%%    Y^j_{11} & Y^j_{12} \\ \hline Y_{12}^{j  T} & Y^j_{22}
%%    \end{array} \right]   
%    \ =: \ \left[ \begin{array}{c | c c}
%      \textcolor[rgb]{1,0,0}{ Y^j_{11, 22} } & Y^j_{12, 21} & 0 
%    \\ \hline
%      & Y^j_{22, 11} & Y^j_{22, 12} 
%    \\
%      & & Y^j_{22, 22} 
%    \end{array} \right]
%\end{align}
%and the admittance matrix $Y^k$ identified from its Kron-reduced matrix $\bar Y^k$ is:
%\begin{align}
%	Y^k & 
%    \ =: \ \left[ \begin{array}{c | c c}
%      \textcolor[rgb]{0,0,1}{ Y^k_{11, 22} } & Y^k_{12, 21} & 0 
%    \\ \hline
%      & Y^k_{22, 11} & Y^k_{22, 12} 
%    \\
%      & & Y^k_{22, 22} 
%    \end{array} \right]
%\end{align}
%Then 
%\label{eq:YjYk}
%\end{subequations}
%\end{comment}
%%
% \begin{comment}
For Case 1, suppose the admittance matrices $Y^j$ and $Y^k$ identified from their Kron-reductions $\bar Y^j$ and $\bar Y^k$ respectively are:
\begin{subequations}
\begin{align}
	Y^j & 
%	\ := \ \left[ \begin{array}{c | c}
%    Y^j_{11} & Y^j_{12} \\ \hline Y_{12}^{j  T} & Y^j_{22}
%    \end{array} \right]   
    \ =: \ \left[ \begin{array}{c | c c}
      \textcolor[rgb]{1,0,0}{ Y^j_{11, 22} } & Y^j_{12, 21} & 0 
    \\ \hline
      & Y^j_{22, 11} & Y^j_{22, 12} 
    \\
      & & Y^j_{22, 22} 
    \end{array} \right]
\\
	Y^k & 
    \ =: \ \left[ \begin{array}{c | c c}
      \textcolor[rgb]{0,0,1}{ Y^k_{11, 22} } & Y^k_{12, 21} & 0 
    \\ \hline
      & Y^k_{22, 11} & Y^k_{22, 12} 
    \\
      & & Y^k_{22, 22} 
    \end{array} \right].
\end{align}
\label{eq:YjYk}
\end{subequations}

% \end{comment}
Then the admittance matrix $Y$ of the overall network is shown in \eqref{eq:overallY},
\begin{figure*}[!]
	\begin{align}
    Y  \ = \ \left[ \begin{array}{c c c | c c c c}
    Y_{11,11} & \multicolumn{2}{c|}{Y_{11, 12}} & 0 & 0 & 0 & 0
    \\ \cline{2-3}
      & \multicolumn{1}{|c}{\textcolor[rgb]{1,0,0}{ \hat Y^j_{11, 22} } } & 0 & Y^j_{12, 21} & 0 & 0 & 0
    \\
    & \multicolumn{1}{|c} 0 & \textcolor[rgb]{0,0,1}{ \hat Y^k_{11, 22} } & 0 & Y^k_{12, 21} & 0 & 0
    \\ \hline
      & & & Y^j_{22, 11} & 0 & Y^j_{22, 12} & 0  \\
      & & & 0 & Y^k_{22, 11}  & 0 & Y^k_{22, 12}  \\
      & & & & & Y^j_{22, 22} & 0 \\
      & & & & & 0 & Y^k_{22, 22}
    \end{array} \right]
\ =: \ \begin{bmatrix} Y_{11} & Y_{12} \\ Y_{12}^{ T} & Y_{22}  \end{bmatrix}.
	\label{eq:overallY}\end{align} 
\end{figure*}
%\!\!\!\!\!\!
where the top-left submatrix $Y_{11}$ of $Y$ has the same structure as $\bar Y$ in \eqref{eq:barY.case1}.
The submatrices $Y_{11,11}$ and $Y_{11,12}$ correspond to the subgraph of \emph{internal}
measured nodes and their connectivity to boundary measured nodes.  They can be identified 
directly from the given Kron-reduced admittance matrix $\bar Y$ of the overall network.
The submatrices $\hat Y^j_{11,22}$ and $\hat Y^k_{11,22}$ are obtain from $Y^j_{11,22}$,
$Y^k_{11,22}$ and $Y_{11,12}$ by modifying the diagonal entries of $Y^j_{11,22}$ and
$Y^k_{11,22}$ respectively so that $Y$ has zero row and column sums, i.e., 
% \begin{strip}
\begin{align*}
\begin{bmatrix} \multicolumn{1}{c}{\textcolor[rgb]{1,0,0}{ \hat Y^j_{11, 22} } }  & 0 \\ 
0 & \textcolor[rgb]{0,0,1}{ \hat Y^k_{11, 22} }  \end{bmatrix} & := 
\begin{bmatrix} \multicolumn{1}{c}{\textcolor[rgb]{1,0,0}{ Y^j_{11, 22} } }  & 0 \\ 
0 & \textcolor[rgb]{0,0,1}{ Y^k_{11, 22} }  \end{bmatrix}  -  
\diag\left( \textbf 1^{ T} Y_{11,12} \right).
\end{align*}
% \end{strip}
% \!\!\!\!\!\!

\vspace{0.1in}\noindent
\textbf{Case 2: Figure \ref{fig:MaximalCliques}(b).}
$\mathcal C_j(\bar Y)$ and $\mathcal C_k(\bar Y)$ are connected by a line in 
	$Y_{11, 22}$ between two boundary observed nodes.
The Kron-reduced matrix $\Bar Y$ is of the form:
	\begin{align}
\Bar Y & \ = \ \left[ \begin{array} {cccc}
\cellcolor[rgb]{1,0.5,0.3}{\times} & \cellcolor[rgb]{1,0.5,0.3} \times & &  \\ 
\cellcolor[rgb]{1,0.5,0.3} \times  & \cellcolor[rgb]{1,0.5,0.3} \times & \times &  \\ 
& \times & \cellcolor[rgb]{0, 0.6, 1} \times & \cellcolor[rgb]{0, 0.6, 1} \times \\ 
& & \cellcolor[rgb]{0, 0.6, 1} \times & \cellcolor[rgb]{0, 0.6, 1} \times \\ 
	\end{array} \right]
\ =: \ \begin{bmatrix} W_{11} & W_{12} \\ W_{12}^{ T} & W_{22} \end{bmatrix}
\label{eq:barY.case2}
\end{align}
where the submatrix $W_{11}$ shaded in red corresponds to the maximal clique 
$\mathcal C_j(\bar Y)$ and the submatrix $W_{22}$ shaded in blue corresponds 
to $\mathcal C_k(\bar Y)$.
The admittance matrices $\bar Y^j$ and $\bar Y^k$ of the maximal cliques 
$\mathcal C_j(\bar Y)$ and $\mathcal C_k(\bar Y)$ respectively are given by 
\eqref{eq:barYjbarYk}.

Suppose the admittance matrices $Y^j$ and $Y^k$
identified from their Kron-reduced matrices $\bar Y^j$ and $\bar Y^k$ respectively are
given by \eqref{eq:YjYk}.
Then the admittance matrix $Y$ of the overall network is

%\begin{strip}
	\begin{align*}
    Y & \ = \ \left[ \begin{array}{c c | c c c c}
    {\textcolor[rgb]{1,0,0}{ \hat Y^j_{11, 22} } } & W_{12} & Y^j_{12, 21} & 0 & 0 & 0
    \\
    W_{12}^{ T}  & \textcolor[rgb]{0,0,1}{ \hat Y^k_{11, 22} } & 0 & Y^k_{12, 21} & 0 & 0
    \\ \hline
    & & Y^j_{22, 11} & 0 & Y^j_{22, 12} & 0  \\
    & & 0 & Y^k_{22, 11}  & 0 & Y^k_{22, 12}  \\
    & & & & Y^j_{22, 22} & 0 \\
    & & & & 0 & Y^k_{22, 22}
    \end{array} \right]\\
&\ =: \ \begin{bmatrix} Y_{11} & Y_{12} \\ Y_{12}^{ T} & Y_{22}  \end{bmatrix},
	\end{align*} 
%\end{strip} \!\!\!\!
where the submatrix $Y_{11}$ of $Y$ has the same structure as $\bar Y$ in \eqref{eq:barY.case2}.
The submatrices $\hat Y^j_{11,22}$ and $\hat Y^k_{11,22}$ are obtain from $Y^j_{11,22}$,
$Y^k_{11,22}$ and $W_{12}$ by modifying the diagonal entries of $Y^j_{11,22}$ and
$Y^k_{11,22}$ respectively so that $Y$ has zero row and column sums, i.e., 
\begin{align*}
{\textcolor[rgb]{1,0,0}{ \hat Y^j_{11, 22} }} & \ := \
{\textcolor[rgb]{1,0,0}{ Y^j_{11, 22} } }  \, - \, \diag\left( W_{12} \textbf 1 \right)
\\
\textcolor[rgb]{0,0,1}{ \hat Y^k_{11, 22} } & \ := \ 
\textcolor[rgb]{0,0,1}{ Y^k_{11, 22} }  \, - \, \diag\left( \textbf 1^{ T} W_{12} \right).
\end{align*}
\vspace{0.1in}\noindent
\textbf{Case 3: Figure \ref{fig:MaximalCliques}(c).}
$\mathcal C_j(\bar Y)$ and $\mathcal C_k(\bar Y)$ are connected by a shared boundary observed node.
The Kron-reduced matrix $\Bar Y$ is of the form:
	\begin{align}
\Bar Y & \ = \ \left[ \begin{array} {cccc}
\cellcolor[rgb]{1,0.5,0.3}{\times} & \cellcolor[rgb]{1,0.5,0.3} \times &  \cellcolor[rgb]{1,0.5,0.3} \times  &  \\ 
\cellcolor[rgb]{1,0.5,0.3} \times  & \cellcolor[rgb]{1,0.5,0.3} \times &  \cellcolor[rgb]{1,0.5,0.3} \times  &  \\ 
 \cellcolor[rgb]{1,0.5,0.3} \times &  \cellcolor[rgb]{1,0.5,0.3} \times  & \cellcolor[rgb]{0.5, 0.6, 0.6} \times & \cellcolor[rgb]{0, 0.6, 1} \times \\ 
& & \cellcolor[rgb]{0, 0.6, 1} \times & \cellcolor[rgb]{0, 0.6, 1} \times \\ 
	\end{array} \right]
\label{eq:barY.case3}
\end{align}
where the submatrix $W_{11}$ shaded in red corresponds to the maximal clique 
$\mathcal C_j(\bar Y)$ and the submatrix $W_{22}$ shaded in blue corresponds 
to $\mathcal C_k(\bar Y)$.
The admittance matrices $\bar Y^j$ and $\bar Y^k$ of the maximal cliques 
$\mathcal C_j(\bar Y)$ and $\mathcal C_k(\bar Y)$ respectively are given by 
\eqref{eq:barYjbarYk}.
		
The two maximal cliques $\mathcal C_j(\bar Y)$ and $\mathcal C_k(\bar Y)$ are
connected by a single shared node.  Suppose without loss of generality that the
last row/column of $\bar Y^j$ and the first row/column of $\bar Y^k$ correspond 
to the shared node.
Suppose the admittance matrices $Y^j$ and $Y^k$
identified from $\bar Y^j$ and $\bar Y^k$ respectively are
\begin{align*}
	Y^j & 
    \ =: \ \left[ \begin{array}{c c | c c}
      \textcolor[rgb]{1,0,0}{ \hat Y^j_{11, 22} } &  \textcolor[rgb]{1,0,0}{ \hat r^{j  T}_{11, 22} } 
      &  Y^j_{12, 21} & 0 \\
      \textcolor[rgb]{1,0,0}{ \hat r^{j}_{11, 22} } & \textcolor[rgb]{1,0,0}{ d^j_{11,22} }
      & r^j_{12, 21} & 0
    \\ \hline
      & & Y^j_{22, 11} & Y^j_{22, 12} 
    \\
      & & & Y^j_{22, 22} 
    \end{array} \right]
\\
	Y^k & 
    \ =: \ \left[ \begin{array}{c c | c c}
    \textcolor[rgb]{0,0,1}{ d^k_{11,22} } & \textcolor[rgb]{0,0, 1}{ \hat r^k_{11, 22} } &  r^k_{12, 21} & 0 \\
     \textcolor[rgb]{0,0,1}{ \hat r^{k  T}_{11, 22} } & \textcolor[rgb]{0,0,1}{ \hat Y^{k}_{11, 22} } &Y^k_{12, 21} & 0
    \\ \hline
     & & Y^k_{22, 11} & Y^k_{22, 12} 
    \\
     & & & Y^k_{22, 22} 
    \end{array} \right],
\end{align*}
where $(\hat r^j_{11,22}, d^j_{11,22}, r^j_{12, 21})$ and 
$(\hat r^k_{11,22}, d^k_{11,22}, r^k_{12, 21})$ 
are the rows corresponding to the shared node.
For the example in Figure \ref{fig:MaximalCliques}(c),
$\hat Y^j_{11,22}$ is $2\times 2$ and $\hat r^j_{11,22}$ is $1\times 2$;
both $\hat Y^k_{11,22}$ and $\hat r^k_{11,22}$ are $1\times 1$.

To construct the admittance matrix $Y$ of the overall network, the rows in $Y^j$ and $Y^k$ 
corresponding to this shared node is combined into a single row:
\begin{align*}
\left[ \begin{array}{ c c c | c c }
\textcolor[rgb]{1,0,0}{ \hat r^{j}_{11, 22} } & 
\textcolor[rgb]{1,0,0}{ d^j_{11,22} } +  \textcolor[rgb]{0,0,1}{ d^k_{11,22} } &
\textcolor[rgb]{0,0,1}{ \hat r^{k}_{11, 22} } & r^j_{12, 21} & r^k_{12, 21} 
\end{array} \right]
\end{align*}
The admittance matrix $Y$  is then
\begin{strip}
	\begin{align*}
    Y & \ = \ \left[ \begin{array}{c c c | c c c c}
    \textcolor[rgb]{1,0,0}{ \hat Y^j_{11, 22} } &  \textcolor[rgb]{1,0,0}{ \hat r^{j  T}_{11, 22} } 
 & 0 & Y^j_{12, 21} & 0 & 0 & 0
    \\
\textcolor[rgb]{1,0,0}{ \hat r^{j}_{11, 22} } & 
\textcolor[rgb]{1,0,0}{ d^j_{11,22} } +  \textcolor[rgb]{0,0,1}{ d^k_{11,22} } &
\textcolor[rgb]{0,0,1}{ \hat r^{k}_{11, 22} } & r^j_{12, 21} &  r^k_{12, 21} & 0 & 0
\\
    0 & \textcolor[rgb]{0,0,1}{ \hat r^{k  T}_{11,22} } &  \textcolor[rgb]{0,0,1}{ \hat Y^k_{11, 22} } &  
    0 & Y^k_{12, 21} & 0 & 0
    \\ \hline
    & & & Y^j_{22, 11} & 0 & Y^j_{22, 12} & 0  \\
    & & & 0 & Y^k_{22, 11}  & 0 & Y^k_{22, 12}  \\
    & & & & & Y^j_{22, 22} & 0 \\
    & & & & & 0 & Y^k_{22, 22}
    \end{array} \right]
\ =: \ \begin{bmatrix} Y_{11} & Y_{12} \\ Y_{12}^{ T} & Y_{22}  \end{bmatrix},
	\end{align*} 
\end{strip} \!\!\!\!
where submatrix $Y_{11}$ of $Y$ has the same structure as $\bar Y$ in \eqref{eq:barY.case3}.

\bibliographystyle{IEEEtran}
\bibliography{thebib}
% \clearpage
\begin{IEEEbiography}
[{\includegraphics[height=1.25in,clip,keepaspectratio]{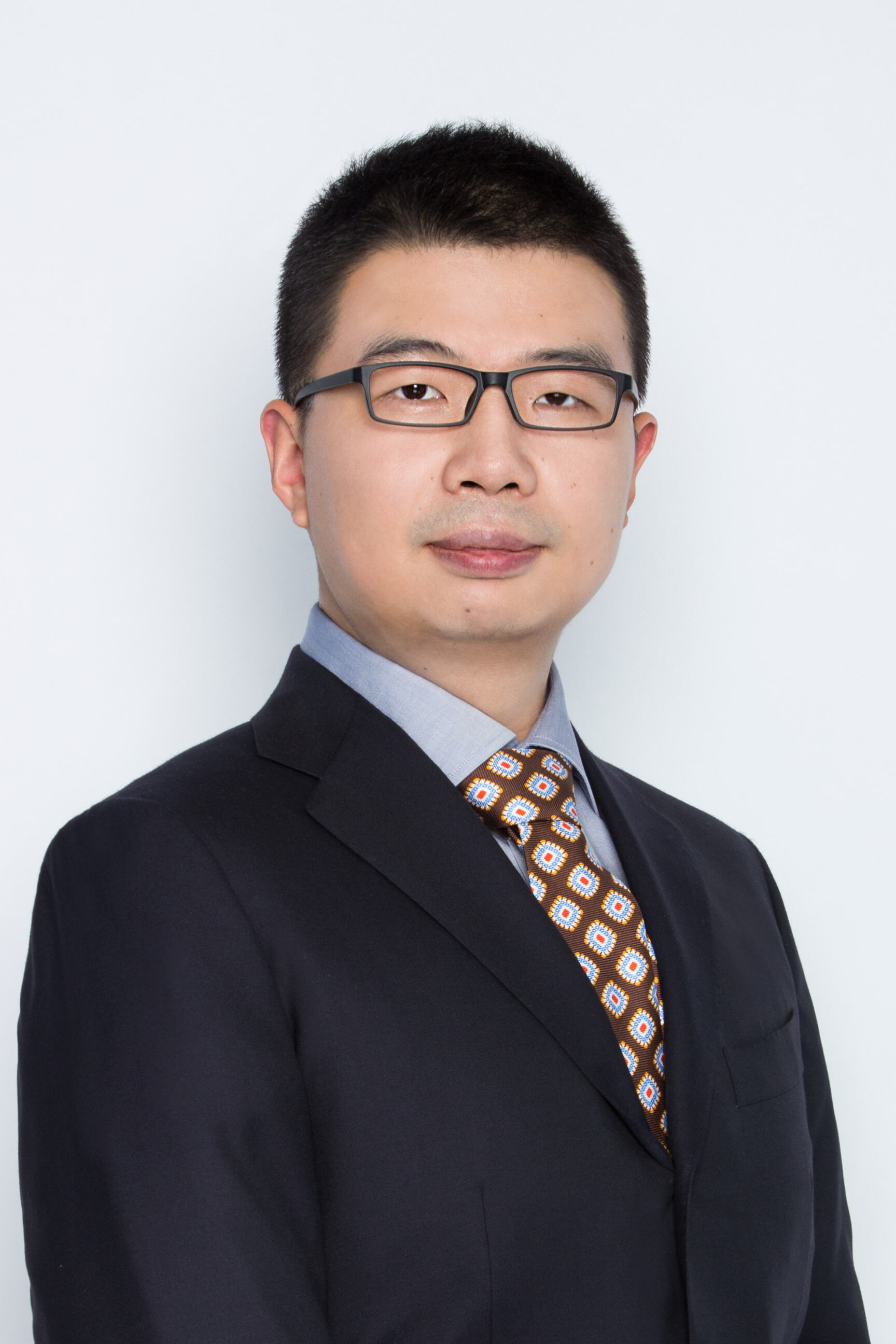}}]
{Ye Yuan}(M'13, SM'20) received the B.Eng. degree from the Department of Automation, Shanghai Jiao Tong University, Shanghai, China, in 2008, and the M.Phil. and Ph.D. degrees from the Department of Engineering, University of Cambridge, Cambridge, U.K., in 2009 and 2012, respectively. He has been a Full Professor at the Huazhong University of Science and Technology, Wuhan, China since 2016. Prior to that, he was a Postdoctoral Researcher at UC Berkeley, a Junior Research Fellow at Darwin College, University of Cambridge. His research interests include system identification and optimization with applications to cyber-physical systems. 
\end{IEEEbiography}

\begin{IEEEbiography}
[{\includegraphics[height=1.25in,clip,keepaspectratio]{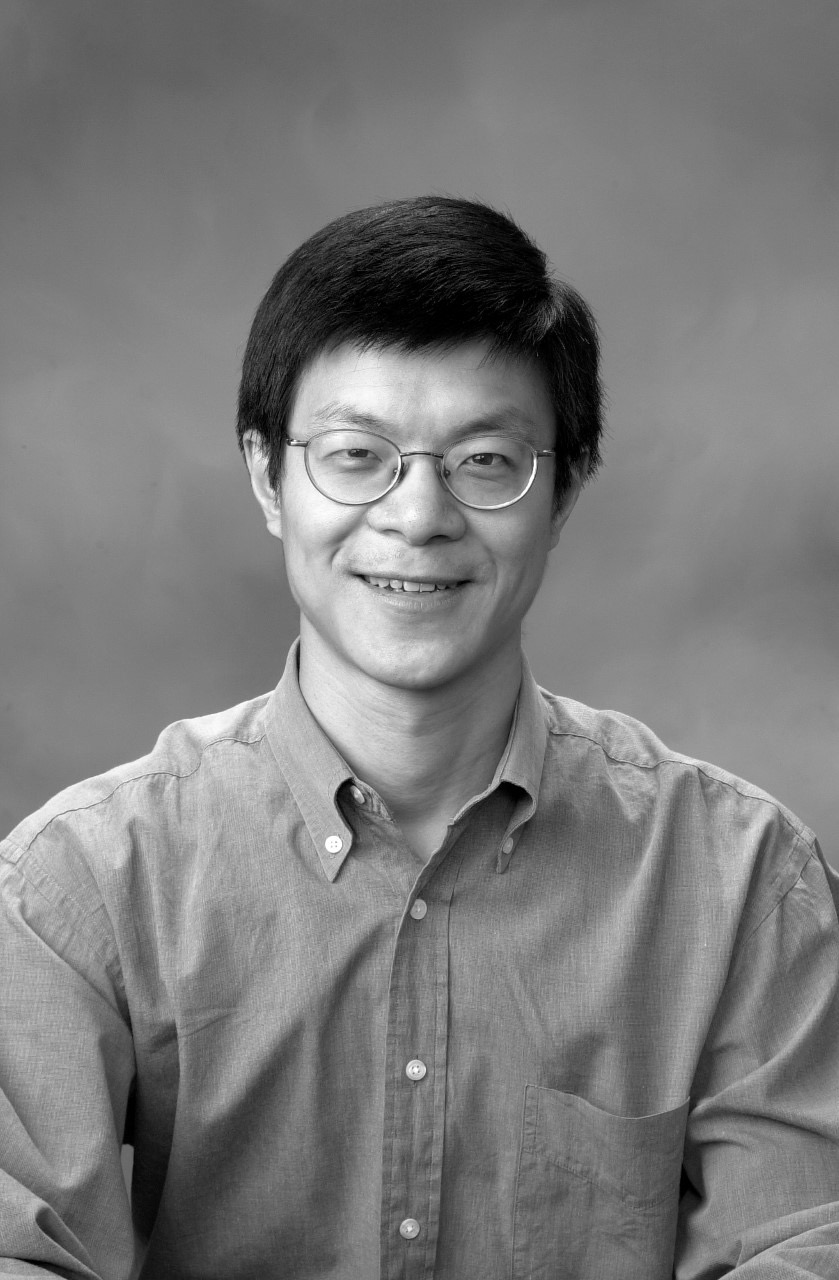}}]
%{Steven Low}(F'08) is the Gilloon Professor at Caltech and an Honorary Professor at the University of Melbourne, Australia. He was an awardee of the IEEE INFOCOM Achievement Award and the ACM SIGMETRICS Test of Time Award, and is a Fellow of IEEE, ACM, and CSEE.  He received his B.S. from Cornell and PhD from Berkeley, both in EE. 
{Steven Low} (F'08) is the F. J. Gilloon Professor of the Department of Computing \& Mathematical Sciences and the Department of Electrical Engineering at Caltech and an Honorary Professor of the University of Melbourne.  Before that, he was with AT\&T Bell Laboratories, Murray Hill, NJ, and the University of Melbourne, Australia.  He has held honorary/chaired professorship in Australia, China and Taiwan.  He was a co-recipient of IEEE best paper awards, an awardee of the IEEE INFOCOM Achievement Award and the ACM SIGMETRICS Test of Time Award, and is a Fellow of IEEE, ACM, and CSEE.  He was well-known for work on Internet congestion control and semidefinite relaxation of optimal power flow problems in smart grid.  His research on networks has been accelerating more than 1TB of Internet traffic every second since 2014.  His research on smart grid is providing large-scale cost effective electric vehicle charging to workplaces.  He received his B.S. from Cornell and PhD from Berkeley, both in EE.
\end{IEEEbiography}

\begin{IEEEbiography}
[{\includegraphics[height=1.25in,clip,keepaspectratio]{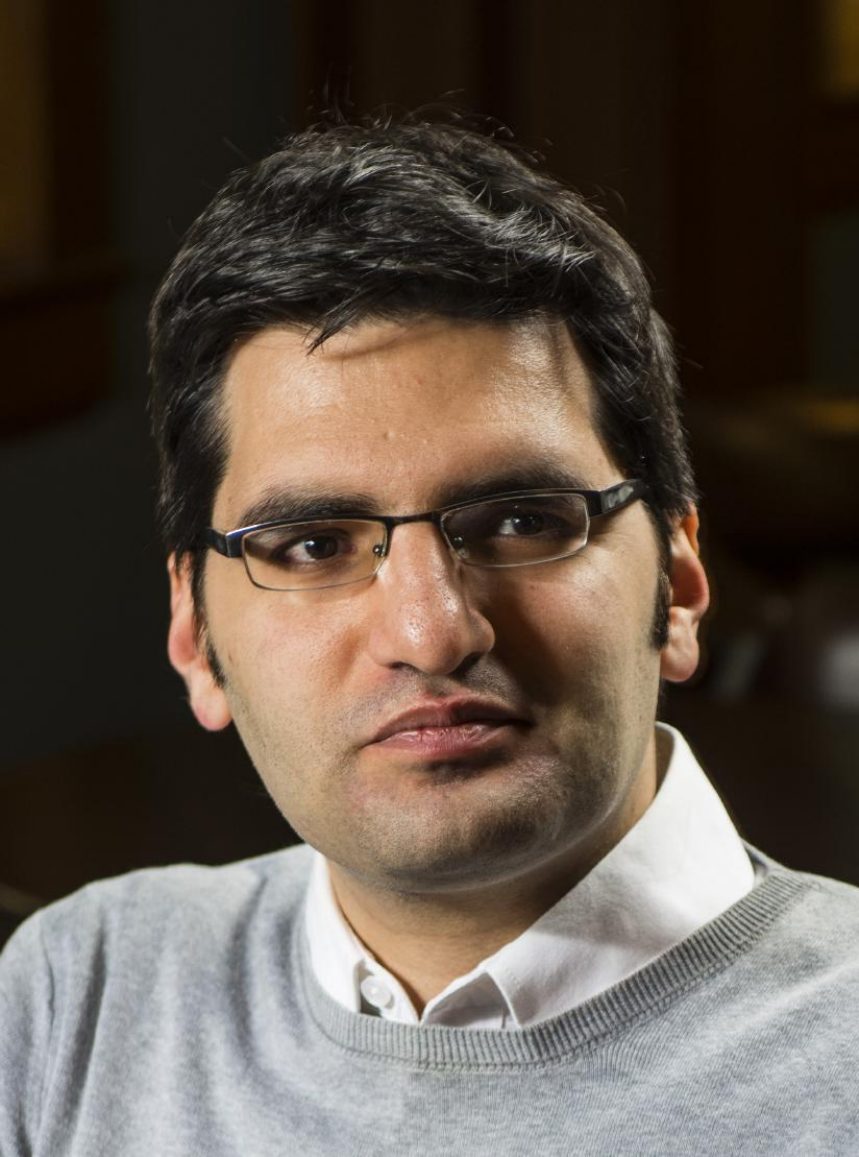}}]
{Omid Ardakanian}(M'15) is an Assistant Professor at the University of Alberta, Canada. He received his B.Sc. from Sharif University of Technology in 2009, and M.Math. and Ph.D. from the University of Waterloo in 2011 and 2015. 
From 2015 to 2017, he was an NSERC Postdoctoral Fellow at UC Berkeley and the University of British Columbia.
He received best paper awards at ACM e-Energy, ACM BuildSys, and IEEE PES General Meeting. 
His research focuses on the design and implementation of intelligent networked systems.
\end{IEEEbiography}

\begin{IEEEbiography}
[{\includegraphics[height=1.25in,clip,keepaspectratio]{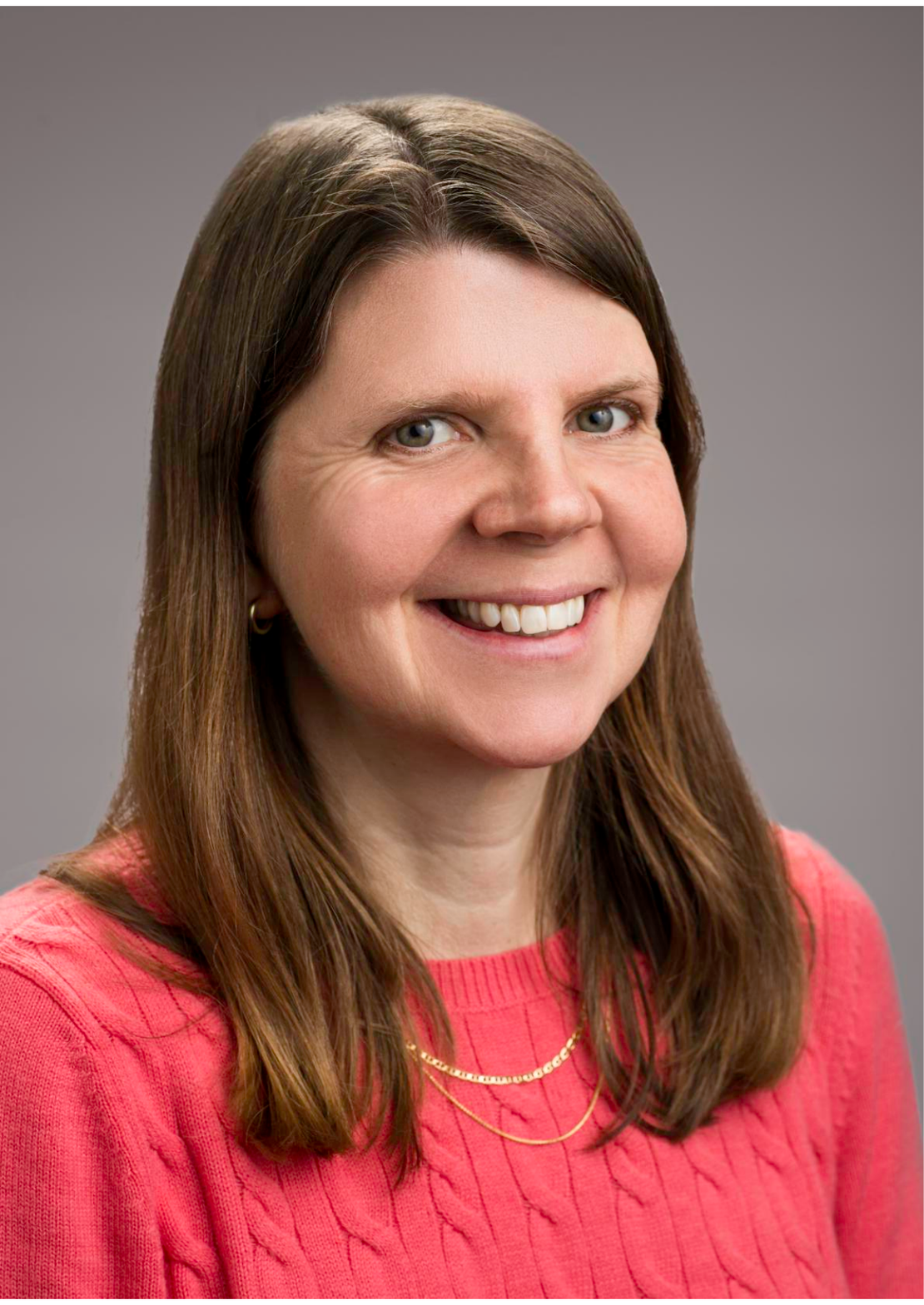}}]
{Claire J. Tomlin}(F'10) is the Charles A. Desoer
Professor of Engineering in the Department of
Electrical Engineering and Computer Sciences
(EECS), University of California Berkeley (UC
Berkeley). She was an Assistant, Associate, and
Full Professor in Aeronautics and Astronautics
at Stanford University from 1998 to 2007, and in
2005, she joined UC Berkeley. Claire works in
the area of control theory and hybrid systems,
with applications to air traffic management, UAV
systems, energy, robotics, and systems biology.
She is a MacArthur Foundation Fellow (2006), an IEEE Fellow (2010),
and in 2017, she was awarded the IEEE Transportation Technologies
Award. In 2019, Claire was elected to the National Academy of Engineering and the American Academy of Arts and Sciences.\end{IEEEbiography}
\end{document}